\documentclass[%
 reprint,
superscriptaddress,
 amsmath,amssymb,
 aps,
]{revtex4-2}

\usepackage{graphicx}
\usepackage{dcolumn}
\usepackage{bm}
\usepackage[version=3]{mhchem} 
\usepackage{xcolor} 

 \newcommand{\Nadd}[1]{\textcolor{black}{#1}}
 \usepackage{ulem}

\begin{document}

\preprint{APS/123-QED}

\title{
Multi-plateau high-harmonic generation in liquids driven by off-site recombination
}

\author{Angana Mondal}
\affiliation{Laboratorium für Physikalische Chemie, ETH Zürich, Zurich, Switzerland}
\affiliation{These authors contributed equally}
\author{Ofer Neufeld}
\email[]{ofern@technion.ac.il}
\affiliation{Technion Israel Institute of Technology, Faculty of Chemistry, Haifa 3200003, Israel}
\affiliation{Max Planck Institute for the Structure and Dynamics of Matter and Center for Free-Electron Laser Science, Luruper Chaussee 149, 22761 Hamburg, Germany}
\affiliation{These authors contributed equally}
\author{Tadas Balciunas}
\affiliation{Laboratorium für Physikalische Chemie, ETH Zürich, Zurich, Switzerland}
\author{Benedikt Waser}
\affiliation{Laboratorium für Physikalische Chemie, ETH Zürich, Zurich, Switzerland}
\author{Serge Müller}
\affiliation{Laboratorium für Physikalische Chemie, ETH Zürich, Zurich, Switzerland}
\author{Mariana Rossi}
\affiliation{Max Planck Institute for the Structure and Dynamics of Matter and Center for Free-Electron Laser Science, Luruper Chaussee 149, 22761 Hamburg, Germany}
\author{Zhong Yin}
\affiliation{Laboratorium für Physikalische Chemie, ETH Zürich, Zurich, Switzerland}
\affiliation{International Center for Synchrotron Radiation Innovation Smart, Tohoku University, Sendai, Japan}
\author{Angel Rubio}
\affiliation{Max Planck Institute for the Structure and Dynamics of Matter and Center for Free-Electron Laser Science, Luruper Chaussee 149, 22761 Hamburg, Germany}
\affiliation{Center for Computational Quantum Physics (CCQ), The Flatiron Institute, 162 Fifth Avenue, New York NY 10010, USA}
\author{Nicolas Tancogne-Dejean}
\email[]{nicolas.tancogne-dejean@mpsd.mpg.de}
\affiliation{Max Planck Institute for the Structure and Dynamics of Matter and Center for Free-Electron Laser Science, Luruper Chaussee 149, 22761 Hamburg, Germany}
\author{Hans Jakob Wörner}
\email[]{hwoerner@ethz.ch}
\affiliation{Laboratorium für Physikalische Chemie, ETH Zürich, Zurich, Switzerland}

\begin{abstract}
Non-perturbative high-harmonic generation (HHG) has recently been observed in the liquid phase, where it was demonstrated to have a different physical mechanism compared to gas and solid phases of matter. The currently best physical picture for liquid HHG eliminates scattered-electron contributions and identifies on-site recombination as the dominant contributor. This mechanism accurately predicts the cut-off energy and its independence of the driving laser wavelength and intensity. However, this implies that additional energy absorbed in the liquid as the driving laser intensity is increased does not result in higher-order non-linearities, which is in contrast to the conventional expectation from most nonlinear media. Here we experimentally observe the formation of a second plateau in HHG from multiple liquids (water, heavy water, propranol, and ethanol), thus explaining the conundrum of the missing higher-order response. We analyze this second plateau with a combination of experimental, state-of-the-art ab-initio numerical (in diverse systems of water, ammonia, and liquid methane), and semi-classical analytical, techniques, and elucidate its physical origin to electrons that recombine on neighboring water molecules rather than at the ionization site, leading to unique HHG ellipticity dependence. Remarkably, we find that the second plateau is dominated by electrons recombining at the second solvation shell, relying on wide hole delocalization. Theory also predicts the appearance of even higher plateaus, indicating a general trend. Our work establishes new physical phenomena in the highly non-linear optical response of liquids, paving the way to attosecond probing of electron dynamics in solutions.   
\end{abstract}

\maketitle


\section{Introduction}

High harmonic generation (HHG) is a non-perturbative nonlinear optical process that occurs when matter is irradiated by intense laser pulses. The phenomenon was initially discovered in noble gases in the 80's \cite{McPherson1987,Ferray1988}, followed by its solid-state version in 2011 \cite{Ghimire2011a}, liquid media in 2018 \cite{luu18b}, and has since been widely studied and explored in a plethora of systems and conditions \cite{paul01a,Popmintchev2012,vampa2015,Luu2018,mondal2023, Schmid2021, Bauer2018, Theidel2024, Jordi2022, Uchida2022, Lv2021,Neufeld2023,Yang2019,Orthodoxou2021, HabiboviA2024}.
In particular, in the gas phase, HHG is well understood as arising from a sequential multi-step mechanistic process in which electrons are: (i) tunnel-ionized from their parent molecule, (ii) accelerated in the continuum by the intense laser, and (iii) coherently recombine with the parent molecule to emit high-energy radiation \cite{Corkum1993, Lewenstein1994}. A hallmark of this simple physical picture is that it correctly predicts the main spectral and temporal features of the emitted spectrum, especially its cutoff - the energy from which emitted harmonic photon yields begin to exponentially decay - which scales parabolically with the driving laser amplitude and wavelength. This physical interpretation has paved way to multiple applications ranging from ultrafast spectroscopy \cite{Corkum2007,Calegari2016, Kraus2015, Pertot2017,  yin2023, wang2022stilbene, Kneller2024}, coherent XUV pulse shaping \cite{yang2021strong, wituschek2020tracking}, and attosecond pulse generation winning the 2023 Physics Nobel prize \cite{Hentschel2001, zhao12a, Gaumnitz2017}. 

More recently, HHG has been established and extensively explored in the solid state \cite{Ghimire2011, luu2015, vampa2015, ghimire2019, Goulielmakis2022, Yue2022},
where an equivalent multi-step picture can be formulated in \textit{k}-space after applying the Bloch theorem \cite{Vampa2015phys, Wu2016, Yue2020, Osika2017}. The \textit{k}-space trajectory picture predicts a cutoff energy that scales linearly with both the driving wavelength and field amplitude \cite{Ghimire2011}. In principle, also real-space trajectories can be formulated in solids\cite{You2017v2,PhysRevA.100.043420,PhysRevResearch.6.043005,Brown24,Parks20}, which include also rescattering channels and recombination on various lattice sites\cite{You2018,wang2020role,neighboursolid}. However, due to the wide delocalization of the Bloch states in real-space that is not always a helpful picture\cite{Li2019,li2021huygens}, e.g. requiring recombination of electrons and holes many unit cells apart\cite{Yue2020}. 
Solids were also shown to produce multiple plateau structures arising from their multi-band nature \cite{Luu2018berry,Uzan2020b, Wu2016, Yu2020,Ndabashimiye2016,you17b,You2017,PhysRevA.95.043416}%
 (just as multiple plateaus can appear in gas-phase HHG if several molecular orbitals are involved) \cite{Luppi2023,Morassut2024, Neufeld2024}.
Quite generally, HHG in both solids and gases follows a universal physical principle: when progressively increasing the intensity while driving a nonlinear medium, a higher-order nonlinear response should appear as long as the system has not entered a saturation regime \cite{Rae1994,Rae1993,Constant1999} 
(i.e. if the ionization rate has not surpassed a certain threshold per molecule, typically on the order of a single electron, or reached the damage threshold where chemical bonds start breaking up \cite{Schouder2022}). 
 This principle should be respected regardless of the specific details of the medium, or even the nature of the interaction and its physical mechanism. Indeed, even in some unique cases such as Cooper minima\cite{Farrell2011} that briefly cause a suppressed response, stronger laser driving eventually leads to appearance of higher harmonic orders.

In contrast, non-perturbative HHG in the liquid phase has been only very recently measured \cite{Luu2018} and successfully theoretically explained as arising from a similar in nature trajectory-based picture, but with electron-molecule scattering playing a dominant role in the dynamics \cite{mondal2023}. It was shown that due to scattering, the cutoff energy is effectively independent of the driving wavelength and intensity \cite{Zheng2020,mondal2023,Ofer2022}, and should depend on the radial distribution of the molecular structure \cite{Li2024}
(also leading to a density dependence \cite{mondal2023}). Note that very similar physics is expected in amorphous solids\cite{you17b} that lack long-range translational symmetry, and are not fully mechanistically understood as HHG sources (with the main difference being that liquids also exhibit strong structural dynamics). This result is in contradiction with the general physical expectation, as the driving parameters are far from the saturation limit, and re-absorption does not change the main HHG spectral features. Thus, it remains unclear where the missing high-order response is, or why it is missing, as the laser intensity and wavelengths are increased.

Here, we experimentally measure HHG in a variety of liquids using methodologies that provide an enhanced signal-to-noise ratio and dynamical range compared to prior measurements \cite{Luu2018,mondal2023,Alexander2023}.
We observe for the first time the emergence of a second plateau in liquid HHG, which accounts for the previously unobserved higher-order response. To support experimental findings, we performed state-of-the-art \textit{ab-initio} calculations and established that it is a robust and generic feature of liquid-phase HHG. We formulate an extended electron-trajectory picture for the laser-driven ultrafast dynamics that qualitatively captures the main features of the second plateau - its spectral range, cutoff scaling, and temporal characteristics - and uncovers the physical origin as a contribution from electrons that recombine with neighboring molecules. Nearest-neighbor recombination trajectories are shown to be the dominant contributor to the emission window in-between the two plateaus, while the next-nearest-neighbor (i.e. electrons recombining at the second solvation shell- shown in Fig. 1(a) inset) dominate the second plateau itself. This mechanism leads to a second plateau cutoff that scales weakly with the laser wavelength and intensity (similar to the first plateau), but also exhibits a characteristic anomalous dependence of the harmonic yields on the driving laser ellipticity, which we confirm with experiments and theory. \textit{Ab-initio} simulations predict yet higher-order plateaus that we suspect are connected to higher-order recombination-scattering processes involving further molecular sites (which currently are not resolvable in the experiment). These HHG mechanisms are different in nature compared to those in both gas and solid phases, thus paving the route towards ultrafast spectroscopy techniques for electron dynamics in solution. In particular, we anticipate that these mechanisms can be exploited to retrieve effective intermolecular separations, probe the delocalization of electronic wavefunctions in amorphous media \cite{gong2022attosecond} and the hybridization of solute and solvent electronic wavefunctions, as well as their attosecond dynamics.

\section{Results}
\begin{figure*} [t!]
\includegraphics[width=\textwidth]{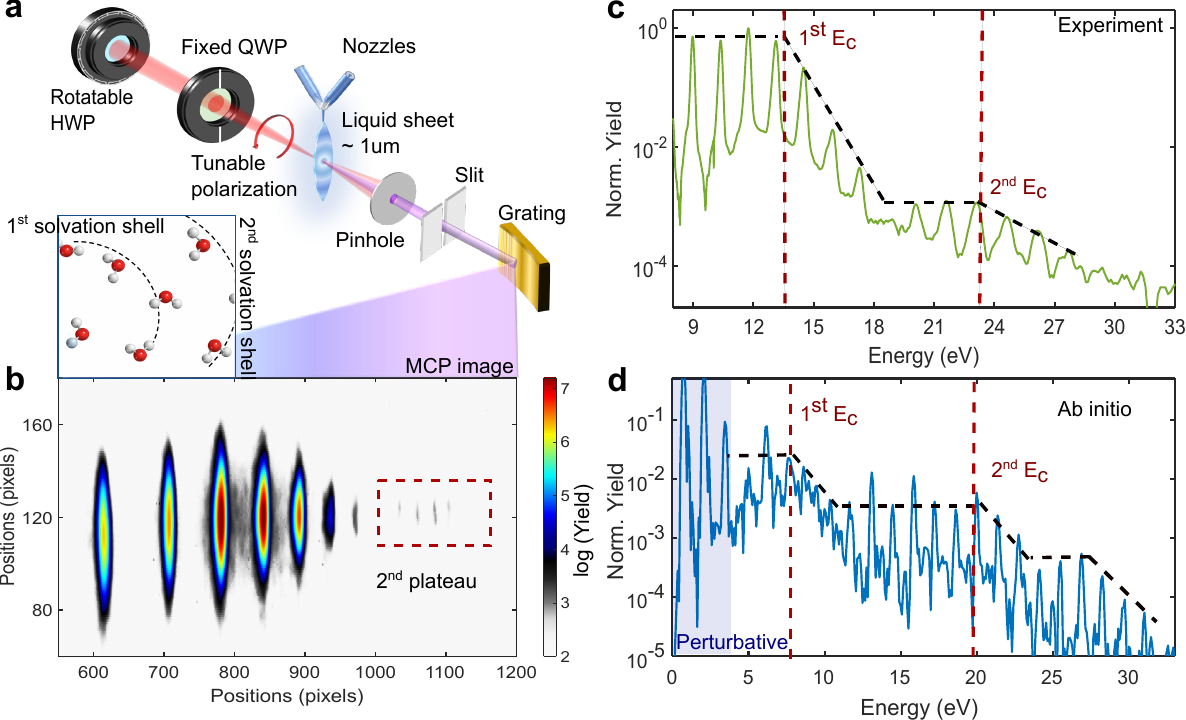}
\caption{
{\bf Experimental setup for liquid-phase HHG and observation of second plateaus in experiment and theory~}
{\textbf{(a)} Schematic of the experimental setup. Laser pulses with central tunable wavelength (here 1800 nm) are focused on the flat liquid jet to generate high harmonics. The generated high harmonics pass through a slit onto a grating that disperses different harmonic orders, which are then detected using an multi-channel plate (MCP) coupled to a phosphor screen and a camera. The inset shows a schematic of the first and second solvation shells marked by the dashed black arcs.}
\textbf{(b)}
Raw MCP image of high-harmonic spectrum observed from D$_2$O liquid acquired at 4.8 $\times$ 10$^{13}$ W/cm$^2$, showing the presence of a second plateau feature. 
\textbf{(c)} 
Experimentally observed high-harmonic spectrum from D$_2$O liquid acquired at 4.8 $\times$ 10$^{13}$ W/cm$^2$. In addition to a first plateau, a second plateau is observed from 19 eV, with a cut-off energy (E$_c$) at $\sim$ 23 eV.
\textbf{(d)} 
Harmonic spectrum of liquid H$_2$O, obtained from ab initio simulations~\cite{neufeld2022} at 1800 nm and 5 $\times$ 10$^{13}$ W/cm$^2$ (with intensity averaging), showing a similar second plateau feature with the second cut-off energy, E$_c$, at $\sim$ 21 eV. The cut-off energies represented by the red dashed lines are assigned at the start of the exponentially decaying harmonic yield region within an error bar of one visible harmonic.}
\end{figure*}

HHG was measured from a liquid flat jet of $\sim$ 1 $\mu$m thickness generated by two colliding cylindrical jets of $\sim$ 50 $\mu$m diameter \cite{yin2020} (see Fig. 1a). The thin flat jets not only allow us to directly measure the bulk-liquid response and to separate contributions to HHG coming from evaporated molecules near the liquid surface (see Methods section for further details), but also to acquire data with high signal-to-noise ratio. HHG was driven in the liquid flat jet with a 1 kHz, Ti:Sa laser coupled with an optical parametric amplifier (HE-Topas), providing tunability in the laser amplitude, wavelength, and polarization state. Harmonics were measured with an XUV spectrometer consisting of a grating and MCP-based detector. All further technical details of the experimental apparatus can be found in the methods section. 

Figure 1(b,c) presents our main experimental observation - in HHG from liquid D$_2$O under ambient conditions \cite{chang2022, buttersack2022}, we observe two distinct HHG plateaus. Similar results have been obtained for H$_2$O and alcohols (see Ext. Data Fig. 1), also at other wavelengths (See Ext. Data Fig. 2). The second plateau appears at two to three orders of magnitude diminished yield compared to the first plateau, and about 5 eV beyond the cutoff energy of the first plateau. In between the plateaus, we observe a $\sim$5 eV-wide exponential-decay region which we refer to as the inter-plateau transition. We note that the cutoff energy of the first plateau is fully consistent with previous measurements and theory \cite{Luu2018, svoboda2021polarization, mondal2023, mondal2023few, neufeld2022}, and is roughly independent of the driving wavelength and intensity. Thus, in the absence of appropriate signal-to-noise contrast, upon driving the medium harder one would get the appearance of a saturated cutoff energy. Only when discerning signals that are about $\sim$500 times weaker is the missing higher-order response discerned. 

\begin{figure*} [htbp!]
\includegraphics[width=\textwidth]{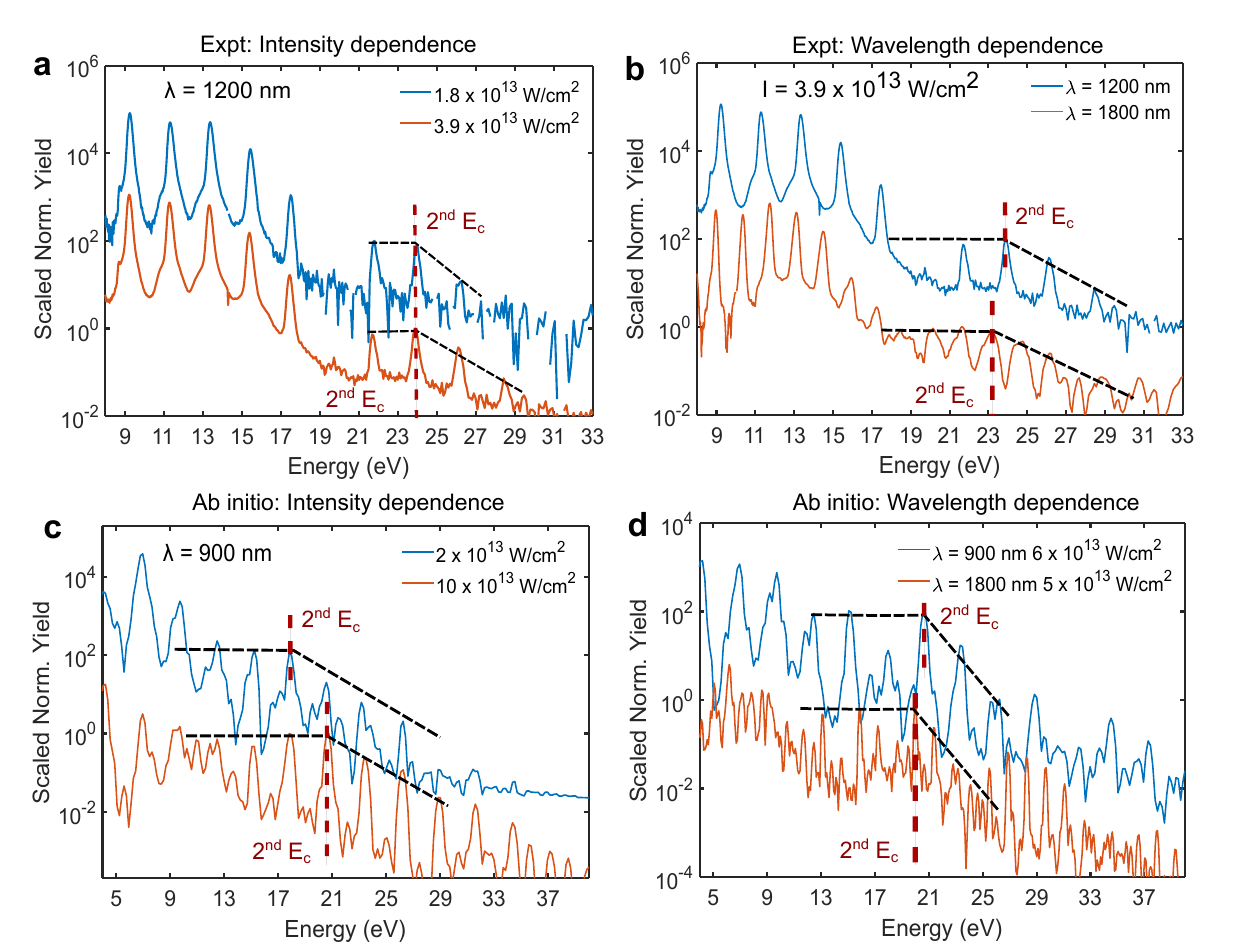}
\caption{
{\bf Intensity and wavelength dependence of the second plateau in HHG from liquids~}
{\textbf{(a)} Experimental HHG measured at varying driving intensity at 1200 nm.\textbf{(b)} Experimental liquid D$_2$O HHG spectrum measured at 1800 nm and 1200 nm at similar driving peak intensity. \textbf{(c)} Ab initio simulations intensity dependent HHG driven at 900 nm in liquid H$_2$O. \textbf{(d)}  \textit{Ab-initio} HHG simulations for 900 nm and 1800 nm wavelength for liquid H$_2$O at the same driving intensity. This data establishes a very weak wavelength and intensity dependence of the second plateau cut-off energy both in experiments and theory. All spectra have been normalized with respect to the maximum within the 18 eV to 35 eV range. Normalized spectra in of lower intensity and lower wavelengths, respectively, have been artificially shifted upwards for visualization purposes. For detailed  dateset please refer to Fig. S8 of SM.}
}
\end{figure*}

Figure 2(a,b) presents further experimental investigations tuning the driving wavelength and intensity, showing that the cutoff of the second plateau is weakly dependent on both, similar to the first plateau characteristics (suggesting a potentially similar in nature mechanism). We also note that an identical nonlinear response exists in liquid water (H$_2$O); however, we have found that D$_2$O provides cleaner spectra that allow a better analysis of the second plateau (most likely due to its higher viscosity \cite{d2oviscosity} and isotopic effects \cite{Baker2006}). For further details on the determination of the second plateau, the second cut-off energy and its associated uncertainty, refer to SM section S7).\\

At this stage, we turn to \textit{ab-initio} simulations of the liquid HHG response. We calculated the liquid HHG spectra relying on time-dependent density functional theory (TDDFT) \cite{marques2012fundamentals} using finite clusters that were previously shown useful \cite{neufeld2022} within the Octopus code\cite{Tancogne-Dejean2020}. 
Note that this theoretical approach assumes that the nuclei remain frozen during their interaction with the laser pulse (which should be valid on femtosecond timescales). This renders \textit{ab-initio} simulations for H$_{2}O$ and D$_{2}O$ essentially equivalent at the microscopic level, which is why \textit{ab-initio} data is only labeled H$_{2}O$ from this point on even when comparing to experiments in D$_{2}O$. All further technical details are given in the Methods Section and Supplementary Material (SM). Figure 1(d) presents exemplary numerical results in similar conditions to the experiment in Fig. 1(c), showing a second HHG plateau appearing with similar attributes and energy scales. In particular, the average yield separation between plateaus and the inter-plateau width are consistent between theory and experiment. Practically, this allows us to rule out macroscopic effects as a possible origin for the second plateau (as has been shown for the first plateau physics \cite{mondal2023}), since the theory is purely microscopic in nature. Further, \textit{ab-initio} calculations agree also with the measured weak scaling properties of the second plateau cutoff with driving wavelength and intensity (Fig. 2(c,d)). These conclusions hold also in several different liquid media and conditions, which have been tested both experimentally in ethanol and isopropanol (Ext. Data Figs. 1 and 2), as well as numerically in liquid methane (Fig. 3b) and ammonia \Nadd{(see SM, Section S4)}. 

Remarkably, theoretical calculations also predict the appearance of an even higher-order third plateau (see Fig. 1(d)), which arises at yet higher photon energies, suggesting a certain universality in the physical mechanism whereby the order of the nonlinear response appears to saturate, but in fact higher harmonics are emitted in exponentially suppressed higher plateaus. However, while \textit{ab-initio} simulations predict this feature, in practice its observation may be hindered by strong dephasing and scattering effects in liquids, which are not fully accounted for in the simulations and are expected to further suppress coherent emission from long trajectories. Consequently, the third plateau yield likely falls below the current signal-to-noise capabilities of our experimental setup. Potential avenues for enhancing detectability include spectrally blocking intense lower-order harmonics to optimize the detector’s dynamic range, using higher repetition-rate sources to improve statistics, and reducing background contributions through helium-assisted jet confinement.

Having established the existence and microscopic origin of a second plateau in liquid HHG for different liquids and laser conditions, we attempt to underpin its physical origin. Given the immense success of electron-trajectory based models in HHG in both gases and solids, as well as in describing the first plateau dynamics in liquid HHG, it stands to reason that the observed second plateau might also be characterized by a set of unique electronic trajectories (which also maintain optical coherence as required from HHG). Due to the lack of translational symmetry and applicability of Bloch theorem in liquids (which prevents a consistent definition of a reciprocal lattice), a real-space picture that can still take into account the local structural arrangement and short-range symmetry of liquids is much preferred to a \textit{k}-space one. Therefore, we focus on real-space trajectories. The main question at hand is then which trajectories could be responsible for it. In what follows, we analyze possible candidates using extended semi-classical models for the electronic dynamics. The goal of such a thorough analysis is to uncover the dominating electron trajectories contributing to the second plateau, provided that such a description exists.

The first possible candidate we analyze are the 'standard' gas-phase-like electron trajectories which were assumed to be suppressed by scattering such that they do not contribute to the first plateau \cite{mondal2023,mondal2023few}. Let us briefly describe this set of trajectories analytically. We follow the three-step model developed for gas-phase HHG \cite{Corkum1993}, while tuning the different steps based on dynamics in the liquid that include scattering effects describable through mean-free-path (MFP) limitations of the electronic motion. The real-space nature of the model is especially important given that the Bloch theorem cannot be consistently applied to liquids, rendering \textit{k}-space trajectories ineffective (which is a result of only short-range order exhibited in liquids, whereas strict translational symmetry is required to define a reciprocal lattice and band structure). We assume that at some initial moment of ionization, $t=t_i$, an electron is excited from its parent molecule to the liquid-dressed continuum. 
Subsequently, that electron propagates as a classical particle under the external laser field, after which it can recombine at a certain moment in time $t=t_f$. We neglect the ionic potential within the strong-field approximation (SFA), which allows for an analytical formulation of the electron's trajectory:
\begin{equation}
x(t) = \frac{qE_0}{m_e\omega^2} [\cos(\omega t)-\cos(\omega t_\mathrm{ion})+\omega(t-t_\mathrm{ion})\sin(\omega t_\mathrm{ion})],
\end{equation}
where $E_0$ is the laser field's peak amplitude, $q$ and $m_e$ are the electron charge and mass, and $\omega$ is the laser carrier frequency. By solving eq. (1) and demanding that $x(t_f)=0$ (for $t_f>t_i$) we obtain the typical sets of short and long trajectories responsible for gas-phase HHG. If we further restrict the solution space to cases where the maximal electron excursion distance is smaller than the typical MFP at those energy scales ($l_{\rm max}$), then we obtain the measured first-plateau cutoff independence on the laser wavelength, and very weak dependence on laser intensity. These are referred to as MFP-limited trajectories, and were established as the dominant mechanism of the first plateau emission in liquid-phase HHG \cite{mondal2023}.

\begin{figure*} [t!]
\includegraphics[width=\textwidth]{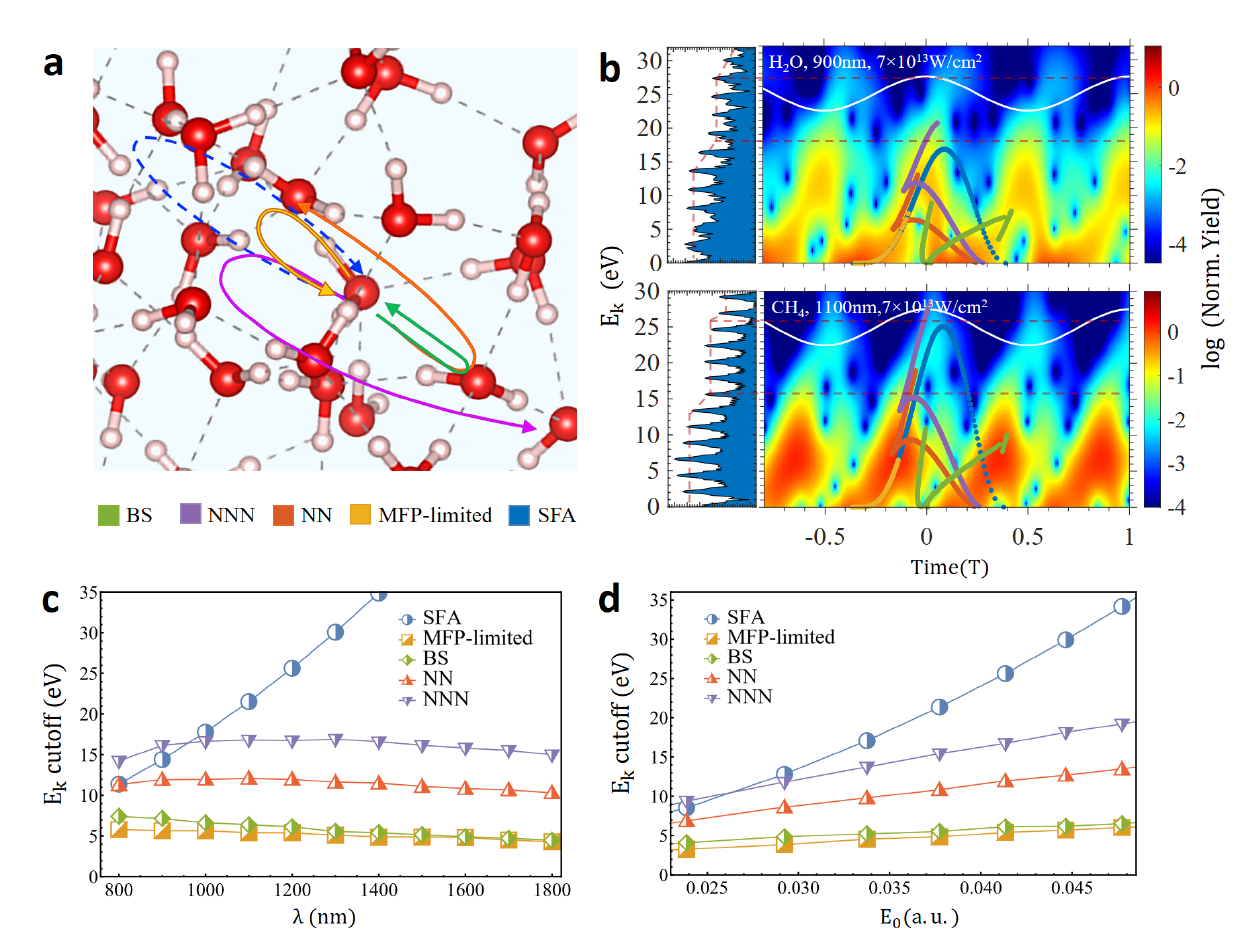}
\caption{
{\bf Electron trajectories contributing to HHG from liquids~}
{\textbf{(a)} Illustration of dominant recombining electron trajectories contributing to liquid HHG. Electronic trajectories that are shorter than the MFP and recombine at the ionization site (orange arrow) dominate the first plateau harmonics. A smaller fraction of electrons with trajectories longer than the MFP can also recombine on neighbor sites with reduced probability (either green arrow for the first solvation shell, or purple arrow for the 2nd solvation shell). Recombination on the first shell contributes mostly to the first plateau and inter-plateau region due to energy constraints. Recombination on the 2nd shell contributes dominantly to the 2nd plateau. Trajectories undergoing back-scattering are not presented as they are shown to negligibly contribute to HHG in all cases. Trajectories longer than the MFP that recombine on-site (labeled SFA) are also possible, but found inconsistent with the second plateau. \textbf{(b)} Time-frequency analysis showing the showing the temporal dependence of harmonic energies for liquid H$_2$O (top, polar liquid), liquid methane (bottom, apolar liquid), for exemplary laser conditions, obtained from \textit{ab-initio} calculations. Insets on the left of panels \textbf{(b)} show the different plateau structures in the harmonic spectra, normalized to electronic kinetic energies upon recombination. Each plot is overlaid with the electron kinetic energies and emission times from the different possible trajectories: MFP-limited contributing to the first plateau (yellow), nearest neighbor recombination (green), next-nearest neighbor recombination at the 2nd solvation shell (purple), back-scattered (orange) and unrestricted gas phase SFA trajectories (blue), calculated with the semi-classical model in the same conditions as the ab initio simulations. \textbf{(c)} Semi-classical scattering model intensity dependence of the second plateau cut-off energy. \textbf{(d)} Semi-classical scattering model wavelength independence of the second cut-off energy.}}
\end{figure*}

Next, in considering the second plateau we can imagine that the set of longer, higher-energy trajectories that were assumed to scatter due to their length exceeding $l_{\rm max}$ might still contribute to HHG, but with a much reduced cross section, leading to a second plateau. However, these trajectories are incompatible with our measurements and calculations for several reasons: (i) The obtained energy ranges for the second plateau do not match those predicted by this set of long trajectories, often differing by $\sim$10~eV. (ii) The cutoff predicted by these trajectories is strongly dependent on the driving wavelength and electric-field amplitude (quadratic in both), inconsistent with the results shown in Fig. 2. (iii) The time-frequency characteristics of these trajectories do not match those coming from \textit{ab-initio} calculations (see Fig. 3(b,d,e) and its deviation from the typical bell-shaped curves obtained from SFA gas-like trajectories). (iv) These trajectories should display a single-Gaussian dependence on the driving laser ellipticity with a width similar or close to that of the gas-phase decay, because the mechanism would be identical \cite{Luu2018,Max2012}. 

We therefore studied the ellipticity dependence of the second-plateau harmonics, both experimentally and through {\it ab-initio} calculations, and compared the liquid- to the gas-phase results in Fig. 4 (with complete data sets shown in Ext. Data Figs. 3-5). Whereas the gas-phase results display the expected and known Gaussian shape, the liquid-phase results differ markedly. Their ellipticity dependence is best captured with a multi-Gaussian fit and a slightly increased width of the central Gaussian. Whereas the position of the side peak in the ellipticity dependence does not strongly depend on the emitted photon energy, the amplitude of the side peak markedly increases with the energy, both in experiments and calculations (Ext. Data Fig. 5). A comparison of the first and second plateau harmonics is presented in Extended Data Figures 3 and 4 and SM figures S9–S11, for both experiments and \textit{ab-initio} simulations, respectively. We observe that while the first plateau harmonics exhibit a single Gaussian dependence of harmonic yield on driver ellipticity—consistent with gas-phase HHG—the second plateau harmonics clearly display the emergence of multi-Gaussian features, indicating additional underlying recombination pathways beyond the central Gaussian distribution.Note that an anomalous ellipticity dependence can also be observed in gas phase systems\cite{Ferre2015}, but with a different characteristic to the side-peaks observed here, and only in a particular energy range close to a resonance (while in our case the effect is robust to intensity and wavelength changes, indicating a non-resonant phenomena).
Overall, the collection of these results allow us to unambiguously rule out 'standard' gas-phase-like trajectories as a source of second-plateau emission. More importantly, the elliptically-driven HHG data and the \textit{ab-initio} time-frequency analysis clearly establish that the second plateau originates from a physical mechanism that is different from the one creating the first plateau. 

\begin{figure*} [t!]
\includegraphics[width=\textwidth]{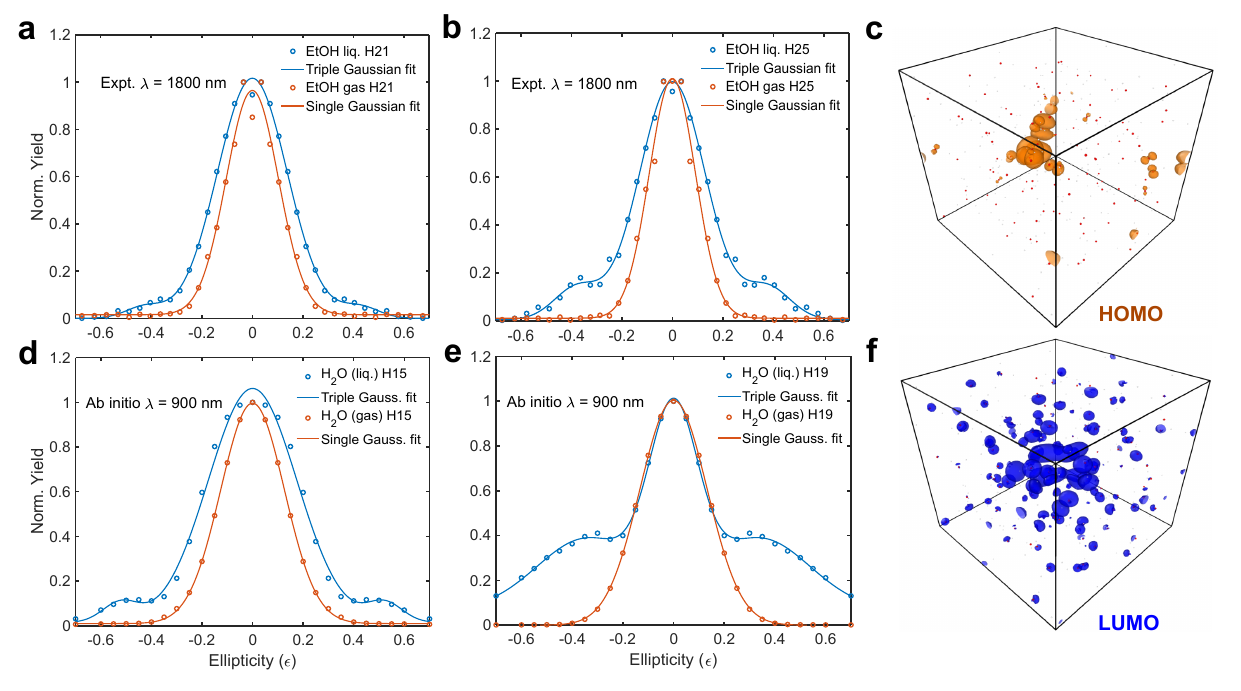}
\caption{
{\bf Ellipticity dependence of harmonics in the second plateau of HHG from liquids~}
\textbf{(a)} and \textbf{(b)} Experimental comparison of harmonic yield as a function of ellipticity for harmonic orders 21 and 25 ($\sim$ 14.5 eV and $\sim$ 17.2 eV, respectively) in gas and liquid phase ethanol at 1800 nm and  2.4$\times$ 10$^{13}$ W/cm$^2$. The gas phase data has been measured by shifting the liquid jet by 300 $\mu$m from the point of collision of the two cylindrical jets. \textbf{(d)} and \textbf{(e)} Comparison of harmonic yield as a function of ellipticity for harmonic orders 15 and 19 ($\sim$ 20.7 eV and $\sim$ 26.2 eV, respectively) in liquid and gas phase H$_2$O, calculated through \textit{ab-initio} simulation at 900 nm and 4$\times$ 10$^{13}$ W/cm$^2$. For gas phase the expected single Gaussian captures the dependence of the harmonic yield as a function of the laser ellipticity (due to on-site recombination). For liquids we observe multiple Gaussian features indicating the off-site recombination for second plateau harmonics. \Nadd{\textbf{c)} and \textbf{f)} are HOMO ad LUMO states obtained from snapshot of molecular dynamics simulations for liquid water, indicating a broad delocalization of the LUMO states which allows for off-site recombination (see SM, Section S1 for details).}}
\end{figure*}

We further consider a set of trajectories whereby the liberated electron is driven in the liquid-dressed continuum, but instead of returning to recombine with its parent molecule, it recombines with a neighboring molecule (see Fig. 3(a) for illustration). This set is denoted as nearest-neighbour (NN) trajectories (somewhat analogous to the real-space trajectory descriptions of solid HHG, but without widely-delocalized wave functions). Due to delocalization of the hole across neighboring molecules, as visible from the results of our molecular-dynamics simulations \Nadd{(see SM S1, and illustration in Fig. 4)}, such recombination on neighboring molecules is allowed; however, typical dipole transition matrix elements are much weaker than those within the same parent molecule, suggesting a suppressed emission. Besides the intuitive exponential suppression, NN trajectories would also exhibit the diminished ellipticity decay with secondary peaks presented in Fig. 4, as neighbouring molecules are generally present in all spatial directions around the parent molecule. Mathematically, we can obtain such trajectories by requiring that $x(t_f)=x_{\rm NN}$, where $x_{\rm NN}$ is a typical inter-molecular separation distance (not the same as the MFP, $l_{\rm max}$). The energy upon recollision can be directly obtained, providing characteristic time-frequency plots. However, in comparing the \textit{ab-initio} calculated time-frequency characteristics shown in Fig. 3(b-d) with the typical time-frequency plots of the NN trajectories (red triangles), we see a typical downshift of $\sim$10eV or more in the expected cutoff energy. This shift cannot be accounted for and generally appears in all liquids under the conditions that we explored. Consequently, NN trajectories seem to mostly contribute to the onset of the second plateau and inter-plateau region, but do not have sufficiently energetic electrons to play a dominant role in the creation of the second plateau harmonics. 

Another option is a set of trajectories denoted as back-scattered (BS) electrons (as explored in certain solid systems\cite{PhysRevA.105.L041101,wang2020role}). Here we consider coherent elastic scattering of electrons after they have traveled the MFP, but whereby the back-scattered electron can recombine with the parent molecule to emit a high-energy photon. Mathematically, these trajectories are an extension of the NN trajectories in a multi-step procedure: (i) The electron is photo-excited into the liquid-dressed continuum at $t=t_i$ (at the origin with zero velocity). (ii) The quasi-free electron is semi-classically accelerated by the laser (within the SFA) up to $l_{\rm max}$, whereby it scatters coherently and elastically, i.e. $v(t_s)\rightarrow -v(t_s)$. (iii) The semi-classical dynamics continues under the laser field until a time $t_f$, where the electron may return to the origin and recombines as a final step. The underlying dynamics can be expressed as:
\begin{equation}
x_{BS}(t) = l_{max} + \frac{qE_0}{m\omega^2} [\cos(\omega t)-\cos(\omega t_\mathrm{s})+\omega(t-t_\mathrm{s})(\sin(\omega t_\mathrm{s})-mv_{s}\omega)],
\end{equation}

\noindent where $t_s$ is the moment in time when scattering occurs (at $l_{max}$) which can be extracted by solving eq. (1), and $v_{s}$ is the velocity of the electron at $t_s$, $v_{s}=x\dot(t_{s})$. The recollision condition is then $x_{BS}(t_{f})=0$. The cross-section for such intricate dynamics is expected to be very small, supporting the exponentially-diminished yield of the second plateau. Moreover, this process would also support a reduced ellipticity decay in the HHG yield. However, in analyzing the BS trajectories the obtained potential HHG emission energies are again substantially too small to account for the second plateau emission (see SM). Moreover, the time-frequency characteristics of BS trajectories extremely differs from the \textit{ab-initio} simulations (see Fig. 3), suggesting BS trajectories play almost no role in HHG emission from the examined liquids. 

We also considered the option that laser-assisted electron scattering may contribute to some of these mechanisms, thereby extending the energy cutoff closer to the measured ranges. However, such laser-assisted mechanisms inherently do not provide enough additional energy to explain the second plateau. Moreover, we remark that our \textit{ab-initio} simulations rely on an adiabatic approximation, which does not include electron-electron scattering effects. While this is usually regarded as a deficiency of the method, it helps here to establish that electron-electron scattering (laser assisted or not) is not responsible for the observed second HHG plateau in liquids.

Lastly, we consider trajectories of electrons recombining on the next-NN (NNN) molecules, i.e. within the second solvation shell. Such trajectories are mathematically equivalent to the NN ones, but demanding recombination at the distance of the second disperse peak of the structure function of the liquid (e.g. extending from about 4 to 5.5~\AA~in water). While the cross section for NNN recombination is weak (exponentially suppressed), it is nonzero (see illustration in Fig. 4 showing molecular dynamics simulations for hole de-localization and further discussion in the SM), and is expected to be enhanced in the strongly laser-driven regime where laser driving and ionization create coherent holes throughout the liquid. Energetically, NNN trajectories account for the correct energy range in the second plateau as the electrons are accelerated for a longer period of time. They also qualitatively match the \textit{ab-initio} time-frequency plots (Fig. 3), though we note that the agreement is not quantitative and there are still ~3 eV missing in the case of water (Fig. 3 (b)) and some features are not perfectly reproduced. Also note that the NNN recombination times should generally depend on the effective mass of the electron, which can differ in the liquid due to screening effects. Thus, we artificially shifted the time axis for these trajectories in Fig. 3 by few hundred attoseconds to optimally match numerics. Importantly, NNN trajectories inherently allow for the reduced yield decay with driving ellipticity and also provide very weak cutoff scaling with respect to the driving wavelength and intensity as measured and observed in \textit{ab-initio} theory, such that overall they make the most reasonable mechanistic explanation for 2nd plateau emission. 
We note that we assumed here fixed holes, which is fully justified in liquids, given that their effective mass is extremely high~\cite{10.1063/1.1940612}.

To further validate this intuition, we developed an extended Lewenstein model \cite{Lewenstein1994}
 (see SM section S3), in which we introduced a delocalized hole that mimics the characteristics of the HOMO state obtained from our molecular-dynamics simulations. This model reveals that the delocalization of the hole directly leads to broader ellipticity-dependent HHG profiles and to secondary peak formation similar to that observed in the experiment and simulations in Fig. 4, confirming our interpretation. 
 
As a whole, these families of electronic trajectories (MFP-limited, NN, NNN, and BS trajectories) paint a more complete picture of potential HHG mechanisms from liquids, as well as higher-order responses combining various scattering and recombination (on-site/off-site) processes (which might explain also the \textit{ab-initio} predictions of higher-order plateaus). We should highlight that all of these options provide cutoff scaling properties consistent with experiments and numerics (i.e. almost independent of laser intensity and wavelength, see Fig. 3). Therefore, separating between the dominant trajectories cannot be performed by simple scaling measurements. However, the fact that only NNN trajectories allow for the correct energy range, in addition to the agreement of ellipticity dependence and time-frequency plots, leads us to pin-point NNN as the most likely explanation for second plateaus in liquid HHG in these conditions. There could, of course, be other physical explanations for the 2nd plateau in liquid HHG, possibly not in real space, or possibly not trajectory-based. However, in our extensive investigation spanning several years we could not identify such alternative explanations, leading us to conclude that NNN trajectories are the most likely culprit.

Overall, a clear physical picture for coherent laser-driven electron dynamics in the liquid, including scattering, emerges: MFP-limited trajectories with on-site recombination dominate the first plateau, up to the first cutoff where electron-molecule scattering starts playing a role in suppressing longer and more energetic trajectories. The exponential decay region in between plateaus is contributed from those suppressed longer trajectories, and from NN trajectories which are three-step-like, but recombine off-site. A second plateau then emerges from coherent NNN trajectories recombining on the second solvation shell. BS trajectories seem to be completely excluded in these conditions (as well as laser-assisted scattering based mechanisms). Such dynamics might in principle involve higher-order neighbors, and higher-order scattering, possibly leading to additional plateaus as observed in our \textit{ab-initio} simulations (e.g. as in Fig. 1(d)).
We note that minor differences are observed between D$_2$O and H$_2$O (see Ext. Data Figs. 1 and 2), in accordance with their slight difference in structure\cite{soper2013radial}.

\section{Discussion and conclusions}
To summarize, we have performed a vast and thorough analysis of HHG in liquids employing experimental, numerical, and theoretical tools. We experimentally observed the appearance of a second, exponentially-suppressed, HHG plateau arising in multiple liquids and laser conditions. We showed that this phenomenon is general to the liquid phase, and universal in that higher-order plateaus are predicted to arise by \textit{ab-initio} calculations performed on all liquids. We have employed semi-classical trajectory-based models for analyzing the laser-driven electron dynamics and resulting HHG, finding that the second plateau is most likely dominated by coherent emission of electrons recombining on neighboring molecules at the second solvation shell. This result was supported by agreement with experimental and numerical data in terms of HHG energy and temporal characteristics, as well as additional experiments and theory in elliptically-driven HHG. We further identified the role of electrons recombining at the first solvation shell, and suppressed on-site recombining long trajectories, as contributing to the inter-plateau decay region, with back-scattered electrons and laser-assisted mechanisms negligible in our examined conditions and liquids. 

Let us now discuss the implications of this work. In terms of the field of HHG, we have unequivocally shown that in the liquid phase HHG can have a different physical mechanism compared to other phases of matter. Practically, this means higher-order scattering and off-site recombination can potentially be probed by tracking HHG yields and cutoff energies, demonstrating a new spectroscopic technique in solution. This could potentially be used to determine effective solvation radii or the extent of spatial delocalization/hybridization of electronic wavefunctions between solute and solvent molecules \cite{gong2022attosecond}.
Exploiting the built-in attosecond temporal resolution of HHG, our results may lead to the development of novel ultrafast spectroscopies for liquid-phase dynamics, including charge migration \cite{Kraus2015}, proton transfer \cite{yin2023} and quantum-nuclear effects \cite{baker06a,Patchkovskii2009,He2018}, e.g. exploiting isotopic substitution as means to separate ultrafast electronic from proton-transfer dynamics. Note that while we explored here liquid phase HHG, we expect similar phenomena to arise also in amorphous solids, where mechanisms and results shown here should translate, though different physical characteristics of the secondary plateaus can be expected. Overall, our work should open up new regimes of ultrafast and nonlinear optical spectroscopies in solution and disordered phases of matter.

\section*{Methods}
Experimental details. The experiments were performed over a wavelength range of 1200 nm - 1800 nm obtained by optical parametric amplification (OPA) of a 1 kHz Ti:Sapphire laser delivering $\sim$ 30 fs pulses centered at 800 nm. The laser beam is focused with a spherical mirror onto a micron-thin liquid flat-jet target to generate high-order harmonics, with details provided in Refs. \cite{Luu2018,yin2020}. 
The emitted harmonics are detected in a custom-built XUV spectrometer composed of an aberration-free flat-field grating (SHIMADZU) and a micro-channel plate (MCP) detector coupled to a phosphor screen. A CCD camera is used to image the phosphor screen. Each spectrum is typically integrated over 200 ms and averaged over 30 measurements.\\
The key methodological improvements in this work lie in the significant reduction of scattered light reaching the MCP detector, thereby enhancing the signal-to-noise ratio and enabling the observation of the second plateau. While the overall experimental setup remains largely consistent with our previous work, several targeted modifications were introduced to suppress background contributions more effectively. The entrance slit was narrowed (Fig. 1a) to reduce stray light and improve spectral resolution. Additionally, a new optical blocker was placed immediately after the grating stage to intercept residual scatter, primarily originating from intense first- and second-order diffraction components. To further improve sensitivity to the weaker high-order harmonics, particularly at longer driving wavelengths where the harmonic yield decreases, the MCP bias voltage was increased to -1.67 kV for 1800 nm and -1.7 kV for 1500 nm, compared to -1.6 kV in previous measurements. Collectively, these refinements substantially improved the detection of higher-order spectral features.

For the ellipticity-dependent measurements, the 1800-nm light is elliptically polarized using a combination of a rotating half-wave plate (HWP) and a fixed quarter-wave plate (QWP). Our experimental geometry ensures a fixed axes of the polarization ellipse during variation from linear polarization ($\epsilon =0$) to circular polarization ($\epsilon =1$)  with the rotation of the HWP axis from 0$^{\circ}$ to 22.5$^{\circ}$, with respect to the QWP axis.

TDDFT simulations. All TDDFT calculations were performed with the open-access code, Octopus \cite{Tancogne-Dejean2020}, using a real-space grid representation. For liquids, we followed the cluster approach, developed and extensively detailed in Ref. \cite{neufeld2022}, in all TDDFT simulations. In these simulations we calculated the liquid's nonlinear response to external intense laser driving, where the driving field duration was eight cycles of the fundamental period. In almost all simulations in the main text, the dipolar-response, \textbf{d}(t), was extracted following linearly-polarized laser excitation, implementing orientation averaging following details in Ref. \cite{neufeld2022}. From \textbf{d}(t) we obtained HHG spectra by taking two time derivatives, and Fourier transforming, following a standard procedure, where we also filtered the dipole acceleration with a super-Gaussian window function. 
For simulations that involved elliptically-polarized driving a similar approach was followed, but as a result of the reduced symmetry of the light-matter Hamiltonian many more orientations needed to be taken into account in the angular-averaging procedure (56 as opposed to 12 in the case of water). From the HHG spectra we extracted the ellipticity-dependent HHG yield upon integrating the yield for each harmonic order.
For the case of gas-phase water we followed the same procedure, but for a single molecule response, as detailed also in Refs. \cite{neufeld2022,Neufeld2024},
and keeping the same level of theory as the liquid system (but with an added self-interaction correction term \cite{Legrand2002}).
For Fig. 3, the Gabor transform was obtained using an exponential window function with a temporal width of a third of an optical cycle of the fundamental.

Semi-classical trajectory simulations. For the various Newtonian equations of motion described in the main text, we numerically solved transcendental equations describing recombination conditions in each case. In all cases we assumed throughout that the electron's effective mass is unity, and that the electron tunnel-ionizes instantaneously and with a zero initial velocity at the molecular center (similar to the gas phase standard approach). For liquid water, NN length was taken as the first sharp peak of the O-O distribution function at $\sim$2.8\r{A}. The NNN distance was taken as the edge of the second broad peak of the distribution function at $\sim$5.3\r{A}. The MFP in water was taken to be $\sim$3.2\r{A}. The time axis was shifted by $\sim$0.1T in each case to compensate for potential effective-mass effects that were not taken into consideration, and to obtain better agreement between the semi-classical theory and \textit{ab-initio} simulations. 
We allowed in all cases the trajectories to also surpass the original parent ionization site once, creating the two branches in the NN and NNN trajectories shown in Fig. 3 (with one branch surpassing the ion, and the other not). 
The kinetic energies upon recombination were obtained directly from the numerical solution of the Newtonian equations of motion, where the \textit{ab-initio} simulations were subtracted by the gap energy in the cluster model. In the SM we further developed a semi-analytic description of the 2nd plateau cut-off energies for NN trajectories, as well as a Lewenstein-like SFA numerical model \cite{Lewenstein1994} for a delocalized system for the the elliptically-driven case.

\section*{Data Availability}
The datasets generated and/or analyzed during the current study are available from the corresponding authors upon reasonable request.

\section*{Code Availability}
The Octopus package used for the TDDFT calculations is publicly available. The remaining computer codes are available from the corresponding authors upon reasonable request.

\bibliography{ref,atto}

\begin{thebibliography}{12}%
\makeatletter
\providecommand \@ifxundefined [1]{%
 \@ifx{#1\undefined}
}%
\providecommand \@ifnum [1]{%
 \ifnum #1\expandafter \@firstoftwo
 \else \expandafter \@secondoftwo
 \fi
}%
\providecommand \@ifx [1]{%
 \ifx #1\expandafter \@firstoftwo
 \else \expandafter \@secondoftwo
 \fi
}%
\providecommand \natexlab [1]{#1}%
\providecommand \enquote  [1]{``#1''}%
\providecommand \bibnamefont  [1]{#1}%
\providecommand \bibfnamefont [1]{#1}%
\providecommand \citenamefont [1]{#1}%
\providecommand \href@noop [0]{\@secondoftwo}%
\providecommand \href [0]{\begingroup \@sanitize@url \@href}%
\providecommand \@href[1]{\@@startlink{#1}\@@href}%
\providecommand \@@href[1]{\endgroup#1\@@endlink}%
\providecommand \@sanitize@url [0]{\catcode `\\12\catcode `\$12\catcode
  `\&12\catcode `\#12\catcode `\^12\catcode `\_12\catcode `\%12\relax}%
\providecommand \@@startlink[1]{}%
\providecommand \@@endlink[0]{}%
\providecommand \url  [0]{\begingroup\@sanitize@url \@url }%
\providecommand \@url [1]{\endgroup\@href {#1}{\urlprefix }}%
\providecommand \urlprefix  [0]{URL }%
\providecommand \Eprint [0]{\href }%
\providecommand \doibase [0]{https://doi.org/}%
\providecommand \selectlanguage [0]{\@gobble}%
\providecommand \bibinfo  [0]{\@secondoftwo}%
\providecommand \bibfield  [0]{\@secondoftwo}%
\providecommand \translation [1]{[#1]}%
\providecommand \BibitemOpen [0]{}%
\providecommand \bibitemStop [0]{}%
\providecommand \bibitemNoStop [0]{.\EOS\space}%
\providecommand \EOS [0]{\spacefactor3000\relax}%
\providecommand \BibitemShut  [1]{\csname bibitem#1\endcsname}%
\let\auto@bib@innerbib\@empty
\bibitem [{\citenamefont {Behler}\ and\ \citenamefont
  {Parrinello}(2007)}]{PhysRevLett.98.146401}%
  \BibitemOpen
  \bibfield  {author} {\bibinfo {author} {\bibfnamefont {J.}~\bibnamefont
  {Behler}}\ and\ \bibinfo {author} {\bibfnamefont {M.}~\bibnamefont
  {Parrinello}},\ }\bibfield  {title} {\bibinfo {title} {Generalized
  neural-network representation of high-dimensional potential-energy
  surfaces},\ }\href {https://doi.org/10.1103/PhysRevLett.98.146401} {\bibfield
   {journal} {\bibinfo  {journal} {Phys. Rev. Lett.}\ }\textbf {\bibinfo
  {volume} {98}},\ \bibinfo {pages} {146401} (\bibinfo {year}
  {2007})}\BibitemShut {NoStop}%
\bibitem [{\citenamefont {O’Neill}\ \emph {et~al.}(2024)\citenamefont
  {O’Neill}, \citenamefont {Shi}, \citenamefont {Fong}, \citenamefont
  {Michaelides},\ and\ \citenamefont {Schran}}]{o2024pair}%
  \BibitemOpen
  \bibfield  {author} {\bibinfo {author} {\bibfnamefont {N.}~\bibnamefont
  {O’Neill}}, \bibinfo {author} {\bibfnamefont {B.~X.}\ \bibnamefont {Shi}},
  \bibinfo {author} {\bibfnamefont {K.}~\bibnamefont {Fong}}, \bibinfo {author}
  {\bibfnamefont {A.}~\bibnamefont {Michaelides}},\ and\ \bibinfo {author}
  {\bibfnamefont {C.}~\bibnamefont {Schran}},\ }\bibfield  {title} {\bibinfo
  {title} {To pair or not to pair? machine-learned explicitly-correlated
  electronic structure for nacl in water},\ }\href@noop {} {\bibfield
  {journal} {\bibinfo  {journal} {The Journal of Physical Chemistry Letters}\
  }\textbf {\bibinfo {volume} {15}},\ \bibinfo {pages} {6081} (\bibinfo {year}
  {2024})}\BibitemShut {NoStop}%
\bibitem [{\citenamefont {Ceriotti}\ \emph {et~al.}(2010)\citenamefont
  {Ceriotti}, \citenamefont {Parrinello}, \citenamefont {Markland},\ and\
  \citenamefont {Manolopoulos}}]{ceriotti2010efficient}%
  \BibitemOpen
  \bibfield  {author} {\bibinfo {author} {\bibfnamefont {M.}~\bibnamefont
  {Ceriotti}}, \bibinfo {author} {\bibfnamefont {M.}~\bibnamefont
  {Parrinello}}, \bibinfo {author} {\bibfnamefont {T.~E.}\ \bibnamefont
  {Markland}},\ and\ \bibinfo {author} {\bibfnamefont {D.~E.}\ \bibnamefont
  {Manolopoulos}},\ }\bibfield  {title} {\bibinfo {title} {Efficient stochastic
  thermostatting of path integral molecular dynamics},\ }\href@noop {}
  {\bibfield  {journal} {\bibinfo  {journal} {The Journal of chemical physics}\
  }\textbf {\bibinfo {volume} {133}} (\bibinfo {year} {2010})}\BibitemShut
  {NoStop}%
\bibitem [{\citenamefont {Litman}\ \emph {et~al.}(2024)\citenamefont {Litman},
  \citenamefont {Kapil}, \citenamefont {Feldman}, \citenamefont {Tisi},
  \citenamefont {Begu{\v{s}}i{\'c}}, \citenamefont {Fidanyan}, \citenamefont
  {Fraux}, \citenamefont {Higer}, \citenamefont {Kellner}, \citenamefont {Li}
  \emph {et~al.}}]{litman2024pi}%
  \BibitemOpen
  \bibfield  {author} {\bibinfo {author} {\bibfnamefont {Y.}~\bibnamefont
  {Litman}}, \bibinfo {author} {\bibfnamefont {V.}~\bibnamefont {Kapil}},
  \bibinfo {author} {\bibfnamefont {Y.~M.}\ \bibnamefont {Feldman}}, \bibinfo
  {author} {\bibfnamefont {D.}~\bibnamefont {Tisi}}, \bibinfo {author}
  {\bibfnamefont {T.}~\bibnamefont {Begu{\v{s}}i{\'c}}}, \bibinfo {author}
  {\bibfnamefont {K.}~\bibnamefont {Fidanyan}}, \bibinfo {author}
  {\bibfnamefont {G.}~\bibnamefont {Fraux}}, \bibinfo {author} {\bibfnamefont
  {J.}~\bibnamefont {Higer}}, \bibinfo {author} {\bibfnamefont
  {M.}~\bibnamefont {Kellner}}, \bibinfo {author} {\bibfnamefont {T.~E.}\
  \bibnamefont {Li}}, \emph {et~al.},\ }\bibfield  {title} {\bibinfo {title}
  {i-pi 3.0: a flexible, efficient framework for advanced atomistic
  simulations},\ }\href@noop {} {\bibfield  {journal} {\bibinfo  {journal}
  {arXiv preprint arXiv:2405.15224}\ } (\bibinfo {year} {2024})}\BibitemShut
  {NoStop}%
\bibitem [{\citenamefont {Schran}\ \emph {et~al.}(2021)\citenamefont {Schran},
  \citenamefont {Thiemann}, \citenamefont {Rowe}, \citenamefont {M{\"u}ller},
  \citenamefont {Marsalek},\ and\ \citenamefont
  {Michaelides}}]{schran2021machine}%
  \BibitemOpen
  \bibfield  {author} {\bibinfo {author} {\bibfnamefont {C.}~\bibnamefont
  {Schran}}, \bibinfo {author} {\bibfnamefont {F.~L.}\ \bibnamefont
  {Thiemann}}, \bibinfo {author} {\bibfnamefont {P.}~\bibnamefont {Rowe}},
  \bibinfo {author} {\bibfnamefont {E.~A.}\ \bibnamefont {M{\"u}ller}},
  \bibinfo {author} {\bibfnamefont {O.}~\bibnamefont {Marsalek}},\ and\
  \bibinfo {author} {\bibfnamefont {A.}~\bibnamefont {Michaelides}},\
  }\bibfield  {title} {\bibinfo {title} {Machine learning potentials for
  complex aqueous systems made simple},\ }\href@noop {} {\bibfield  {journal}
  {\bibinfo  {journal} {Proceedings of the National Academy of Sciences}\
  }\textbf {\bibinfo {volume} {118}},\ \bibinfo {pages} {e2110077118} (\bibinfo
  {year} {2021})}\BibitemShut {NoStop}%
\bibitem [{\citenamefont {Blum}\ \emph {et~al.}(2009)\citenamefont {Blum},
  \citenamefont {Gehrke}, \citenamefont {Hanke}, \citenamefont {Havu},
  \citenamefont {Havu}, \citenamefont {Ren}, \citenamefont {Reuter},\ and\
  \citenamefont {Scheffler}}]{blum2009ab}%
  \BibitemOpen
  \bibfield  {author} {\bibinfo {author} {\bibfnamefont {V.}~\bibnamefont
  {Blum}}, \bibinfo {author} {\bibfnamefont {R.}~\bibnamefont {Gehrke}},
  \bibinfo {author} {\bibfnamefont {F.}~\bibnamefont {Hanke}}, \bibinfo
  {author} {\bibfnamefont {P.}~\bibnamefont {Havu}}, \bibinfo {author}
  {\bibfnamefont {V.}~\bibnamefont {Havu}}, \bibinfo {author} {\bibfnamefont
  {X.}~\bibnamefont {Ren}}, \bibinfo {author} {\bibfnamefont {K.}~\bibnamefont
  {Reuter}},\ and\ \bibinfo {author} {\bibfnamefont {M.}~\bibnamefont
  {Scheffler}},\ }\bibfield  {title} {\bibinfo {title} {Ab initio molecular
  simulations with numeric atom-centered orbitals},\ }\href@noop {} {\bibfield
  {journal} {\bibinfo  {journal} {Computer Physics Communications}\ }\textbf
  {\bibinfo {volume} {180}},\ \bibinfo {pages} {2175} (\bibinfo {year}
  {2009})}\BibitemShut {NoStop}%
\bibitem [{\citenamefont {Gong}\ \emph {et~al.}(2022)\citenamefont {Gong},
  \citenamefont {Heck}, \citenamefont {Jelovina}, \citenamefont {Perry},
  \citenamefont {Zinchenko}, \citenamefont {Lucchese},\ and\ \citenamefont
  {W{\"o}rner}}]{gong2022attosecond}%
  \BibitemOpen
  \bibfield  {author} {\bibinfo {author} {\bibfnamefont {X.}~\bibnamefont
  {Gong}}, \bibinfo {author} {\bibfnamefont {S.}~\bibnamefont {Heck}}, \bibinfo
  {author} {\bibfnamefont {D.}~\bibnamefont {Jelovina}}, \bibinfo {author}
  {\bibfnamefont {C.}~\bibnamefont {Perry}}, \bibinfo {author} {\bibfnamefont
  {K.}~\bibnamefont {Zinchenko}}, \bibinfo {author} {\bibfnamefont
  {R.}~\bibnamefont {Lucchese}},\ and\ \bibinfo {author} {\bibfnamefont
  {H.~J.}\ \bibnamefont {W{\"o}rner}},\ }\bibfield  {title} {\bibinfo {title}
  {Attosecond spectroscopy of size-resolved water clusters},\ }\href@noop {}
  {\bibfield  {journal} {\bibinfo  {journal} {Nature}\ }\textbf {\bibinfo
  {volume} {609}},\ \bibinfo {pages} {507} (\bibinfo {year}
  {2022})}\BibitemShut {NoStop}%
\bibitem [{\citenamefont {Nourbakhsh}\ \emph {et~al.}(2022)\citenamefont
  {Nourbakhsh}, \citenamefont {Neufeld}, \citenamefont {Tancogne-Dejean},\ and\
  \citenamefont {Rubio}}]{2212.04177}%
  \BibitemOpen
  \bibfield  {author} {\bibinfo {author} {\bibfnamefont {Z.}~\bibnamefont
  {Nourbakhsh}}, \bibinfo {author} {\bibfnamefont {O.}~\bibnamefont {Neufeld}},
  \bibinfo {author} {\bibfnamefont {N.}~\bibnamefont {Tancogne-Dejean}},\ and\
  \bibinfo {author} {\bibfnamefont {A.}~\bibnamefont {Rubio}},\ }\href@noop {}
  {\bibinfo {title} {An ab initio supercell approach for high-harmonic
  generation in liquids}} (\bibinfo {year} {2022}),\ \Eprint
  {https://arxiv.org/abs/arXiv:2212.04177} {arXiv:2212.04177} \BibitemShut
  {NoStop}%
\bibitem [{\citenamefont {Xu}\ and\ \citenamefont {Meng}(2025)}]{xu2025high}%
  \BibitemOpen
  \bibfield  {author} {\bibinfo {author} {\bibfnamefont {J.}~\bibnamefont
  {Xu}}\ and\ \bibinfo {author} {\bibfnamefont {S.}~\bibnamefont {Meng}},\
  }\bibfield  {title} {\bibinfo {title} {High-harmonic generation and
  femtosecond-resolved ultrafast dynamics in liquid water},\ }\href@noop {}
  {\bibfield  {journal} {\bibinfo  {journal} {The Journal of Physical Chemistry
  Letters}\ }\textbf {\bibinfo {volume} {16}},\ \bibinfo {pages} {5295}
  (\bibinfo {year} {2025})}\BibitemShut {NoStop}%
\bibitem [{\citenamefont {Lewenstein}\ \emph {et~al.}(1994)\citenamefont
  {Lewenstein}, \citenamefont {Balcou}, \citenamefont {Ivanov}, \citenamefont
  {L'Huillier},\ and\ \citenamefont {Corkum}}]{PhysRevA.49.2117}%
  \BibitemOpen
  \bibfield  {author} {\bibinfo {author} {\bibfnamefont {M.}~\bibnamefont
  {Lewenstein}}, \bibinfo {author} {\bibfnamefont {P.}~\bibnamefont {Balcou}},
  \bibinfo {author} {\bibfnamefont {M.~Y.}\ \bibnamefont {Ivanov}}, \bibinfo
  {author} {\bibfnamefont {A.}~\bibnamefont {L'Huillier}},\ and\ \bibinfo
  {author} {\bibfnamefont {P.~B.}\ \bibnamefont {Corkum}},\ }\bibfield  {title}
  {\bibinfo {title} {Theory of high-harmonic generation by low-frequency laser
  fields},\ }\href {https://doi.org/10.1103/PhysRevA.49.2117} {\bibfield
  {journal} {\bibinfo  {journal} {Phys. Rev. A}\ }\textbf {\bibinfo {volume}
  {49}},\ \bibinfo {pages} {2117} (\bibinfo {year} {1994})}\BibitemShut
  {NoStop}%
\bibitem [{\citenamefont {Ndabashimiye}\ \emph {et~al.}(2016)\citenamefont
  {Ndabashimiye}, \citenamefont {Ghimire}, \citenamefont {Wu}, \citenamefont
  {Browne}, \citenamefont {Schafer}, \citenamefont {Gaarde},\ and\
  \citenamefont {Reis}}]{Ndabashimiye2016}%
  \BibitemOpen
  \bibfield  {author} {\bibinfo {author} {\bibfnamefont {G.}~\bibnamefont
  {Ndabashimiye}}, \bibinfo {author} {\bibfnamefont {S.}~\bibnamefont
  {Ghimire}}, \bibinfo {author} {\bibfnamefont {M.}~\bibnamefont {Wu}},
  \bibinfo {author} {\bibfnamefont {D.~A.}\ \bibnamefont {Browne}}, \bibinfo
  {author} {\bibfnamefont {K.~J.}\ \bibnamefont {Schafer}}, \bibinfo {author}
  {\bibfnamefont {M.~B.}\ \bibnamefont {Gaarde}},\ and\ \bibinfo {author}
  {\bibfnamefont {D.~A.}\ \bibnamefont {Reis}},\ }\bibfield  {title} {\bibinfo
  {title} {{Solid-state harmonics beyond the atomic limit}},\ }\bibfield
  {journal} {\bibinfo  {journal} {Nature}\ }\textbf {\bibinfo {volume} {534}},\
  \href {https://doi.org/10.1038/nature17660} {10.1038/nature17660} (\bibinfo
  {year} {2016})\BibitemShut {NoStop}%
\bibitem [{\citenamefont {Luu}\ and\ \citenamefont
  {W{\"o}rner}(2018)}]{luu18b}%
  \BibitemOpen
  \bibfield  {author} {\bibinfo {author} {\bibfnamefont {T.~T.}\ \bibnamefont
  {Luu}}\ and\ \bibinfo {author} {\bibfnamefont {H.~J.}\ \bibnamefont
  {W{\"o}rner}},\ }\bibfield  {title} {\bibinfo {title} {Measurement of the
  berry curvature of solids using high-harmonic spectroscopy},\ }\href@noop {}
  {\bibfield  {journal} {\bibinfo  {journal} {Nature Communications}\ }\textbf
  {\bibinfo {volume} {9}},\ \bibinfo {pages} {916} (\bibinfo {year}
  {2018})}\BibitemShut {NoStop}%
\end{thebibliography}%


\begin{thebibliography}{100}%
\makeatletter
\providecommand \@ifxundefined [1]{%
 \@ifx{#1\undefined}
}%
\providecommand \@ifnum [1]{%
 \ifnum #1\expandafter \@firstoftwo
 \else \expandafter \@secondoftwo
 \fi
}%
\providecommand \@ifx [1]{%
 \ifx #1\expandafter \@firstoftwo
 \else \expandafter \@secondoftwo
 \fi
}%
\providecommand \natexlab [1]{#1}%
\providecommand \enquote  [1]{``#1''}%
\providecommand \bibnamefont  [1]{#1}%
\providecommand \bibfnamefont [1]{#1}%
\providecommand \citenamefont [1]{#1}%
\providecommand \href@noop [0]{\@secondoftwo}%
\providecommand \href [0]{\begingroup \@sanitize@url \@href}%
\providecommand \@href[1]{\@@startlink{#1}\@@href}%
\providecommand \@@href[1]{\endgroup#1\@@endlink}%
\providecommand \@sanitize@url [0]{\catcode `\\12\catcode `\$12\catcode
  `\&12\catcode `\#12\catcode `\^12\catcode `\_12\catcode `\%12\relax}%
\providecommand \@@startlink[1]{}%
\providecommand \@@endlink[0]{}%
\providecommand \url  [0]{\begingroup\@sanitize@url \@url }%
\providecommand \@url [1]{\endgroup\@href {#1}{\urlprefix }}%
\providecommand \urlprefix  [0]{URL }%
\providecommand \Eprint [0]{\href }%
\providecommand \doibase [0]{https://doi.org/}%
\providecommand \selectlanguage [0]{\@gobble}%
\providecommand \bibinfo  [0]{\@secondoftwo}%
\providecommand \bibfield  [0]{\@secondoftwo}%
\providecommand \translation [1]{[#1]}%
\providecommand \BibitemOpen [0]{}%
\providecommand \bibitemStop [0]{}%
\providecommand \bibitemNoStop [0]{.\EOS\space}%
\providecommand \EOS [0]{\spacefactor3000\relax}%
\providecommand \BibitemShut  [1]{\csname bibitem#1\endcsname}%
\let\auto@bib@innerbib\@empty
\bibitem [{\citenamefont {McPherson}\ \emph {et~al.}(1987)\citenamefont
  {McPherson}, \citenamefont {Gibson}, \citenamefont {Jara}, \citenamefont
  {Johann}, \citenamefont {Luk}, \citenamefont {McIntyre}, \citenamefont
  {Boyer},\ and\ \citenamefont {Rhodes}}]{McPherson1987}%
  \BibitemOpen
  \bibfield  {author} {\bibinfo {author} {\bibfnamefont {A.}~\bibnamefont
  {McPherson}}, \bibinfo {author} {\bibfnamefont {G.}~\bibnamefont {Gibson}},
  \bibinfo {author} {\bibfnamefont {H.}~\bibnamefont {Jara}}, \bibinfo {author}
  {\bibfnamefont {U.}~\bibnamefont {Johann}}, \bibinfo {author} {\bibfnamefont
  {T.~S.}\ \bibnamefont {Luk}}, \bibinfo {author} {\bibfnamefont {I.~A.}\
  \bibnamefont {McIntyre}}, \bibinfo {author} {\bibfnamefont {K.}~\bibnamefont
  {Boyer}},\ and\ \bibinfo {author} {\bibfnamefont {C.~K.}\ \bibnamefont
  {Rhodes}},\ }\bibfield  {title} {\bibinfo {title} {{Studies of multiphoton
  production of vacuum-ultraviolet radiation in the rare gases}},\ }\href
  {https://doi.org/10.1364/JOSAB.4.000595} {\bibfield  {journal} {\bibinfo
  {journal} {Journal of the Optical Society of America B}\ }\textbf {\bibinfo
  {volume} {4}},\ \bibinfo {pages} {595} (\bibinfo {year} {1987})}\BibitemShut
  {NoStop}%
\bibitem [{\citenamefont {Ferray}\ \emph {et~al.}(1988)\citenamefont {Ferray},
  \citenamefont {L'Huillier}, \citenamefont {Li}, \citenamefont {Lompre},
  \citenamefont {Mainfray},\ and\ \citenamefont {Manus}}]{Ferray1988}%
  \BibitemOpen
  \bibfield  {author} {\bibinfo {author} {\bibfnamefont {M.}~\bibnamefont
  {Ferray}}, \bibinfo {author} {\bibfnamefont {A.}~\bibnamefont {L'Huillier}},
  \bibinfo {author} {\bibfnamefont {X.~F.}\ \bibnamefont {Li}}, \bibinfo
  {author} {\bibfnamefont {L.~A.}\ \bibnamefont {Lompre}}, \bibinfo {author}
  {\bibfnamefont {G.}~\bibnamefont {Mainfray}},\ and\ \bibinfo {author}
  {\bibfnamefont {C.}~\bibnamefont {Manus}},\ }\bibfield  {title} {\bibinfo
  {title} {{Multiple-harmonic conversion of 1064 nm radiation in rare gases}},\
  }\href {https://doi.org/10.1088/0953-4075/21/3/001} {\bibfield  {journal}
  {\bibinfo  {journal} {Journal of Physics B: Atomic, Molecular and Optical
  Physics}\ }\textbf {\bibinfo {volume} {21}},\ \bibinfo {pages} {L31}
  (\bibinfo {year} {1988})}\BibitemShut {NoStop}%
\bibitem [{\citenamefont {Ghimire}\ \emph
  {et~al.}(2011{\natexlab{a}})\citenamefont {Ghimire}, \citenamefont
  {Dichiara}, \citenamefont {Sistrunk}, \citenamefont {Agostini}, \citenamefont
  {Dimauro},\ and\ \citenamefont {Reis}}]{Ghimire2011a}%
  \BibitemOpen
  \bibfield  {author} {\bibinfo {author} {\bibfnamefont {S.}~\bibnamefont
  {Ghimire}}, \bibinfo {author} {\bibfnamefont {A.~D.}\ \bibnamefont
  {Dichiara}}, \bibinfo {author} {\bibfnamefont {E.}~\bibnamefont {Sistrunk}},
  \bibinfo {author} {\bibfnamefont {P.}~\bibnamefont {Agostini}}, \bibinfo
  {author} {\bibfnamefont {L.~F.}\ \bibnamefont {Dimauro}},\ and\ \bibinfo
  {author} {\bibfnamefont {D.~A.}\ \bibnamefont {Reis}},\ }\bibfield  {title}
  {\bibinfo {title} {{Observation of high-order harmonic generation in a bulk
  crystal}},\ }\href {https://doi.org/10.1038/nphys1847} {\bibfield  {journal}
  {\bibinfo  {journal} {Nature Physics}\ }\textbf {\bibinfo {volume} {7}},\
  \bibinfo {pages} {138} (\bibinfo {year} {2011}{\natexlab{a}})}\BibitemShut
  {NoStop}%
\bibitem [{\citenamefont {Luu}\ and\ \citenamefont
  {W{\"o}rner}(2018)}]{luu18b}%
  \BibitemOpen
  \bibfield  {author} {\bibinfo {author} {\bibfnamefont {T.~T.}\ \bibnamefont
  {Luu}}\ and\ \bibinfo {author} {\bibfnamefont {H.~J.}\ \bibnamefont
  {W{\"o}rner}},\ }\bibfield  {title} {\bibinfo {title} {Measurement of the
  berry curvature of solids using high-harmonic spectroscopy},\ }\href@noop {}
  {\bibfield  {journal} {\bibinfo  {journal} {Nature Communications}\ }\textbf
  {\bibinfo {volume} {9}},\ \bibinfo {pages} {916} (\bibinfo {year}
  {2018})}\BibitemShut {NoStop}%
\bibitem [{\citenamefont {Paul}\ \emph {et~al.}(2001)\citenamefont {Paul},
  \citenamefont {Toma}, \citenamefont {Breger}, \citenamefont {Mullot},
  \citenamefont {Aug\'e}, \citenamefont {Balcou}, \citenamefont {Muller},\ and\
  \citenamefont {Agostini}}]{paul01a}%
  \BibitemOpen
  \bibfield  {author} {\bibinfo {author} {\bibfnamefont {P.~M.}\ \bibnamefont
  {Paul}}, \bibinfo {author} {\bibfnamefont {E.~S.}\ \bibnamefont {Toma}},
  \bibinfo {author} {\bibfnamefont {P.}~\bibnamefont {Breger}}, \bibinfo
  {author} {\bibfnamefont {G.}~\bibnamefont {Mullot}}, \bibinfo {author}
  {\bibfnamefont {F.}~\bibnamefont {Aug\'e}}, \bibinfo {author} {\bibfnamefont
  {P.}~\bibnamefont {Balcou}}, \bibinfo {author} {\bibfnamefont {H.~G.}\
  \bibnamefont {Muller}},\ and\ \bibinfo {author} {\bibfnamefont
  {P.}~\bibnamefont {Agostini}},\ }\bibfield  {title} {\bibinfo {title}
  {Observation of a train of attosecond pulses from high harmonic generation},\
  }\href@noop {} {\bibfield  {journal} {\bibinfo  {journal} {Science}\ }\textbf
  {\bibinfo {volume} {292}},\ \bibinfo {pages} {1689} (\bibinfo {year}
  {2001})}\BibitemShut {NoStop}%
\bibitem [{\citenamefont {Popmintchev}\ \emph {et~al.}(2012)\citenamefont
  {Popmintchev}, \citenamefont {Chen}, \citenamefont {Popmintchev},
  \citenamefont {Arpin}, \citenamefont {Brown}, \citenamefont {Alisauskas},
  \citenamefont {Andriukaitis}, \citenamefont {Balciunas}, \citenamefont
  {Mucke}, \citenamefont {Pugzlys}, \citenamefont {Baltuska}, \citenamefont
  {Shim}, \citenamefont {Schrauth}, \citenamefont {Gaeta}, \citenamefont
  {Hernandez-Garcia}, \citenamefont {Plaja}, \citenamefont {Becker},
  \citenamefont {Jaron-Becker}, \citenamefont {Murnane},\ and\ \citenamefont
  {Kapteyn}}]{Popmintchev2012}%
  \BibitemOpen
  \bibfield  {author} {\bibinfo {author} {\bibfnamefont {T.}~\bibnamefont
  {Popmintchev}}, \bibinfo {author} {\bibfnamefont {M.-C.}\ \bibnamefont
  {Chen}}, \bibinfo {author} {\bibfnamefont {D.}~\bibnamefont {Popmintchev}},
  \bibinfo {author} {\bibfnamefont {P.}~\bibnamefont {Arpin}}, \bibinfo
  {author} {\bibfnamefont {S.}~\bibnamefont {Brown}}, \bibinfo {author}
  {\bibfnamefont {S.}~\bibnamefont {Alisauskas}}, \bibinfo {author}
  {\bibfnamefont {G.}~\bibnamefont {Andriukaitis}}, \bibinfo {author}
  {\bibfnamefont {T.}~\bibnamefont {Balciunas}}, \bibinfo {author}
  {\bibfnamefont {O.~D.}\ \bibnamefont {Mucke}}, \bibinfo {author}
  {\bibfnamefont {A.}~\bibnamefont {Pugzlys}}, \bibinfo {author} {\bibfnamefont
  {A.}~\bibnamefont {Baltuska}}, \bibinfo {author} {\bibfnamefont
  {B.}~\bibnamefont {Shim}}, \bibinfo {author} {\bibfnamefont {S.~E.}\
  \bibnamefont {Schrauth}}, \bibinfo {author} {\bibfnamefont {A.}~\bibnamefont
  {Gaeta}}, \bibinfo {author} {\bibfnamefont {C.}~\bibnamefont
  {Hernandez-Garcia}}, \bibinfo {author} {\bibfnamefont {L.}~\bibnamefont
  {Plaja}}, \bibinfo {author} {\bibfnamefont {A.}~\bibnamefont {Becker}},
  \bibinfo {author} {\bibfnamefont {A.}~\bibnamefont {Jaron-Becker}}, \bibinfo
  {author} {\bibfnamefont {M.~M.}\ \bibnamefont {Murnane}},\ and\ \bibinfo
  {author} {\bibfnamefont {H.~C.}\ \bibnamefont {Kapteyn}},\ }\bibfield
  {title} {\bibinfo {title} {{Bright Coherent Ultrahigh Harmonics in the keV
  X-ray Regime from Mid-Infrared Femtosecond Lasers}},\ }\href
  {https://doi.org/10.1126/science.1218497} {\bibfield  {journal} {\bibinfo
  {journal} {Science}\ }\textbf {\bibinfo {volume} {336}},\ \bibinfo {pages}
  {1287} (\bibinfo {year} {2012})}\BibitemShut {NoStop}%
\bibitem [{\citenamefont {Vampa}\ \emph
  {et~al.}(2015{\natexlab{a}})\citenamefont {Vampa}, \citenamefont {Hammond},
  \citenamefont {Thir{\'e}}, \citenamefont {Schmidt}, \citenamefont
  {L{\'e}gar{\'e}}, \citenamefont {McDonald}, \citenamefont {Brabec},\ and\
  \citenamefont {Corkum}}]{vampa2015}%
  \BibitemOpen
  \bibfield  {author} {\bibinfo {author} {\bibfnamefont {G.}~\bibnamefont
  {Vampa}}, \bibinfo {author} {\bibfnamefont {T.}~\bibnamefont {Hammond}},
  \bibinfo {author} {\bibfnamefont {N.}~\bibnamefont {Thir{\'e}}}, \bibinfo
  {author} {\bibfnamefont {B.}~\bibnamefont {Schmidt}}, \bibinfo {author}
  {\bibfnamefont {F.}~\bibnamefont {L{\'e}gar{\'e}}}, \bibinfo {author}
  {\bibfnamefont {C.}~\bibnamefont {McDonald}}, \bibinfo {author}
  {\bibfnamefont {T.}~\bibnamefont {Brabec}},\ and\ \bibinfo {author}
  {\bibfnamefont {P.}~\bibnamefont {Corkum}},\ }\bibfield  {title} {\bibinfo
  {title} {Linking high harmonics from gases and solids},\ }\href@noop {}
  {\bibfield  {journal} {\bibinfo  {journal} {Nature}\ }\textbf {\bibinfo
  {volume} {522}},\ \bibinfo {pages} {462} (\bibinfo {year}
  {2015}{\natexlab{a}})}\BibitemShut {NoStop}%
\bibitem [{\citenamefont {Luu}\ \emph {et~al.}(2018)\citenamefont {Luu},
  \citenamefont {Yin}, \citenamefont {Jain}, \citenamefont {Gaumnitz},
  \citenamefont {Pertot}, \citenamefont {Ma},\ and\ \citenamefont
  {W{\"{o}}rner}}]{Luu2018}%
  \BibitemOpen
  \bibfield  {author} {\bibinfo {author} {\bibfnamefont {T.~T.}\ \bibnamefont
  {Luu}}, \bibinfo {author} {\bibfnamefont {Z.}~\bibnamefont {Yin}}, \bibinfo
  {author} {\bibfnamefont {A.}~\bibnamefont {Jain}}, \bibinfo {author}
  {\bibfnamefont {T.}~\bibnamefont {Gaumnitz}}, \bibinfo {author}
  {\bibfnamefont {Y.}~\bibnamefont {Pertot}}, \bibinfo {author} {\bibfnamefont
  {J.}~\bibnamefont {Ma}},\ and\ \bibinfo {author} {\bibfnamefont {H.~J.}\
  \bibnamefont {W{\"{o}}rner}},\ }\bibfield  {title} {\bibinfo {title}
  {{Extreme–ultraviolet high–harmonic generation in liquids}},\ }\href
  {https://doi.org/10.1038/s41467-018-06040-4} {\bibfield  {journal} {\bibinfo
  {journal} {Nature Communications}\ }\textbf {\bibinfo {volume} {9}},\
  \bibinfo {pages} {3723} (\bibinfo {year} {2018})}\BibitemShut {NoStop}%
\bibitem [{\citenamefont {Mondal}\ \emph
  {et~al.}(2023{\natexlab{a}})\citenamefont {Mondal}, \citenamefont {Neufeld},
  \citenamefont {Yin}, \citenamefont {Nourbakhsh}, \citenamefont {Svoboda},
  \citenamefont {Rubio}, \citenamefont {Tancogne-Dejean},\ and\ \citenamefont
  {W{\"o}rner}}]{mondal2023}%
  \BibitemOpen
  \bibfield  {author} {\bibinfo {author} {\bibfnamefont {A.}~\bibnamefont
  {Mondal}}, \bibinfo {author} {\bibfnamefont {O.}~\bibnamefont {Neufeld}},
  \bibinfo {author} {\bibfnamefont {Z.}~\bibnamefont {Yin}}, \bibinfo {author}
  {\bibfnamefont {Z.}~\bibnamefont {Nourbakhsh}}, \bibinfo {author}
  {\bibfnamefont {V.}~\bibnamefont {Svoboda}}, \bibinfo {author} {\bibfnamefont
  {A.}~\bibnamefont {Rubio}}, \bibinfo {author} {\bibfnamefont
  {N.}~\bibnamefont {Tancogne-Dejean}},\ and\ \bibinfo {author} {\bibfnamefont
  {H.~J.}\ \bibnamefont {W{\"o}rner}},\ }\bibfield  {title} {\bibinfo {title}
  {High-harmonic spectroscopy of low-energy electron-scattering dynamics in
  liquids},\ }\href@noop {} {\bibfield  {journal} {\bibinfo  {journal} {Nature
  Physics}\ }\textbf {\bibinfo {volume} {19}},\ \bibinfo {pages} {1813}
  (\bibinfo {year} {2023}{\natexlab{a}})}\BibitemShut {NoStop}%
\bibitem [{\citenamefont {Schmid}\ \emph {et~al.}(2021)\citenamefont {Schmid},
  \citenamefont {Weigl}, \citenamefont {Gr\"{o}ssing}, \citenamefont {Junk},
  \citenamefont {Gorini}, \citenamefont {Schlauderer}, \citenamefont {Ito},
  \citenamefont {Meierhofer}, \citenamefont {Hofmann}, \citenamefont
  {Afanasiev}, \citenamefont {Crewse}, \citenamefont {Kokh}, \citenamefont
  {Tereshchenko}, \citenamefont {G\"{u}dde}, \citenamefont {Evers},
  \citenamefont {Wilhelm}, \citenamefont {Richter}, \citenamefont {H\"{o}fer},\
  and\ \citenamefont {Huber}}]{Schmid2021}%
  \BibitemOpen
  \bibfield  {author} {\bibinfo {author} {\bibfnamefont {C.~P.}\ \bibnamefont
  {Schmid}}, \bibinfo {author} {\bibfnamefont {L.}~\bibnamefont {Weigl}},
  \bibinfo {author} {\bibfnamefont {P.}~\bibnamefont {Gr\"{o}ssing}}, \bibinfo
  {author} {\bibfnamefont {V.}~\bibnamefont {Junk}}, \bibinfo {author}
  {\bibfnamefont {C.}~\bibnamefont {Gorini}}, \bibinfo {author} {\bibfnamefont
  {S.}~\bibnamefont {Schlauderer}}, \bibinfo {author} {\bibfnamefont
  {S.}~\bibnamefont {Ito}}, \bibinfo {author} {\bibfnamefont {M.}~\bibnamefont
  {Meierhofer}}, \bibinfo {author} {\bibfnamefont {N.}~\bibnamefont {Hofmann}},
  \bibinfo {author} {\bibfnamefont {D.}~\bibnamefont {Afanasiev}}, \bibinfo
  {author} {\bibfnamefont {J.}~\bibnamefont {Crewse}}, \bibinfo {author}
  {\bibfnamefont {K.~A.}\ \bibnamefont {Kokh}}, \bibinfo {author}
  {\bibfnamefont {O.~E.}\ \bibnamefont {Tereshchenko}}, \bibinfo {author}
  {\bibfnamefont {J.}~\bibnamefont {G\"{u}dde}}, \bibinfo {author}
  {\bibfnamefont {F.}~\bibnamefont {Evers}}, \bibinfo {author} {\bibfnamefont
  {J.}~\bibnamefont {Wilhelm}}, \bibinfo {author} {\bibfnamefont
  {K.}~\bibnamefont {Richter}}, \bibinfo {author} {\bibfnamefont
  {U.}~\bibnamefont {H\"{o}fer}},\ and\ \bibinfo {author} {\bibfnamefont
  {R.}~\bibnamefont {Huber}},\ }\bibfield  {title} {\bibinfo {title} {Tunable
  non-integer high-harmonic generation in a topological insulator},\ }\href
  {https://doi.org/10.1038/s41586-021-03466-7} {\bibfield  {journal} {\bibinfo
  {journal} {Nature}\ }\textbf {\bibinfo {volume} {593}},\ \bibinfo {pages}
  {385} (\bibinfo {year} {2021})}\BibitemShut {NoStop}%
\bibitem [{\citenamefont {Bauer}\ and\ \citenamefont
  {Hansen}(2018)}]{Bauer2018}%
  \BibitemOpen
  \bibfield  {author} {\bibinfo {author} {\bibfnamefont {D.}~\bibnamefont
  {Bauer}}\ and\ \bibinfo {author} {\bibfnamefont {K.~K.}\ \bibnamefont
  {Hansen}},\ }\bibfield  {title} {\bibinfo {title} {High-harmonic generation
  in solids with and without topological edge states},\ }\href
  {https://doi.org/10.1103/PhysRevLett.120.177401} {\bibfield  {journal}
  {\bibinfo  {journal} {Phys. Rev. Lett.}\ }\textbf {\bibinfo {volume} {120}},\
  \bibinfo {pages} {177401} (\bibinfo {year} {2018})}\BibitemShut {NoStop}%
\bibitem [{\citenamefont {Theidel}\ \emph {et~al.}(2024)\citenamefont
  {Theidel}, \citenamefont {Cotte}, \citenamefont {Sondenheimer}, \citenamefont
  {Shiriaeva}, \citenamefont {Froidevaux}, \citenamefont {Severin},
  \citenamefont {Merdji-Larue}, \citenamefont {Mosel}, \citenamefont
  {Fr\"ohlich}, \citenamefont {Weber}, \citenamefont {Morgner}, \citenamefont
  {Kovacev}, \citenamefont {Biegert},\ and\ \citenamefont
  {Merdji}}]{Theidel2024}%
  \BibitemOpen
  \bibfield  {author} {\bibinfo {author} {\bibfnamefont {D.}~\bibnamefont
  {Theidel}}, \bibinfo {author} {\bibfnamefont {V.}~\bibnamefont {Cotte}},
  \bibinfo {author} {\bibfnamefont {R.}~\bibnamefont {Sondenheimer}}, \bibinfo
  {author} {\bibfnamefont {V.}~\bibnamefont {Shiriaeva}}, \bibinfo {author}
  {\bibfnamefont {M.}~\bibnamefont {Froidevaux}}, \bibinfo {author}
  {\bibfnamefont {V.}~\bibnamefont {Severin}}, \bibinfo {author} {\bibfnamefont
  {A.}~\bibnamefont {Merdji-Larue}}, \bibinfo {author} {\bibfnamefont
  {P.}~\bibnamefont {Mosel}}, \bibinfo {author} {\bibfnamefont
  {S.}~\bibnamefont {Fr\"ohlich}}, \bibinfo {author} {\bibfnamefont {K.-A.}\
  \bibnamefont {Weber}}, \bibinfo {author} {\bibfnamefont {U.}~\bibnamefont
  {Morgner}}, \bibinfo {author} {\bibfnamefont {M.}~\bibnamefont {Kovacev}},
  \bibinfo {author} {\bibfnamefont {J.}~\bibnamefont {Biegert}},\ and\ \bibinfo
  {author} {\bibfnamefont {H.}~\bibnamefont {Merdji}},\ }\bibfield  {title}
  {\bibinfo {title} {Evidence of the quantum optical nature of high-harmonic
  generation},\ }\href {https://doi.org/10.1103/PRXQuantum.5.040319} {\bibfield
   {journal} {\bibinfo  {journal} {PRX Quantum}\ }\textbf {\bibinfo {volume}
  {5}},\ \bibinfo {pages} {040319} (\bibinfo {year} {2024})}\BibitemShut
  {NoStop}%
\bibitem [{\citenamefont {Alcalà}\ \emph {et~al.}(2022)\citenamefont
  {Alcalà}, \citenamefont {Bhattacharya}, \citenamefont {Biegert},
  \citenamefont {Ciappina}, \citenamefont {Elu}, \citenamefont {Graß},
  \citenamefont {Grochowski}, \citenamefont {Lewenstein}, \citenamefont
  {Palau}, \citenamefont {Sidiropoulos}, \citenamefont {Steinle},\ and\
  \citenamefont {Tyulnev}}]{Jordi2022}%
  \BibitemOpen
  \bibfield  {author} {\bibinfo {author} {\bibfnamefont {J.}~\bibnamefont
  {Alcalà}}, \bibinfo {author} {\bibfnamefont {U.}~\bibnamefont
  {Bhattacharya}}, \bibinfo {author} {\bibfnamefont {J.}~\bibnamefont
  {Biegert}}, \bibinfo {author} {\bibfnamefont {M.}~\bibnamefont {Ciappina}},
  \bibinfo {author} {\bibfnamefont {U.}~\bibnamefont {Elu}}, \bibinfo {author}
  {\bibfnamefont {T.}~\bibnamefont {Graß}}, \bibinfo {author} {\bibfnamefont
  {P.~T.}\ \bibnamefont {Grochowski}}, \bibinfo {author} {\bibfnamefont
  {M.}~\bibnamefont {Lewenstein}}, \bibinfo {author} {\bibfnamefont
  {A.}~\bibnamefont {Palau}}, \bibinfo {author} {\bibfnamefont {T.~P.~H.}\
  \bibnamefont {Sidiropoulos}}, \bibinfo {author} {\bibfnamefont
  {T.}~\bibnamefont {Steinle}},\ and\ \bibinfo {author} {\bibfnamefont
  {I.}~\bibnamefont {Tyulnev}},\ }\bibfield  {title} {\bibinfo {title}
  {High-harmonic spectroscopy of quantum phase transitions in a high-tc
  superconductor},\ }\href {https://doi.org/10.1073/pnas.2207766119} {\bibfield
   {journal} {\bibinfo  {journal} {Proceedings of the National Academy of
  Sciences}\ }\textbf {\bibinfo {volume} {119}},\ \bibinfo {pages}
  {e2207766119} (\bibinfo {year} {2022})}\BibitemShut {NoStop}%
\bibitem [{\citenamefont {Uchida}\ \emph {et~al.}(2022)\citenamefont {Uchida},
  \citenamefont {Mattoni}, \citenamefont {Yonezawa}, \citenamefont {Nakamura},
  \citenamefont {Maeno},\ and\ \citenamefont {Tanaka}}]{Uchida2022}%
  \BibitemOpen
  \bibfield  {author} {\bibinfo {author} {\bibfnamefont {K.}~\bibnamefont
  {Uchida}}, \bibinfo {author} {\bibfnamefont {G.}~\bibnamefont {Mattoni}},
  \bibinfo {author} {\bibfnamefont {S.}~\bibnamefont {Yonezawa}}, \bibinfo
  {author} {\bibfnamefont {F.}~\bibnamefont {Nakamura}}, \bibinfo {author}
  {\bibfnamefont {Y.}~\bibnamefont {Maeno}},\ and\ \bibinfo {author}
  {\bibfnamefont {K.}~\bibnamefont {Tanaka}},\ }\bibfield  {title} {\bibinfo
  {title} {High-order harmonic generation and its unconventional scaling law in
  the mott-insulating ${\mathrm{ca}}_{2}{\mathrm{ruo}}_{4}$},\ }\href
  {https://doi.org/10.1103/PhysRevLett.128.127401} {\bibfield  {journal}
  {\bibinfo  {journal} {Phys. Rev. Lett.}\ }\textbf {\bibinfo {volume} {128}},\
  \bibinfo {pages} {127401} (\bibinfo {year} {2022})}\BibitemShut {NoStop}%
\bibitem [{\citenamefont {Lv}\ \emph {et~al.}(2021)\citenamefont {Lv},
  \citenamefont {Xu}, \citenamefont {Han}, \citenamefont {Zhang}, \citenamefont
  {Han}, \citenamefont {Zhou}, \citenamefont {Yao}, \citenamefont {Liu},
  \citenamefont {Lu}, \citenamefont {Weng}, \citenamefont {Xie}, \citenamefont
  {Chen}, \citenamefont {Hu}, \citenamefont {Chen},\ and\ \citenamefont
  {Zhu}}]{Lv2021}%
  \BibitemOpen
  \bibfield  {author} {\bibinfo {author} {\bibfnamefont {Y.-Y.}\ \bibnamefont
  {Lv}}, \bibinfo {author} {\bibfnamefont {J.}~\bibnamefont {Xu}}, \bibinfo
  {author} {\bibfnamefont {S.}~\bibnamefont {Han}}, \bibinfo {author}
  {\bibfnamefont {C.}~\bibnamefont {Zhang}}, \bibinfo {author} {\bibfnamefont
  {Y.}~\bibnamefont {Han}}, \bibinfo {author} {\bibfnamefont {J.}~\bibnamefont
  {Zhou}}, \bibinfo {author} {\bibfnamefont {S.-H.}\ \bibnamefont {Yao}},
  \bibinfo {author} {\bibfnamefont {X.-P.}\ \bibnamefont {Liu}}, \bibinfo
  {author} {\bibfnamefont {M.-H.}\ \bibnamefont {Lu}}, \bibinfo {author}
  {\bibfnamefont {H.}~\bibnamefont {Weng}}, \bibinfo {author} {\bibfnamefont
  {Z.}~\bibnamefont {Xie}}, \bibinfo {author} {\bibfnamefont {Y.~B.}\
  \bibnamefont {Chen}}, \bibinfo {author} {\bibfnamefont {J.}~\bibnamefont
  {Hu}}, \bibinfo {author} {\bibfnamefont {Y.-F.}\ \bibnamefont {Chen}},\ and\
  \bibinfo {author} {\bibfnamefont {S.}~\bibnamefont {Zhu}},\ }\bibfield
  {title} {\bibinfo {title} {High-harmonic generation in weyl semimetal
  {\^i}{\texttwosuperior}-wp2 crystals},\ }\href
  {https://doi.org/10.1038/s41467-021-26766-y} {\bibfield  {journal} {\bibinfo
  {journal} {Nature Communications}\ }\textbf {\bibinfo {volume} {12}},\
  \bibinfo {pages} {6437} (\bibinfo {year} {2021})}\BibitemShut {NoStop}%
\bibitem [{\citenamefont {Neufeld}\ \emph {et~al.}(2023)\citenamefont
  {Neufeld}, \citenamefont {Tancogne-Dejean}, \citenamefont {H\"ubener},
  \citenamefont {De~Giovannini},\ and\ \citenamefont {Rubio}}]{Neufeld2023}%
  \BibitemOpen
  \bibfield  {author} {\bibinfo {author} {\bibfnamefont {O.}~\bibnamefont
  {Neufeld}}, \bibinfo {author} {\bibfnamefont {N.}~\bibnamefont
  {Tancogne-Dejean}}, \bibinfo {author} {\bibfnamefont {H.}~\bibnamefont
  {H\"ubener}}, \bibinfo {author} {\bibfnamefont {U.}~\bibnamefont
  {De~Giovannini}},\ and\ \bibinfo {author} {\bibfnamefont {A.}~\bibnamefont
  {Rubio}},\ }\bibfield  {title} {\bibinfo {title} {Are there universal
  signatures of topological phases in high-harmonic generation? probably
  not.},\ }\href {https://doi.org/10.1103/PhysRevX.13.031011} {\bibfield
  {journal} {\bibinfo  {journal} {Phys. Rev. X}\ }\textbf {\bibinfo {volume}
  {13}},\ \bibinfo {pages} {031011} (\bibinfo {year} {2023})}\BibitemShut
  {NoStop}%
\bibitem [{\citenamefont {Yang}\ \emph {et~al.}(2019)\citenamefont {Yang},
  \citenamefont {Lu}, \citenamefont {Manjavacas}, \citenamefont {Luk},
  \citenamefont {Liu}, \citenamefont {Kelley}, \citenamefont {Maria},
  \citenamefont {Runnerstrom}, \citenamefont {Sinclair}, \citenamefont
  {Ghimire},\ and\ \citenamefont {Brener}}]{Yang2019}%
  \BibitemOpen
  \bibfield  {author} {\bibinfo {author} {\bibfnamefont {Y.}~\bibnamefont
  {Yang}}, \bibinfo {author} {\bibfnamefont {J.}~\bibnamefont {Lu}}, \bibinfo
  {author} {\bibfnamefont {A.}~\bibnamefont {Manjavacas}}, \bibinfo {author}
  {\bibfnamefont {T.~S.}\ \bibnamefont {Luk}}, \bibinfo {author} {\bibfnamefont
  {H.}~\bibnamefont {Liu}}, \bibinfo {author} {\bibfnamefont {K.}~\bibnamefont
  {Kelley}}, \bibinfo {author} {\bibfnamefont {J.-P.}\ \bibnamefont {Maria}},
  \bibinfo {author} {\bibfnamefont {E.~L.}\ \bibnamefont {Runnerstrom}},
  \bibinfo {author} {\bibfnamefont {M.~B.}\ \bibnamefont {Sinclair}}, \bibinfo
  {author} {\bibfnamefont {S.}~\bibnamefont {Ghimire}},\ and\ \bibinfo {author}
  {\bibfnamefont {I.}~\bibnamefont {Brener}},\ }\bibfield  {title} {\bibinfo
  {title} {High-harmonic generation from an epsilon-near-zero material},\
  }\href {https://doi.org/10.1038/s41567-019-0584-7} {\bibfield  {journal}
  {\bibinfo  {journal} {Nature Physics}\ }\textbf {\bibinfo {volume} {15}},\
  \bibinfo {pages} {1022} (\bibinfo {year} {2019})}\BibitemShut {NoStop}%
\bibitem [{\citenamefont {Orthodoxou}\ \emph {et~al.}(2021)\citenamefont
  {Orthodoxou}, \citenamefont {Za\"{i}r},\ and\ \citenamefont
  {Booth}}]{Orthodoxou2021}%
  \BibitemOpen
  \bibfield  {author} {\bibinfo {author} {\bibfnamefont {C.}~\bibnamefont
  {Orthodoxou}}, \bibinfo {author} {\bibfnamefont {A.}~\bibnamefont
  {Za\"{i}r}},\ and\ \bibinfo {author} {\bibfnamefont {G.~H.}\ \bibnamefont
  {Booth}},\ }\bibfield  {title} {\bibinfo {title} {High harmonic generation in
  two-dimensional mott insulators},\ }\href
  {https://doi.org/10.1038/s41535-021-00377-8} {\bibfield  {journal} {\bibinfo
  {journal} {npj Quantum Materials}\ }\textbf {\bibinfo {volume} {6}},\
  \bibinfo {pages} {76} (\bibinfo {year} {2021})}\BibitemShut {NoStop}%
\bibitem [{\citenamefont {Habibovi\'{c}}\ \emph {et~al.}(2024)\citenamefont
  {Habibovi\'{c}}, \citenamefont {Hamilton}, \citenamefont {Neufeld},\ and\
  \citenamefont {Rego}}]{HabiboviA2024}%
  \BibitemOpen
  \bibfield  {author} {\bibinfo {author} {\bibfnamefont {D.}~\bibnamefont
  {Habibovi\'{c}}}, \bibinfo {author} {\bibfnamefont {K.~R.}\ \bibnamefont
  {Hamilton}}, \bibinfo {author} {\bibfnamefont {O.}~\bibnamefont {Neufeld}},\
  and\ \bibinfo {author} {\bibfnamefont {L.}~\bibnamefont {Rego}},\ }\bibfield
  {title} {\bibinfo {title} {Emerging tailored light sources for studying
  chirality and symmetry},\ }\href {https://doi.org/10.1038/s42254-024-00763-8}
  {\bibfield  {journal} {\bibinfo  {journal} {Nature Reviews Physics}\ }\textbf
  {\bibinfo {volume} {6}},\ \bibinfo {pages} {663} (\bibinfo {year}
  {2024})}\BibitemShut {NoStop}%
\bibitem [{\citenamefont {Corkum}(1993)}]{Corkum1993}%
  \BibitemOpen
  \bibfield  {author} {\bibinfo {author} {\bibfnamefont {P.~B.}\ \bibnamefont
  {Corkum}},\ }\bibfield  {title} {\bibinfo {title} {{Plasma perspective on
  strong field multiphoton ionization}},\ }\href
  {https://doi.org/10.1103/PhysRevLett.71.1994} {\bibfield  {journal} {\bibinfo
   {journal} {Physical Review Letters}\ }\textbf {\bibinfo {volume} {71}},\
  \bibinfo {pages} {1994} (\bibinfo {year} {1993})}\BibitemShut {NoStop}%
\bibitem [{\citenamefont {Lewenstein}\ \emph
  {et~al.}(1994{\natexlab{a}})\citenamefont {Lewenstein}, \citenamefont
  {Balcou}, \citenamefont {Ivanov}, \citenamefont {L'Huillier},\ and\
  \citenamefont {Corkum}}]{Lewenstein1994}%
  \BibitemOpen
  \bibfield  {author} {\bibinfo {author} {\bibfnamefont {M.}~\bibnamefont
  {Lewenstein}}, \bibinfo {author} {\bibfnamefont {P.}~\bibnamefont {Balcou}},
  \bibinfo {author} {\bibfnamefont {M.~Y.}\ \bibnamefont {Ivanov}}, \bibinfo
  {author} {\bibfnamefont {A.}~\bibnamefont {L'Huillier}},\ and\ \bibinfo
  {author} {\bibfnamefont {P.~B.}\ \bibnamefont {Corkum}},\ }\bibfield  {title}
  {\bibinfo {title} {{Theory of high-harmonic generation by low-frequency laser
  fields}},\ }\href {https://doi.org/10.1103/PhysRevA.49.2117} {\bibfield
  {journal} {\bibinfo  {journal} {Physical Review A}\ }\textbf {\bibinfo
  {volume} {49}},\ \bibinfo {pages} {2117} (\bibinfo {year}
  {1994}{\natexlab{a}})}\BibitemShut {NoStop}%
\bibitem [{\citenamefont {Corkum}\ and\ \citenamefont
  {Krausz}(2007)}]{Corkum2007}%
  \BibitemOpen
  \bibfield  {author} {\bibinfo {author} {\bibfnamefont {P.~B.}\ \bibnamefont
  {Corkum}}\ and\ \bibinfo {author} {\bibfnamefont {F.}~\bibnamefont
  {Krausz}},\ }\bibfield  {title} {\bibinfo {title} {Attosecond science},\
  }\href {https://doi.org/10.1038/nphys620} {\bibfield  {journal} {\bibinfo
  {journal} {Nature Physics}\ }\textbf {\bibinfo {volume} {3}},\ \bibinfo
  {pages} {381} (\bibinfo {year} {2007})}\BibitemShut {NoStop}%
\bibitem [{\citenamefont {Calegari}\ \emph {et~al.}(2016)\citenamefont
  {Calegari}, \citenamefont {Sansone}, \citenamefont {Stagira}, \citenamefont
  {Vozzi},\ and\ \citenamefont {Nisoli}}]{Calegari2016}%
  \BibitemOpen
  \bibfield  {author} {\bibinfo {author} {\bibfnamefont {F.}~\bibnamefont
  {Calegari}}, \bibinfo {author} {\bibfnamefont {G.}~\bibnamefont {Sansone}},
  \bibinfo {author} {\bibfnamefont {S.}~\bibnamefont {Stagira}}, \bibinfo
  {author} {\bibfnamefont {C.}~\bibnamefont {Vozzi}},\ and\ \bibinfo {author}
  {\bibfnamefont {M.}~\bibnamefont {Nisoli}},\ }\bibfield  {title} {\bibinfo
  {title} {Advances in attosecond science},\ }\href
  {https://doi.org/10.1088/0953-4075/49/6/062001} {\bibfield  {journal}
  {\bibinfo  {journal} {Journal of Physics B: Atomic, Molecular and Optical
  Physics}\ }\textbf {\bibinfo {volume} {49}},\ \bibinfo {pages} {062001}
  (\bibinfo {year} {2016})}\BibitemShut {NoStop}%
\bibitem [{\citenamefont {Kraus}\ \emph {et~al.}(2015)\citenamefont {Kraus},
  \citenamefont {Mignolet}, \citenamefont {Baykusheva}, \citenamefont
  {Rupenyan}, \citenamefont {Horn{\'{y}}}, \citenamefont {Penka}, \citenamefont
  {Grassi}, \citenamefont {Tolstikhin}, \citenamefont {Schneider},
  \citenamefont {Jensen}, \citenamefont {Madsen}, \citenamefont {Bandrauk},
  \citenamefont {Remacle},\ and\ \citenamefont {W{\"{o}}rner}}]{Kraus2015}%
  \BibitemOpen
  \bibfield  {author} {\bibinfo {author} {\bibfnamefont {P.~M.}\ \bibnamefont
  {Kraus}}, \bibinfo {author} {\bibfnamefont {B.}~\bibnamefont {Mignolet}},
  \bibinfo {author} {\bibfnamefont {D.}~\bibnamefont {Baykusheva}}, \bibinfo
  {author} {\bibfnamefont {A.}~\bibnamefont {Rupenyan}}, \bibinfo {author}
  {\bibfnamefont {L.}~\bibnamefont {Horn{\'{y}}}}, \bibinfo {author}
  {\bibfnamefont {E.~F.}\ \bibnamefont {Penka}}, \bibinfo {author}
  {\bibfnamefont {G.}~\bibnamefont {Grassi}}, \bibinfo {author} {\bibfnamefont
  {O.~I.}\ \bibnamefont {Tolstikhin}}, \bibinfo {author} {\bibfnamefont
  {J.}~\bibnamefont {Schneider}}, \bibinfo {author} {\bibfnamefont
  {F.}~\bibnamefont {Jensen}}, \bibinfo {author} {\bibfnamefont {L.~B.}\
  \bibnamefont {Madsen}}, \bibinfo {author} {\bibfnamefont {A.~D.}\
  \bibnamefont {Bandrauk}}, \bibinfo {author} {\bibfnamefont {F.}~\bibnamefont
  {Remacle}},\ and\ \bibinfo {author} {\bibfnamefont {H.~J.}\ \bibnamefont
  {W{\"{o}}rner}},\ }\bibfield  {title} {\bibinfo {title} {{Measurement and
  laser control of attosecond charge migration in ionized iodoacetylene.}},\
  }\href {https://doi.org/10.1126/science.aab2160} {\bibfield  {journal}
  {\bibinfo  {journal} {Science (New York, N.Y.)}\ }\textbf {\bibinfo {volume}
  {350}},\ \bibinfo {pages} {790} (\bibinfo {year} {2015})}\BibitemShut
  {NoStop}%
\bibitem [{\citenamefont {Pertot}\ \emph {et~al.}(2017)\citenamefont {Pertot},
  \citenamefont {Schmidt}, \citenamefont {Matthews}, \citenamefont {Chauvet},
  \citenamefont {Huppert}, \citenamefont {Svoboda}, \citenamefont {von Conta},
  \citenamefont {Tehlar}, \citenamefont {Baykusheva}, \citenamefont {Wolf},\
  and\ \citenamefont {W{\"{o}}rner}}]{Pertot2017}%
  \BibitemOpen
  \bibfield  {author} {\bibinfo {author} {\bibfnamefont {Y.}~\bibnamefont
  {Pertot}}, \bibinfo {author} {\bibfnamefont {C.}~\bibnamefont {Schmidt}},
  \bibinfo {author} {\bibfnamefont {M.}~\bibnamefont {Matthews}}, \bibinfo
  {author} {\bibfnamefont {A.}~\bibnamefont {Chauvet}}, \bibinfo {author}
  {\bibfnamefont {M.}~\bibnamefont {Huppert}}, \bibinfo {author} {\bibfnamefont
  {V.}~\bibnamefont {Svoboda}}, \bibinfo {author} {\bibfnamefont
  {A.}~\bibnamefont {von Conta}}, \bibinfo {author} {\bibfnamefont
  {A.}~\bibnamefont {Tehlar}}, \bibinfo {author} {\bibfnamefont
  {D.}~\bibnamefont {Baykusheva}}, \bibinfo {author} {\bibfnamefont {J.-P.~P.}\
  \bibnamefont {Wolf}},\ and\ \bibinfo {author} {\bibfnamefont {H.~J.}\
  \bibnamefont {W{\"{o}}rner}},\ }\bibfield  {title} {\bibinfo {title}
  {{Time-resolved x-ray absorption spectroscopy with a water window
  high-harmonic source}},\ }\href {https://doi.org/10.1126/science.aah6114}
  {\bibfield  {journal} {\bibinfo  {journal} {Science}\ }\textbf {\bibinfo
  {volume} {355}},\ \bibinfo {pages} {264} (\bibinfo {year}
  {2017})}\BibitemShut {NoStop}%
\bibitem [{\citenamefont {Yin}\ \emph {et~al.}(2023)\citenamefont {Yin},
  \citenamefont {Chang}, \citenamefont {Bal{\v{c}}i{\=u}nas}, \citenamefont
  {Shakya}, \citenamefont {Djorovi{\'c}}, \citenamefont {Gaulier},
  \citenamefont {Fazio}, \citenamefont {Santra}, \citenamefont {Inhester},
  \citenamefont {Wolf} \emph {et~al.}}]{yin2023}%
  \BibitemOpen
  \bibfield  {author} {\bibinfo {author} {\bibfnamefont {Z.}~\bibnamefont
  {Yin}}, \bibinfo {author} {\bibfnamefont {Y.-P.}\ \bibnamefont {Chang}},
  \bibinfo {author} {\bibfnamefont {T.}~\bibnamefont {Bal{\v{c}}i{\=u}nas}},
  \bibinfo {author} {\bibfnamefont {Y.}~\bibnamefont {Shakya}}, \bibinfo
  {author} {\bibfnamefont {A.}~\bibnamefont {Djorovi{\'c}}}, \bibinfo {author}
  {\bibfnamefont {G.}~\bibnamefont {Gaulier}}, \bibinfo {author} {\bibfnamefont
  {G.}~\bibnamefont {Fazio}}, \bibinfo {author} {\bibfnamefont
  {R.}~\bibnamefont {Santra}}, \bibinfo {author} {\bibfnamefont
  {L.}~\bibnamefont {Inhester}}, \bibinfo {author} {\bibfnamefont {J.-P.}\
  \bibnamefont {Wolf}}, \emph {et~al.},\ }\bibfield  {title} {\bibinfo {title}
  {Femtosecond proton transfer in urea solutions probed by x-ray
  spectroscopy},\ }\href@noop {} {\bibfield  {journal} {\bibinfo  {journal}
  {Nature}\ }\textbf {\bibinfo {volume} {619}},\ \bibinfo {pages} {749}
  (\bibinfo {year} {2023})}\BibitemShut {NoStop}%
\bibitem [{\citenamefont {Wang}\ \emph {et~al.}(2022)\citenamefont {Wang},
  \citenamefont {Waters}, \citenamefont {Zhang}, \citenamefont {Suchan},
  \citenamefont {Svoboda}, \citenamefont {Luu}, \citenamefont {Perry},
  \citenamefont {Yin}, \citenamefont {Slav{\'\i}{\v{c}}ek},\ and\ \citenamefont
  {W{\"o}rner}}]{wang2022stilbene}%
  \BibitemOpen
  \bibfield  {author} {\bibinfo {author} {\bibfnamefont {C.}~\bibnamefont
  {Wang}}, \bibinfo {author} {\bibfnamefont {M.~D.}\ \bibnamefont {Waters}},
  \bibinfo {author} {\bibfnamefont {P.}~\bibnamefont {Zhang}}, \bibinfo
  {author} {\bibfnamefont {J.}~\bibnamefont {Suchan}}, \bibinfo {author}
  {\bibfnamefont {V.}~\bibnamefont {Svoboda}}, \bibinfo {author} {\bibfnamefont
  {T.~T.}\ \bibnamefont {Luu}}, \bibinfo {author} {\bibfnamefont
  {C.}~\bibnamefont {Perry}}, \bibinfo {author} {\bibfnamefont
  {Z.}~\bibnamefont {Yin}}, \bibinfo {author} {\bibfnamefont {P.}~\bibnamefont
  {Slav{\'\i}{\v{c}}ek}},\ and\ \bibinfo {author} {\bibfnamefont {H.~J.}\
  \bibnamefont {W{\"o}rner}},\ }\bibfield  {title} {\bibinfo {title} {Different
  timescales during ultrafast stilbene isomerization in the gas and liquid
  phases revealed using time-resolved photoelectron spectroscopy},\ }\href@noop
  {} {\bibfield  {journal} {\bibinfo  {journal} {Nature chemistry}\ }\textbf
  {\bibinfo {volume} {14}},\ \bibinfo {pages} {1126} (\bibinfo {year}
  {2022})}\BibitemShut {NoStop}%
\bibitem [{\citenamefont {Kneller}\ \emph {et~al.}(2024)\citenamefont
  {Kneller}, \citenamefont {Mor}, \citenamefont {Klimkin}, \citenamefont
  {Yaffe}, \citenamefont {Kr\"{u}ger}, \citenamefont {Azoury}, \citenamefont
  {Uzan-Narovlansky}, \citenamefont {Federman}, \citenamefont {Rajak},
  \citenamefont {Bruner}, \citenamefont {Smirnova}, \citenamefont
  {Patchkovskii}, \citenamefont {Mairesse}, \citenamefont {Ivanov},\ and\
  \citenamefont {Dudovich}}]{Kneller2024}%
  \BibitemOpen
  \bibfield  {author} {\bibinfo {author} {\bibfnamefont {O.}~\bibnamefont
  {Kneller}}, \bibinfo {author} {\bibfnamefont {C.}~\bibnamefont {Mor}},
  \bibinfo {author} {\bibfnamefont {N.~D.}\ \bibnamefont {Klimkin}}, \bibinfo
  {author} {\bibfnamefont {N.}~\bibnamefont {Yaffe}}, \bibinfo {author}
  {\bibfnamefont {M.}~\bibnamefont {Kr\"{u}ger}}, \bibinfo {author}
  {\bibfnamefont {D.}~\bibnamefont {Azoury}}, \bibinfo {author} {\bibfnamefont
  {A.~J.}\ \bibnamefont {Uzan-Narovlansky}}, \bibinfo {author} {\bibfnamefont
  {Y.}~\bibnamefont {Federman}}, \bibinfo {author} {\bibfnamefont
  {D.}~\bibnamefont {Rajak}}, \bibinfo {author} {\bibfnamefont {B.~D.}\
  \bibnamefont {Bruner}}, \bibinfo {author} {\bibfnamefont {O.}~\bibnamefont
  {Smirnova}}, \bibinfo {author} {\bibfnamefont {S.}~\bibnamefont
  {Patchkovskii}}, \bibinfo {author} {\bibfnamefont {Y.}~\bibnamefont
  {Mairesse}}, \bibinfo {author} {\bibfnamefont {M.}~\bibnamefont {Ivanov}},\
  and\ \bibinfo {author} {\bibfnamefont {N.}~\bibnamefont {Dudovich}},\
  }\bibfield  {title} {\bibinfo {title} {Attosecond transient interferometry},\
  }\bibfield  {journal} {\bibinfo  {journal} {Nature Photonics}\ }\href
  {https://doi.org/10.1038/s41566-024-01556-2} {10.1038/s41566-024-01556-2}
  (\bibinfo {year} {2024})\BibitemShut {NoStop}%
\bibitem [{\citenamefont {Yang}\ \emph {et~al.}(2021)\citenamefont {Yang},
  \citenamefont {Mainz}, \citenamefont {Rossi}, \citenamefont {Scheiba},
  \citenamefont {Silva-Toledo}, \citenamefont {Keathley}, \citenamefont
  {Cirmi},\ and\ \citenamefont {K{\"a}rtner}}]{yang2021strong}%
  \BibitemOpen
  \bibfield  {author} {\bibinfo {author} {\bibfnamefont {Y.}~\bibnamefont
  {Yang}}, \bibinfo {author} {\bibfnamefont {R.~E.}\ \bibnamefont {Mainz}},
  \bibinfo {author} {\bibfnamefont {G.~M.}\ \bibnamefont {Rossi}}, \bibinfo
  {author} {\bibfnamefont {F.}~\bibnamefont {Scheiba}}, \bibinfo {author}
  {\bibfnamefont {M.~A.}\ \bibnamefont {Silva-Toledo}}, \bibinfo {author}
  {\bibfnamefont {P.~D.}\ \bibnamefont {Keathley}}, \bibinfo {author}
  {\bibfnamefont {G.}~\bibnamefont {Cirmi}},\ and\ \bibinfo {author}
  {\bibfnamefont {F.~X.}\ \bibnamefont {K{\"a}rtner}},\ }\bibfield  {title}
  {\bibinfo {title} {Strong-field coherent control of isolated attosecond pulse
  generation},\ }\href@noop {} {\bibfield  {journal} {\bibinfo  {journal}
  {Nature Communications}\ }\textbf {\bibinfo {volume} {12}},\ \bibinfo {pages}
  {6641} (\bibinfo {year} {2021})}\BibitemShut {NoStop}%
\bibitem [{\citenamefont {Wituschek}\ \emph {et~al.}(2020)\citenamefont
  {Wituschek}, \citenamefont {Bruder}, \citenamefont {Allaria}, \citenamefont
  {Bangert}, \citenamefont {Binz}, \citenamefont {Borghes}, \citenamefont
  {Callegari}, \citenamefont {Cerullo}, \citenamefont {Cinquegrana},
  \citenamefont {Giannessi} \emph {et~al.}}]{wituschek2020tracking}%
  \BibitemOpen
  \bibfield  {author} {\bibinfo {author} {\bibfnamefont {A.}~\bibnamefont
  {Wituschek}}, \bibinfo {author} {\bibfnamefont {L.}~\bibnamefont {Bruder}},
  \bibinfo {author} {\bibfnamefont {E.}~\bibnamefont {Allaria}}, \bibinfo
  {author} {\bibfnamefont {U.}~\bibnamefont {Bangert}}, \bibinfo {author}
  {\bibfnamefont {M.}~\bibnamefont {Binz}}, \bibinfo {author} {\bibfnamefont
  {R.}~\bibnamefont {Borghes}}, \bibinfo {author} {\bibfnamefont
  {C.}~\bibnamefont {Callegari}}, \bibinfo {author} {\bibfnamefont
  {G.}~\bibnamefont {Cerullo}}, \bibinfo {author} {\bibfnamefont
  {P.}~\bibnamefont {Cinquegrana}}, \bibinfo {author} {\bibfnamefont
  {L.}~\bibnamefont {Giannessi}}, \emph {et~al.},\ }\bibfield  {title}
  {\bibinfo {title} {Tracking attosecond electronic coherences using
  phase-manipulated extreme ultraviolet pulses},\ }\href@noop {} {\bibfield
  {journal} {\bibinfo  {journal} {Nature communications}\ }\textbf {\bibinfo
  {volume} {11}},\ \bibinfo {pages} {883} (\bibinfo {year} {2020})}\BibitemShut
  {NoStop}%
\bibitem [{\citenamefont {Hentschel}\ \emph {et~al.}(2001)\citenamefont
  {Hentschel}, \citenamefont {Kienberger}, \citenamefont {Spielmann},
  \citenamefont {Reider}, \citenamefont {Milosevic}, \citenamefont {Brabec},
  \citenamefont {Corkum}, \citenamefont {Heinzmann}, \citenamefont {Drescher},\
  and\ \citenamefont {Krausz}}]{Hentschel2001}%
  \BibitemOpen
  \bibfield  {author} {\bibinfo {author} {\bibfnamefont {M.}~\bibnamefont
  {Hentschel}}, \bibinfo {author} {\bibfnamefont {R.}~\bibnamefont
  {Kienberger}}, \bibinfo {author} {\bibfnamefont {C.}~\bibnamefont
  {Spielmann}}, \bibinfo {author} {\bibfnamefont {G.~A.}\ \bibnamefont
  {Reider}}, \bibinfo {author} {\bibfnamefont {N.}~\bibnamefont {Milosevic}},
  \bibinfo {author} {\bibfnamefont {T.}~\bibnamefont {Brabec}}, \bibinfo
  {author} {\bibfnamefont {P.}~\bibnamefont {Corkum}}, \bibinfo {author}
  {\bibfnamefont {U.}~\bibnamefont {Heinzmann}}, \bibinfo {author}
  {\bibfnamefont {M.}~\bibnamefont {Drescher}},\ and\ \bibinfo {author}
  {\bibfnamefont {F.}~\bibnamefont {Krausz}},\ }\bibfield  {title} {\bibinfo
  {title} {{Attosecond metrology}},\ }\href {https://doi.org/10.1038/35107000}
  {\bibfield  {journal} {\bibinfo  {journal} {Nature}\ }\textbf {\bibinfo
  {volume} {414}},\ \bibinfo {pages} {509} (\bibinfo {year}
  {2001})}\BibitemShut {NoStop}%
\bibitem [{\citenamefont {Zhao}\ \emph {et~al.}(2012)\citenamefont {Zhao},
  \citenamefont {Zhang}, \citenamefont {Chini}, \citenamefont {Wu},
  \citenamefont {Wang},\ and\ \citenamefont {Chang}}]{zhao12a}%
  \BibitemOpen
  \bibfield  {author} {\bibinfo {author} {\bibfnamefont {K.}~\bibnamefont
  {Zhao}}, \bibinfo {author} {\bibfnamefont {Q.}~\bibnamefont {Zhang}},
  \bibinfo {author} {\bibfnamefont {M.}~\bibnamefont {Chini}}, \bibinfo
  {author} {\bibfnamefont {Y.}~\bibnamefont {Wu}}, \bibinfo {author}
  {\bibfnamefont {X.}~\bibnamefont {Wang}},\ and\ \bibinfo {author}
  {\bibfnamefont {Z.}~\bibnamefont {Chang}},\ }\bibfield  {title} {\bibinfo
  {title} {Tailoring a 67 attosecond pulse through advantageous
  phase-mismatch},\ }\href {http://ol.osa.org/abstract.cfm?URI=ol-37-18-3891}
  {\bibfield  {journal} {\bibinfo  {journal} {Optics Letters}\ }\textbf
  {\bibinfo {volume} {37}},\ \bibinfo {pages} {3891} (\bibinfo {year}
  {2012})}\BibitemShut {NoStop}%
\bibitem [{\citenamefont {Gaumnitz}\ \emph {et~al.}(2017)\citenamefont
  {Gaumnitz}, \citenamefont {Jain}, \citenamefont {Pertot}, \citenamefont
  {Huppert}, \citenamefont {Jordan}, \citenamefont {Ardana-Lamas},\ and\
  \citenamefont {W{\"{o}}rner}}]{Gaumnitz2017}%
  \BibitemOpen
  \bibfield  {author} {\bibinfo {author} {\bibfnamefont {T.}~\bibnamefont
  {Gaumnitz}}, \bibinfo {author} {\bibfnamefont {A.}~\bibnamefont {Jain}},
  \bibinfo {author} {\bibfnamefont {Y.}~\bibnamefont {Pertot}}, \bibinfo
  {author} {\bibfnamefont {M.}~\bibnamefont {Huppert}}, \bibinfo {author}
  {\bibfnamefont {I.}~\bibnamefont {Jordan}}, \bibinfo {author} {\bibfnamefont
  {F.}~\bibnamefont {Ardana-Lamas}},\ and\ \bibinfo {author} {\bibfnamefont
  {H.~J.}\ \bibnamefont {W{\"{o}}rner}},\ }\bibfield  {title} {\bibinfo {title}
  {{Streaking of 43-attosecond soft-X-ray pulses generated by a passively
  CEP-stable mid-infrared driver}},\ }\href
  {https://doi.org/10.1364/OE.25.027506} {\bibfield  {journal} {\bibinfo
  {journal} {Opt. Express}\ }\textbf {\bibinfo {volume} {25}},\ \bibinfo
  {pages} {27506} (\bibinfo {year} {2017})}\BibitemShut {NoStop}%
\bibitem [{\citenamefont {Ghimire}\ \emph
  {et~al.}(2011{\natexlab{b}})\citenamefont {Ghimire}, \citenamefont
  {DiChiara}, \citenamefont {Sistrunk}, \citenamefont {Szafruga}, \citenamefont
  {Agostini}, \citenamefont {DiMauro},\ and\ \citenamefont
  {Reis}}]{Ghimire2011}%
  \BibitemOpen
  \bibfield  {author} {\bibinfo {author} {\bibfnamefont {S.}~\bibnamefont
  {Ghimire}}, \bibinfo {author} {\bibfnamefont {A.~D.}\ \bibnamefont
  {DiChiara}}, \bibinfo {author} {\bibfnamefont {E.}~\bibnamefont {Sistrunk}},
  \bibinfo {author} {\bibfnamefont {U.~B.}\ \bibnamefont {Szafruga}}, \bibinfo
  {author} {\bibfnamefont {P.}~\bibnamefont {Agostini}}, \bibinfo {author}
  {\bibfnamefont {L.~F.}\ \bibnamefont {DiMauro}},\ and\ \bibinfo {author}
  {\bibfnamefont {D.~A.}\ \bibnamefont {Reis}},\ }\bibfield  {title} {\bibinfo
  {title} {Redshift in the optical absorption of zno single crystals in the
  presence of an intense midinfrared laser field},\ }\href@noop {} {\bibfield
  {journal} {\bibinfo  {journal} {Physical review letters}\ }\textbf {\bibinfo
  {volume} {107}},\ \bibinfo {pages} {167407} (\bibinfo {year}
  {2011}{\natexlab{b}})}\BibitemShut {NoStop}%
\bibitem [{\citenamefont {Luu}\ \emph {et~al.}(2015)\citenamefont {Luu},
  \citenamefont {Garg}, \citenamefont {Kruchinin}, \citenamefont {Moulet},
  \citenamefont {Hassan},\ and\ \citenamefont {Goulielmakis}}]{luu2015}%
  \BibitemOpen
  \bibfield  {author} {\bibinfo {author} {\bibfnamefont {T.~T.}\ \bibnamefont
  {Luu}}, \bibinfo {author} {\bibfnamefont {M.}~\bibnamefont {Garg}}, \bibinfo
  {author} {\bibfnamefont {S.~Y.}\ \bibnamefont {Kruchinin}}, \bibinfo {author}
  {\bibfnamefont {A.}~\bibnamefont {Moulet}}, \bibinfo {author} {\bibfnamefont
  {M.~T.}\ \bibnamefont {Hassan}},\ and\ \bibinfo {author} {\bibfnamefont
  {E.}~\bibnamefont {Goulielmakis}},\ }\bibfield  {title} {\bibinfo {title}
  {Extreme ultraviolet high-harmonic spectroscopy of solids},\ }\href@noop {}
  {\bibfield  {journal} {\bibinfo  {journal} {Nature}\ }\textbf {\bibinfo
  {volume} {521}},\ \bibinfo {pages} {498} (\bibinfo {year}
  {2015})}\BibitemShut {NoStop}%
\bibitem [{\citenamefont {Ghimire}\ and\ \citenamefont
  {Reis}(2019)}]{ghimire2019}%
  \BibitemOpen
  \bibfield  {author} {\bibinfo {author} {\bibfnamefont {S.}~\bibnamefont
  {Ghimire}}\ and\ \bibinfo {author} {\bibfnamefont {D.~A.}\ \bibnamefont
  {Reis}},\ }\bibfield  {title} {\bibinfo {title} {High-harmonic generation
  from solids},\ }\href@noop {} {\bibfield  {journal} {\bibinfo  {journal}
  {Nature physics}\ }\textbf {\bibinfo {volume} {15}},\ \bibinfo {pages} {10}
  (\bibinfo {year} {2019})}\BibitemShut {NoStop}%
\bibitem [{\citenamefont {Goulielmakis}\ and\ \citenamefont
  {Brabec}(2022)}]{Goulielmakis2022}%
  \BibitemOpen
  \bibfield  {author} {\bibinfo {author} {\bibfnamefont {E.}~\bibnamefont
  {Goulielmakis}}\ and\ \bibinfo {author} {\bibfnamefont {T.}~\bibnamefont
  {Brabec}},\ }\bibfield  {title} {\bibinfo {title} {High harmonic generation
  in condensed matter},\ }\href {https://doi.org/10.1038/s41566-022-00988-y}
  {\bibfield  {journal} {\bibinfo  {journal} {Nature Photonics}\ }\textbf
  {\bibinfo {volume} {16}},\ \bibinfo {pages} {411} (\bibinfo {year}
  {2022})}\BibitemShut {NoStop}%
\bibitem [{\citenamefont {Yue}\ and\ \citenamefont {Gaarde}(2022)}]{Yue2022}%
  \BibitemOpen
  \bibfield  {author} {\bibinfo {author} {\bibfnamefont {L.}~\bibnamefont
  {Yue}}\ and\ \bibinfo {author} {\bibfnamefont {M.~B.}\ \bibnamefont
  {Gaarde}},\ }\bibfield  {title} {\bibinfo {title} {Introduction to theory of
  high-harmonic generation in solids: tutorial},\ }\href
  {https://doi.org/10.1364/JOSAB.448602} {\bibfield  {journal} {\bibinfo
  {journal} {Journal of the Optical Society of America B}\ }\textbf {\bibinfo
  {volume} {39}},\ \bibinfo {pages} {535} (\bibinfo {year} {2022})}\BibitemShut
  {NoStop}%
\bibitem [{\citenamefont {Vampa}\ \emph
  {et~al.}(2015{\natexlab{b}})\citenamefont {Vampa}, \citenamefont {McDonald},
  \citenamefont {Orlando}, \citenamefont {Corkum},\ and\ \citenamefont
  {Brabec}}]{Vampa2015phys}%
  \BibitemOpen
  \bibfield  {author} {\bibinfo {author} {\bibfnamefont {G.}~\bibnamefont
  {Vampa}}, \bibinfo {author} {\bibfnamefont {C.~R.}\ \bibnamefont {McDonald}},
  \bibinfo {author} {\bibfnamefont {G.}~\bibnamefont {Orlando}}, \bibinfo
  {author} {\bibfnamefont {P.~B.}\ \bibnamefont {Corkum}},\ and\ \bibinfo
  {author} {\bibfnamefont {T.}~\bibnamefont {Brabec}},\ }\bibfield  {title}
  {\bibinfo {title} {Semiclassical analysis of high harmonic generation in bulk
  crystals},\ }\href {https://doi.org/10.1103/PhysRevB.91.064302} {\bibfield
  {journal} {\bibinfo  {journal} {Physical Review B}\ }\textbf {\bibinfo
  {volume} {91}},\ \bibinfo {pages} {064302} (\bibinfo {year}
  {2015}{\natexlab{b}})}\BibitemShut {NoStop}%
\bibitem [{\citenamefont {Wu}\ \emph {et~al.}(2016)\citenamefont {Wu},
  \citenamefont {Browne}, \citenamefont {Schafer},\ and\ \citenamefont
  {Gaarde}}]{Wu2016}%
  \BibitemOpen
  \bibfield  {author} {\bibinfo {author} {\bibfnamefont {M.}~\bibnamefont
  {Wu}}, \bibinfo {author} {\bibfnamefont {D.~A.}\ \bibnamefont {Browne}},
  \bibinfo {author} {\bibfnamefont {K.~J.}\ \bibnamefont {Schafer}},\ and\
  \bibinfo {author} {\bibfnamefont {M.~B.}\ \bibnamefont {Gaarde}},\ }\bibfield
   {title} {\bibinfo {title} {Multilevel perspective on high-order harmonic
  generation in solids},\ }\href {https://doi.org/10.1103/PhysRevA.94.063403}
  {\bibfield  {journal} {\bibinfo  {journal} {Phys. Rev. A}\ }\textbf {\bibinfo
  {volume} {94}},\ \bibinfo {pages} {063403} (\bibinfo {year}
  {2016})}\BibitemShut {NoStop}%
\bibitem [{\citenamefont {Yue}\ and\ \citenamefont {Gaarde}(2020)}]{Yue2020}%
  \BibitemOpen
  \bibfield  {author} {\bibinfo {author} {\bibfnamefont {L.}~\bibnamefont
  {Yue}}\ and\ \bibinfo {author} {\bibfnamefont {M.~B.}\ \bibnamefont
  {Gaarde}},\ }\bibfield  {title} {\bibinfo {title} {Imperfect recollisions in
  high-harmonic generation in solids},\ }\href
  {https://doi.org/10.1103/PhysRevLett.124.153204} {\bibfield  {journal}
  {\bibinfo  {journal} {Phys. Rev. Lett.}\ }\textbf {\bibinfo {volume} {124}},\
  \bibinfo {pages} {153204} (\bibinfo {year} {2020})}\BibitemShut {NoStop}%
\bibitem [{\citenamefont {Osika}\ \emph {et~al.}(2017)\citenamefont {Osika},
  \citenamefont {Chac\'on}, \citenamefont {Ortmann}, \citenamefont {Su\'arez},
  \citenamefont {P\'erez-Hern\'andez}, \citenamefont {Szafran}, \citenamefont
  {Ciappina}, \citenamefont {Sols}, \citenamefont {Landsman},\ and\
  \citenamefont {Lewenstein}}]{Osika2017}%
  \BibitemOpen
  \bibfield  {author} {\bibinfo {author} {\bibfnamefont {E.~N.}\ \bibnamefont
  {Osika}}, \bibinfo {author} {\bibfnamefont {A.}~\bibnamefont {Chac\'on}},
  \bibinfo {author} {\bibfnamefont {L.}~\bibnamefont {Ortmann}}, \bibinfo
  {author} {\bibfnamefont {N.}~\bibnamefont {Su\'arez}}, \bibinfo {author}
  {\bibfnamefont {J.~A.}\ \bibnamefont {P\'erez-Hern\'andez}}, \bibinfo
  {author} {\bibfnamefont {B.}~\bibnamefont {Szafran}}, \bibinfo {author}
  {\bibfnamefont {M.~F.}\ \bibnamefont {Ciappina}}, \bibinfo {author}
  {\bibfnamefont {F.}~\bibnamefont {Sols}}, \bibinfo {author} {\bibfnamefont
  {A.~S.}\ \bibnamefont {Landsman}},\ and\ \bibinfo {author} {\bibfnamefont
  {M.}~\bibnamefont {Lewenstein}},\ }\bibfield  {title} {\bibinfo {title}
  {Wannier-bloch approach to localization in high-harmonics generation in
  solids},\ }\href {https://doi.org/10.1103/PhysRevX.7.021017} {\bibfield
  {journal} {\bibinfo  {journal} {Phys. Rev. X}\ }\textbf {\bibinfo {volume}
  {7}},\ \bibinfo {pages} {021017} (\bibinfo {year} {2017})}\BibitemShut
  {NoStop}%
\bibitem [{\citenamefont {You}\ \emph {et~al.}(2017{\natexlab{a}})\citenamefont
  {You}, \citenamefont {Reis},\ and\ \citenamefont {Ghimire}}]{You2017v2}%
  \BibitemOpen
  \bibfield  {author} {\bibinfo {author} {\bibfnamefont {Y.~S.}\ \bibnamefont
  {You}}, \bibinfo {author} {\bibfnamefont {D.~A.~A.}\ \bibnamefont {Reis}},\
  and\ \bibinfo {author} {\bibfnamefont {S.}~\bibnamefont {Ghimire}},\
  }\bibfield  {title} {\bibinfo {title} {{Anisotropic high-harmonic generation
  in bulk crystals}},\ }\href {https://doi.org/10.1038/nphys3955} {\bibfield
  {journal} {\bibinfo  {journal} {Nature Physics}\ }\textbf {\bibinfo {volume}
  {13}},\ \bibinfo {pages} {345} (\bibinfo {year}
  {2017}{\natexlab{a}})}\BibitemShut {NoStop}%
\bibitem [{\citenamefont {M.~S.}\ \emph {et~al.}(2019)\citenamefont {M.~S.},
  \citenamefont {Pattanayak}, \citenamefont {Ivanov},\ and\ \citenamefont
  {Dixit}}]{PhysRevA.100.043420}%
  \BibitemOpen
  \bibfield  {author} {\bibinfo {author} {\bibfnamefont {M.}~\bibnamefont
  {M.~S.}}, \bibinfo {author} {\bibfnamefont {A.}~\bibnamefont {Pattanayak}},
  \bibinfo {author} {\bibfnamefont {M.}~\bibnamefont {Ivanov}},\ and\ \bibinfo
  {author} {\bibfnamefont {G.}~\bibnamefont {Dixit}},\ }\bibfield  {title}
  {\bibinfo {title} {Direct numerical observation of real-space recollision in
  high-order harmonic generation from solids},\ }\href
  {https://doi.org/10.1103/PhysRevA.100.043420} {\bibfield  {journal} {\bibinfo
   {journal} {Phys. Rev. A}\ }\textbf {\bibinfo {volume} {100}},\ \bibinfo
  {pages} {043420} (\bibinfo {year} {2019})}\BibitemShut {NoStop}%
\bibitem [{\citenamefont {Brown}\ \emph
  {et~al.}(2024{\natexlab{a}})\citenamefont {Brown}, \citenamefont
  {Jim\'enez-Gal\'an}, \citenamefont {Silva},\ and\ \citenamefont
  {Ivanov}}]{PhysRevResearch.6.043005}%
  \BibitemOpen
  \bibfield  {author} {\bibinfo {author} {\bibfnamefont {G.~G.}\ \bibnamefont
  {Brown}}, \bibinfo {author} {\bibfnamefont {A.}~\bibnamefont
  {Jim\'enez-Gal\'an}}, \bibinfo {author} {\bibfnamefont {R.~E.~F.}\
  \bibnamefont {Silva}},\ and\ \bibinfo {author} {\bibfnamefont
  {M.}~\bibnamefont {Ivanov}},\ }\bibfield  {title} {\bibinfo {title}
  {Real-space perspective on dephasing in solid-state high harmonic
  generation},\ }\href {https://doi.org/10.1103/PhysRevResearch.6.043005}
  {\bibfield  {journal} {\bibinfo  {journal} {Phys. Rev. Res.}\ }\textbf
  {\bibinfo {volume} {6}},\ \bibinfo {pages} {043005} (\bibinfo {year}
  {2024}{\natexlab{a}})}\BibitemShut {NoStop}%
\bibitem [{\citenamefont {Brown}\ \emph
  {et~al.}(2024{\natexlab{b}})\citenamefont {Brown}, \citenamefont
  {Jim{\'{e}}nez-Gal{\'{a}}n}, \citenamefont {Silva},\ and\ \citenamefont
  {Ivanov}}]{Brown24}%
  \BibitemOpen
  \bibfield  {author} {\bibinfo {author} {\bibfnamefont {G.~G.}\ \bibnamefont
  {Brown}}, \bibinfo {author} {\bibfnamefont {{\'{A}}.}~\bibnamefont
  {Jim{\'{e}}nez-Gal{\'{a}}n}}, \bibinfo {author} {\bibfnamefont {R.~E.~F.}\
  \bibnamefont {Silva}},\ and\ \bibinfo {author} {\bibfnamefont
  {M.}~\bibnamefont {Ivanov}},\ }\bibfield  {title} {\bibinfo {title}
  {{Ultrafast dephasing in solid-state high harmonic generation: macroscopic
  origin revealed by real-space dynamics \[Invited\]}},\ }\href
  {https://doi.org/10.1364/JOSAB.513543} {\bibfield  {journal} {\bibinfo
  {journal} {J. Opt. Soc. Am. B}\ }\textbf {\bibinfo {volume} {41}},\ \bibinfo
  {pages} {B40} (\bibinfo {year} {2024}{\natexlab{b}})}\BibitemShut {NoStop}%
\bibitem [{\citenamefont {Parks}\ \emph {et~al.}(2020)\citenamefont {Parks},
  \citenamefont {Ernotte}, \citenamefont {Thorpe}, \citenamefont {McDonald},
  \citenamefont {Corkum}, \citenamefont {Taucer},\ and\ \citenamefont
  {Brabec}}]{Parks20}%
  \BibitemOpen
  \bibfield  {author} {\bibinfo {author} {\bibfnamefont {A.~M.}\ \bibnamefont
  {Parks}}, \bibinfo {author} {\bibfnamefont {G.}~\bibnamefont {Ernotte}},
  \bibinfo {author} {\bibfnamefont {A.}~\bibnamefont {Thorpe}}, \bibinfo
  {author} {\bibfnamefont {C.~R.}\ \bibnamefont {McDonald}}, \bibinfo {author}
  {\bibfnamefont {P.~B.}\ \bibnamefont {Corkum}}, \bibinfo {author}
  {\bibfnamefont {M.}~\bibnamefont {Taucer}},\ and\ \bibinfo {author}
  {\bibfnamefont {T.}~\bibnamefont {Brabec}},\ }\bibfield  {title} {\bibinfo
  {title} {{Wannier quasi-classical approach to high harmonic generation in
  semiconductors}},\ }\href {https://doi.org/10.1364/OPTICA.402393} {\bibfield
  {journal} {\bibinfo  {journal} {Optica}\ }\textbf {\bibinfo {volume} {7}},\
  \bibinfo {pages} {1764} (\bibinfo {year} {2020})}\BibitemShut {NoStop}%
\bibitem [{\citenamefont {You}\ \emph {et~al.}(2018)\citenamefont {You},
  \citenamefont {Cunningham}, \citenamefont {Reis},\ and\ \citenamefont
  {Ghimire}}]{You2018}%
  \BibitemOpen
  \bibfield  {author} {\bibinfo {author} {\bibfnamefont {Y.~S.}\ \bibnamefont
  {You}}, \bibinfo {author} {\bibfnamefont {E.}~\bibnamefont {Cunningham}},
  \bibinfo {author} {\bibfnamefont {D.~A.}\ \bibnamefont {Reis}},\ and\
  \bibinfo {author} {\bibfnamefont {S.}~\bibnamefont {Ghimire}},\ }\bibfield
  {title} {\bibinfo {title} {{Probing periodic potential of crystals via
  strong-field re-scattering}},\ }\href
  {https://doi.org/10.1088/1361-6455/aac11d} {\bibfield  {journal} {\bibinfo
  {journal} {Journal of Physics B: Atomic, Molecular and Optical Physics}\
  }\textbf {\bibinfo {volume} {51}},\ \bibinfo {pages} {114002} (\bibinfo
  {year} {2018})}\BibitemShut {NoStop}%
\bibitem [{\citenamefont {Wang}\ \emph {et~al.}(2020)\citenamefont {Wang},
  \citenamefont {Tancogne-Dejean}, \citenamefont {Altarelli}, \citenamefont
  {Rubio},\ and\ \citenamefont {Sato}}]{wang2020role}%
  \BibitemOpen
  \bibfield  {author} {\bibinfo {author} {\bibfnamefont {C.-M.}\ \bibnamefont
  {Wang}}, \bibinfo {author} {\bibfnamefont {N.}~\bibnamefont
  {Tancogne-Dejean}}, \bibinfo {author} {\bibfnamefont {M.}~\bibnamefont
  {Altarelli}}, \bibinfo {author} {\bibfnamefont {A.}~\bibnamefont {Rubio}},\
  and\ \bibinfo {author} {\bibfnamefont {S.~A.}\ \bibnamefont {Sato}},\
  }\bibfield  {title} {\bibinfo {title} {Role of electron scattering on the
  high-order harmonic generation from solids},\ }\href@noop {} {\bibfield
  {journal} {\bibinfo  {journal} {Physical Review Research}\ }\textbf {\bibinfo
  {volume} {2}},\ \bibinfo {pages} {033333} (\bibinfo {year}
  {2020})}\BibitemShut {NoStop}%
\bibitem [{\citenamefont {Zuo}\ \emph {et~al.}(2021)\citenamefont {Zuo},
  \citenamefont {Trautmann}, \citenamefont {Wang}, \citenamefont {Hannes},
  \citenamefont {Yang}, \citenamefont {Song}, \citenamefont {Meier},
  \citenamefont {Ciappina}, \citenamefont {Duc},\ and\ \citenamefont
  {Yang}}]{neighboursolid}%
  \BibitemOpen
  \bibfield  {author} {\bibinfo {author} {\bibfnamefont {R.}~\bibnamefont
  {Zuo}}, \bibinfo {author} {\bibfnamefont {A.}~\bibnamefont {Trautmann}},
  \bibinfo {author} {\bibfnamefont {G.}~\bibnamefont {Wang}}, \bibinfo {author}
  {\bibfnamefont {W.-R.}\ \bibnamefont {Hannes}}, \bibinfo {author}
  {\bibfnamefont {S.}~\bibnamefont {Yang}}, \bibinfo {author} {\bibfnamefont
  {X.}~\bibnamefont {Song}}, \bibinfo {author} {\bibfnamefont {T.}~\bibnamefont
  {Meier}}, \bibinfo {author} {\bibfnamefont {M.}~\bibnamefont {Ciappina}},
  \bibinfo {author} {\bibfnamefont {H.~T.}\ \bibnamefont {Duc}},\ and\ \bibinfo
  {author} {\bibfnamefont {W.}~\bibnamefont {Yang}},\ }\bibfield  {title}
  {\bibinfo {title} {{Neighboring Atom Collisions in Solid-State High Harmonic
  Generation}},\ }\bibfield  {journal} {\bibinfo  {journal} {Ultrafast
  Science}\ }\textbf {\bibinfo {volume} {2021}},\ \href
  {https://doi.org/10.34133/2021/9861923} {10.34133/2021/9861923} (\bibinfo
  {year} {2021})\BibitemShut {NoStop}%
\bibitem [{\citenamefont {Li}\ \emph {et~al.}(2019)\citenamefont {Li},
  \citenamefont {Lan}, \citenamefont {Zhu}, \citenamefont {Huang},
  \citenamefont {Zhang}, \citenamefont {Lein},\ and\ \citenamefont
  {Lu}}]{Li2019}%
  \BibitemOpen
  \bibfield  {author} {\bibinfo {author} {\bibfnamefont {L.}~\bibnamefont
  {Li}}, \bibinfo {author} {\bibfnamefont {P.}~\bibnamefont {Lan}}, \bibinfo
  {author} {\bibfnamefont {X.}~\bibnamefont {Zhu}}, \bibinfo {author}
  {\bibfnamefont {T.}~\bibnamefont {Huang}}, \bibinfo {author} {\bibfnamefont
  {Q.}~\bibnamefont {Zhang}}, \bibinfo {author} {\bibfnamefont
  {M.}~\bibnamefont {Lein}},\ and\ \bibinfo {author} {\bibfnamefont
  {P.}~\bibnamefont {Lu}},\ }\bibfield  {title} {\bibinfo {title}
  {{Reciprocal-Space-Trajectory Perspective on High-Harmonic Generation in
  Solids}},\ }\href {https://doi.org/10.1103/PhysRevLett.122.193901} {\bibfield
   {journal} {\bibinfo  {journal} {Physical Review Letters}\ }\textbf {\bibinfo
  {volume} {122}},\ \bibinfo {pages} {193901} (\bibinfo {year}
  {2019})}\BibitemShut {NoStop}%
\bibitem [{\citenamefont {Li}\ \emph {et~al.}(2021)\citenamefont {Li},
  \citenamefont {Lan}, \citenamefont {Zhu},\ and\ \citenamefont
  {Lu}}]{li2021huygens}%
  \BibitemOpen
  \bibfield  {author} {\bibinfo {author} {\bibfnamefont {L.}~\bibnamefont
  {Li}}, \bibinfo {author} {\bibfnamefont {P.}~\bibnamefont {Lan}}, \bibinfo
  {author} {\bibfnamefont {X.}~\bibnamefont {Zhu}},\ and\ \bibinfo {author}
  {\bibfnamefont {P.}~\bibnamefont {Lu}},\ }\bibfield  {title} {\bibinfo
  {title} {Huygens-fresnel picture for high harmonic generation in solids},\
  }\href@noop {} {\bibfield  {journal} {\bibinfo  {journal} {Physical review
  letters}\ }\textbf {\bibinfo {volume} {127}},\ \bibinfo {pages} {223201}
  (\bibinfo {year} {2021})}\BibitemShut {NoStop}%
\bibitem [{\citenamefont {Luu}\ and\ \citenamefont
  {W\"{o}rner}(2018)}]{Luu2018berry}%
  \BibitemOpen
  \bibfield  {author} {\bibinfo {author} {\bibfnamefont {T.~T.}\ \bibnamefont
  {Luu}}\ and\ \bibinfo {author} {\bibfnamefont {H.~J.}\ \bibnamefont
  {W\"{o}rner}},\ }\bibfield  {title} {\bibinfo {title} {Measurement of the
  berry curvature of solids using high-harmonic spectroscopy},\ }\href
  {https://doi.org/10.1038/s41467-018-03397-4} {\bibfield  {journal} {\bibinfo
  {journal} {Nature Communications}\ }\textbf {\bibinfo {volume} {9}},\
  \bibinfo {pages} {916} (\bibinfo {year} {2018})}\BibitemShut {NoStop}%
\bibitem [{\citenamefont {Uzan}\ \emph {et~al.}(2020)\citenamefont {Uzan},
  \citenamefont {Orenstein}, \citenamefont {Jim\'{e}nez-Gal\'{a}n},
  \citenamefont {McDonald}, \citenamefont {Silva}, \citenamefont {Bruner},
  \citenamefont {Klimkin}, \citenamefont {Blanchet}, \citenamefont
  {Arusi-Parpar}, \citenamefont {Kr\"{u}ger}, \citenamefont {Rubtsov},
  \citenamefont {Smirnova}, \citenamefont {Ivanov}, \citenamefont {Yan},
  \citenamefont {Brabec},\ and\ \citenamefont {Dudovich}}]{Uzan2020b}%
  \BibitemOpen
  \bibfield  {author} {\bibinfo {author} {\bibfnamefont {A.~J.}\ \bibnamefont
  {Uzan}}, \bibinfo {author} {\bibfnamefont {G.}~\bibnamefont {Orenstein}},
  \bibinfo {author} {\bibfnamefont {A.}~\bibnamefont {Jim\'{e}nez-Gal\'{a}n}},
  \bibinfo {author} {\bibfnamefont {C.}~\bibnamefont {McDonald}}, \bibinfo
  {author} {\bibfnamefont {R.~E.~F.}\ \bibnamefont {Silva}}, \bibinfo {author}
  {\bibfnamefont {B.~D.}\ \bibnamefont {Bruner}}, \bibinfo {author}
  {\bibfnamefont {N.~D.}\ \bibnamefont {Klimkin}}, \bibinfo {author}
  {\bibfnamefont {V.}~\bibnamefont {Blanchet}}, \bibinfo {author}
  {\bibfnamefont {T.}~\bibnamefont {Arusi-Parpar}}, \bibinfo {author}
  {\bibfnamefont {M.}~\bibnamefont {Kr\"{u}ger}}, \bibinfo {author}
  {\bibfnamefont {A.~N.}\ \bibnamefont {Rubtsov}}, \bibinfo {author}
  {\bibfnamefont {O.}~\bibnamefont {Smirnova}}, \bibinfo {author}
  {\bibfnamefont {M.}~\bibnamefont {Ivanov}}, \bibinfo {author} {\bibfnamefont
  {B.}~\bibnamefont {Yan}}, \bibinfo {author} {\bibfnamefont {T.}~\bibnamefont
  {Brabec}},\ and\ \bibinfo {author} {\bibfnamefont {N.}~\bibnamefont
  {Dudovich}},\ }\bibfield  {title} {\bibinfo {title} {Attosecond spectral
  singularities in solid-state high-harmonic generation},\ }\href
  {https://doi.org/10.1038/s41566-019-0574-4} {\bibfield  {journal} {\bibinfo
  {journal} {Nature Photonics}\ }\textbf {\bibinfo {volume} {14}},\ \bibinfo
  {pages} {183} (\bibinfo {year} {2020})}\BibitemShut {NoStop}%
\bibitem [{\citenamefont {Yu}\ \emph {et~al.}(2020)\citenamefont {Yu},
  \citenamefont {Iravani},\ and\ \citenamefont {Madsen}}]{Yu2020}%
  \BibitemOpen
  \bibfield  {author} {\bibinfo {author} {\bibfnamefont {C.}~\bibnamefont
  {Yu}}, \bibinfo {author} {\bibfnamefont {H.}~\bibnamefont {Iravani}},\ and\
  \bibinfo {author} {\bibfnamefont {L.~B.}\ \bibnamefont {Madsen}},\ }\bibfield
   {title} {\bibinfo {title} {Crystal-momentum-resolved contributions to
  multiple plateaus of high-order harmonic generation from band-gap
  materials},\ }\href {https://doi.org/10.1103/PhysRevA.102.033105} {\bibfield
  {journal} {\bibinfo  {journal} {Phys. Rev. A}\ }\textbf {\bibinfo {volume}
  {102}},\ \bibinfo {pages} {033105} (\bibinfo {year} {2020})}\BibitemShut
  {NoStop}%
\bibitem [{\citenamefont {Ndabashimiye}\ \emph {et~al.}(2016)\citenamefont
  {Ndabashimiye}, \citenamefont {Ghimire}, \citenamefont {Wu}, \citenamefont
  {Browne}, \citenamefont {Schafer}, \citenamefont {Gaarde},\ and\
  \citenamefont {Reis}}]{Ndabashimiye2016}%
  \BibitemOpen
  \bibfield  {author} {\bibinfo {author} {\bibfnamefont {G.}~\bibnamefont
  {Ndabashimiye}}, \bibinfo {author} {\bibfnamefont {S.}~\bibnamefont
  {Ghimire}}, \bibinfo {author} {\bibfnamefont {M.}~\bibnamefont {Wu}},
  \bibinfo {author} {\bibfnamefont {D.~A.}\ \bibnamefont {Browne}}, \bibinfo
  {author} {\bibfnamefont {K.~J.}\ \bibnamefont {Schafer}}, \bibinfo {author}
  {\bibfnamefont {M.~B.}\ \bibnamefont {Gaarde}},\ and\ \bibinfo {author}
  {\bibfnamefont {D.~A.}\ \bibnamefont {Reis}},\ }\bibfield  {title} {\bibinfo
  {title} {{Solid-state harmonics beyond the atomic limit}},\ }\bibfield
  {journal} {\bibinfo  {journal} {Nature}\ }\textbf {\bibinfo {volume} {534}},\
  \href {https://doi.org/10.1038/nature17660} {10.1038/nature17660} (\bibinfo
  {year} {2016})\BibitemShut {NoStop}%
\bibitem [{\citenamefont {You}\ \emph {et~al.}(2017{\natexlab{b}})\citenamefont
  {You}, \citenamefont {Yin}, \citenamefont {Wu}, \citenamefont {Chew},
  \citenamefont {Ren}, \citenamefont {Zhuang}, \citenamefont {Gholam-Mirzaei},
  \citenamefont {Chini}, \citenamefont {Chang},\ and\ \citenamefont
  {Ghimire}}]{you17b}%
  \BibitemOpen
  \bibfield  {author} {\bibinfo {author} {\bibfnamefont {Y.~S.}\ \bibnamefont
  {You}}, \bibinfo {author} {\bibfnamefont {Y.}~\bibnamefont {Yin}}, \bibinfo
  {author} {\bibfnamefont {Y.}~\bibnamefont {Wu}}, \bibinfo {author}
  {\bibfnamefont {A.}~\bibnamefont {Chew}}, \bibinfo {author} {\bibfnamefont
  {X.}~\bibnamefont {Ren}}, \bibinfo {author} {\bibfnamefont {F.}~\bibnamefont
  {Zhuang}}, \bibinfo {author} {\bibfnamefont {S.}~\bibnamefont
  {Gholam-Mirzaei}}, \bibinfo {author} {\bibfnamefont {M.}~\bibnamefont
  {Chini}}, \bibinfo {author} {\bibfnamefont {Z.}~\bibnamefont {Chang}},\ and\
  \bibinfo {author} {\bibfnamefont {S.}~\bibnamefont {Ghimire}},\ }\bibfield
  {title} {\bibinfo {title} {High-harmonic generation in amorphous solids},\
  }\href@noop {} {\bibfield  {journal} {\bibinfo  {journal} {Nature
  communications}\ }\textbf {\bibinfo {volume} {8}},\ \bibinfo {pages} {1}
  (\bibinfo {year} {2017}{\natexlab{b}})}\BibitemShut {NoStop}%
\bibitem [{\citenamefont {You}\ \emph {et~al.}(2017{\natexlab{c}})\citenamefont
  {You}, \citenamefont {Wu}, \citenamefont {Yin}, \citenamefont {Chew},
  \citenamefont {Ren}, \citenamefont {Gholam-Mirzaei}, \citenamefont {Browne},
  \citenamefont {Chini}, \citenamefont {Chang}, \citenamefont {Schafer},
  \citenamefont {Gaarde},\ and\ \citenamefont {Ghimire}}]{You2017}%
  \BibitemOpen
  \bibfield  {author} {\bibinfo {author} {\bibfnamefont {Y.~S.}\ \bibnamefont
  {You}}, \bibinfo {author} {\bibfnamefont {M.}~\bibnamefont {Wu}}, \bibinfo
  {author} {\bibfnamefont {Y.}~\bibnamefont {Yin}}, \bibinfo {author}
  {\bibfnamefont {A.}~\bibnamefont {Chew}}, \bibinfo {author} {\bibfnamefont
  {X.}~\bibnamefont {Ren}}, \bibinfo {author} {\bibfnamefont {S.}~\bibnamefont
  {Gholam-Mirzaei}}, \bibinfo {author} {\bibfnamefont {D.~A.}\ \bibnamefont
  {Browne}}, \bibinfo {author} {\bibfnamefont {M.}~\bibnamefont {Chini}},
  \bibinfo {author} {\bibfnamefont {Z.}~\bibnamefont {Chang}}, \bibinfo
  {author} {\bibfnamefont {K.~J.}\ \bibnamefont {Schafer}}, \bibinfo {author}
  {\bibfnamefont {M.~B.}\ \bibnamefont {Gaarde}},\ and\ \bibinfo {author}
  {\bibfnamefont {S.}~\bibnamefont {Ghimire}},\ }\bibfield  {title} {\bibinfo
  {title} {{Laser waveform control of extreme ultraviolet high harmonics from
  solids}},\ }\href {https://doi.org/10.1364/OL.42.001816} {\bibfield
  {journal} {\bibinfo  {journal} {Optics Letters}\ }\textbf {\bibinfo {volume}
  {42}},\ \bibinfo {pages} {1816} (\bibinfo {year}
  {2017}{\natexlab{c}})}\BibitemShut {NoStop}%
\bibitem [{\citenamefont {Ikemachi}\ \emph {et~al.}(2017)\citenamefont
  {Ikemachi}, \citenamefont {Shinohara}, \citenamefont {Sato}, \citenamefont
  {Yumoto}, \citenamefont {Kuwata-Gonokami},\ and\ \citenamefont
  {Ishikawa}}]{PhysRevA.95.043416}%
  \BibitemOpen
  \bibfield  {author} {\bibinfo {author} {\bibfnamefont {T.}~\bibnamefont
  {Ikemachi}}, \bibinfo {author} {\bibfnamefont {Y.}~\bibnamefont {Shinohara}},
  \bibinfo {author} {\bibfnamefont {T.}~\bibnamefont {Sato}}, \bibinfo {author}
  {\bibfnamefont {J.}~\bibnamefont {Yumoto}}, \bibinfo {author} {\bibfnamefont
  {M.}~\bibnamefont {Kuwata-Gonokami}},\ and\ \bibinfo {author} {\bibfnamefont
  {K.~L.}\ \bibnamefont {Ishikawa}},\ }\bibfield  {title} {\bibinfo {title}
  {Trajectory analysis of high-order-harmonic generation from periodic
  crystals},\ }\href {https://doi.org/10.1103/PhysRevA.95.043416} {\bibfield
  {journal} {\bibinfo  {journal} {Phys. Rev. A}\ }\textbf {\bibinfo {volume}
  {95}},\ \bibinfo {pages} {043416} (\bibinfo {year} {2017})}\BibitemShut
  {NoStop}%
\bibitem [{\citenamefont {Luppi}\ and\ \citenamefont
  {Coccia}(2023)}]{Luppi2023}%
  \BibitemOpen
  \bibfield  {author} {\bibinfo {author} {\bibfnamefont {E.}~\bibnamefont
  {Luppi}}\ and\ \bibinfo {author} {\bibfnamefont {E.}~\bibnamefont {Coccia}},\
  }\bibfield  {title} {\bibinfo {title} {Role of inner molecular orbitals in
  high-harmonic generation spectra of aligned uracil},\ }\href
  {https://doi.org/10.1021/acs.jpca.3c03990} {\bibfield  {journal} {\bibinfo
  {journal} {The Journal of Physical Chemistry A}\ }\textbf {\bibinfo {volume}
  {127}},\ \bibinfo {pages} {7335} (\bibinfo {year} {2023})},\ \bibinfo {note}
  {pMID: 37640677}\BibitemShut {NoStop}%
\bibitem [{\citenamefont {Morassut}\ \emph {et~al.}(2024)\citenamefont
  {Morassut}, \citenamefont {Ravindran}, \citenamefont {Ciavardini},
  \citenamefont {Luppi}, \citenamefont {De~Ninno},\ and\ \citenamefont
  {Coccia}}]{Morassut2024}%
  \BibitemOpen
  \bibfield  {author} {\bibinfo {author} {\bibfnamefont {C.}~\bibnamefont
  {Morassut}}, \bibinfo {author} {\bibfnamefont {A.}~\bibnamefont {Ravindran}},
  \bibinfo {author} {\bibfnamefont {A.}~\bibnamefont {Ciavardini}}, \bibinfo
  {author} {\bibfnamefont {E.}~\bibnamefont {Luppi}}, \bibinfo {author}
  {\bibfnamefont {G.}~\bibnamefont {De~Ninno}},\ and\ \bibinfo {author}
  {\bibfnamefont {E.}~\bibnamefont {Coccia}},\ }\bibfield  {title} {\bibinfo
  {title} {High-harmonic generation spectroscopy of gas-phase bromoform},\
  }\href {https://doi.org/10.1021/acs.jpca.3c07699} {\bibfield  {journal}
  {\bibinfo  {journal} {The Journal of Physical Chemistry A}\ }\textbf
  {\bibinfo {volume} {128}},\ \bibinfo {pages} {2015} (\bibinfo {year}
  {2024})}\BibitemShut {NoStop}%
\bibitem [{\citenamefont {Neufeld}\ \emph {et~al.}(2024)\citenamefont
  {Neufeld}, \citenamefont {Tancogne-Dejean},\ and\ \citenamefont
  {Rubio}}]{Neufeld2024}%
  \BibitemOpen
  \bibfield  {author} {\bibinfo {author} {\bibfnamefont {O.}~\bibnamefont
  {Neufeld}}, \bibinfo {author} {\bibfnamefont {N.}~\bibnamefont
  {Tancogne-Dejean}},\ and\ \bibinfo {author} {\bibfnamefont {A.}~\bibnamefont
  {Rubio}},\ }\bibfield  {title} {\bibinfo {title} {Benchmarking functionals
  for strong-field light-matter interactions in adiabatic time-dependent
  density functional theory},\ }\href
  {https://doi.org/10.1021/acs.jpclett.4c01383} {\bibfield  {journal} {\bibinfo
   {journal} {The Journal of Physical Chemistry Letters}\ }\textbf {\bibinfo
  {volume} {15}},\ \bibinfo {pages} {7254} (\bibinfo {year}
  {2024})}\BibitemShut {NoStop}%
\bibitem [{\citenamefont {Rae}\ \emph {et~al.}(1994)\citenamefont {Rae},
  \citenamefont {Chen},\ and\ \citenamefont {Burnett}}]{Rae1994}%
  \BibitemOpen
  \bibfield  {author} {\bibinfo {author} {\bibfnamefont {S.~C.}\ \bibnamefont
  {Rae}}, \bibinfo {author} {\bibfnamefont {X.}~\bibnamefont {Chen}},\ and\
  \bibinfo {author} {\bibfnamefont {K.}~\bibnamefont {Burnett}},\ }\bibfield
  {title} {\bibinfo {title} {Saturation of harmonic generation in one- and
  three-dimensional atoms},\ }\href {https://doi.org/10.1103/PhysRevA.50.1946}
  {\bibfield  {journal} {\bibinfo  {journal} {Phys. Rev. A}\ }\textbf {\bibinfo
  {volume} {50}},\ \bibinfo {pages} {1946} (\bibinfo {year}
  {1994})}\BibitemShut {NoStop}%
\bibitem [{\citenamefont {Rae}\ and\ \citenamefont {Burnett}(1993)}]{Rae1993}%
  \BibitemOpen
  \bibfield  {author} {\bibinfo {author} {\bibfnamefont {S.~C.}\ \bibnamefont
  {Rae}}\ and\ \bibinfo {author} {\bibfnamefont {K.}~\bibnamefont {Burnett}},\
  }\bibfield  {title} {\bibinfo {title} {Calculations of high-order-harmonic
  generation in the strongly ionizing regime},\ }\href
  {https://doi.org/10.1103/PhysRevA.48.2490} {\bibfield  {journal} {\bibinfo
  {journal} {Phys. Rev. A}\ }\textbf {\bibinfo {volume} {48}},\ \bibinfo
  {pages} {2490} (\bibinfo {year} {1993})}\BibitemShut {NoStop}%
\bibitem [{\citenamefont {Constant}\ \emph {et~al.}(1999)\citenamefont
  {Constant}, \citenamefont {Garzella}, \citenamefont {Breger}, \citenamefont
  {M\'evel}, \citenamefont {Dorrer}, \citenamefont {Le~Blanc}, \citenamefont
  {Salin},\ and\ \citenamefont {Agostini}}]{Constant1999}%
  \BibitemOpen
  \bibfield  {author} {\bibinfo {author} {\bibfnamefont {E.}~\bibnamefont
  {Constant}}, \bibinfo {author} {\bibfnamefont {D.}~\bibnamefont {Garzella}},
  \bibinfo {author} {\bibfnamefont {P.}~\bibnamefont {Breger}}, \bibinfo
  {author} {\bibfnamefont {E.}~\bibnamefont {M\'evel}}, \bibinfo {author}
  {\bibfnamefont {C.}~\bibnamefont {Dorrer}}, \bibinfo {author} {\bibfnamefont
  {C.}~\bibnamefont {Le~Blanc}}, \bibinfo {author} {\bibfnamefont
  {F.}~\bibnamefont {Salin}},\ and\ \bibinfo {author} {\bibfnamefont
  {P.}~\bibnamefont {Agostini}},\ }\bibfield  {title} {\bibinfo {title}
  {Optimizing high harmonic generation in absorbing gases: Model and
  experiment},\ }\href {https://doi.org/10.1103/PhysRevLett.82.1668} {\bibfield
   {journal} {\bibinfo  {journal} {Phys. Rev. Lett.}\ }\textbf {\bibinfo
  {volume} {82}},\ \bibinfo {pages} {1668} (\bibinfo {year}
  {1999})}\BibitemShut {NoStop}%
\bibitem [{\citenamefont {Schouder}\ \emph {et~al.}(2022)\citenamefont
  {Schouder}, \citenamefont {Chatterley}, \citenamefont {Pickering},\ and\
  \citenamefont {Stapelfeldt}}]{Schouder2022}%
  \BibitemOpen
  \bibfield  {author} {\bibinfo {author} {\bibfnamefont {C.~A.}\ \bibnamefont
  {Schouder}}, \bibinfo {author} {\bibfnamefont {A.~S.}\ \bibnamefont
  {Chatterley}}, \bibinfo {author} {\bibfnamefont {J.~D.}\ \bibnamefont
  {Pickering}},\ and\ \bibinfo {author} {\bibfnamefont {H.}~\bibnamefont
  {Stapelfeldt}},\ }\bibfield  {title} {\bibinfo {title} {Laser-induced coulomb
  explosion imaging of aligned molecules and molecular dimers},\ }\href
  {https://doi.org/https://doi.org/10.1146/annurev-physchem-090419-053627}
  {\bibfield  {journal} {\bibinfo  {journal} {Annual Review of Physical
  Chemistry}\ }\textbf {\bibinfo {volume} {73}},\ \bibinfo {pages} {323}
  (\bibinfo {year} {2022})}\BibitemShut {NoStop}%
\bibitem [{\citenamefont {Farrell}\ \emph {et~al.}(2011)\citenamefont
  {Farrell}, \citenamefont {Spector}, \citenamefont {McFarland}, \citenamefont
  {Bucksbaum}, \citenamefont {G{\"{u}}hr}, \citenamefont {Gaarde},\ and\
  \citenamefont {Schafer}}]{Farrell2011}%
  \BibitemOpen
  \bibfield  {author} {\bibinfo {author} {\bibfnamefont {J.~P.}\ \bibnamefont
  {Farrell}}, \bibinfo {author} {\bibfnamefont {L.~S.}\ \bibnamefont
  {Spector}}, \bibinfo {author} {\bibfnamefont {B.~K.}\ \bibnamefont
  {McFarland}}, \bibinfo {author} {\bibfnamefont {P.~H.}\ \bibnamefont
  {Bucksbaum}}, \bibinfo {author} {\bibfnamefont {M.}~\bibnamefont
  {G{\"{u}}hr}}, \bibinfo {author} {\bibfnamefont {M.~B.}\ \bibnamefont
  {Gaarde}},\ and\ \bibinfo {author} {\bibfnamefont {K.~J.}\ \bibnamefont
  {Schafer}},\ }\bibfield  {title} {\bibinfo {title} {{Influence of phase
  matching on the Cooper minimum in Ar high-order harmonic spectra}},\ }\href
  {https://doi.org/10.1103/PhysRevA.83.023420} {\bibfield  {journal} {\bibinfo
  {journal} {Physical Review A - Atomic, Molecular, and Optical Physics}\
  }\textbf {\bibinfo {volume} {83}},\ \bibinfo {pages} {1} (\bibinfo {year}
  {2011})},\ \Eprint {https://arxiv.org/abs/1011.1297} {arXiv:1011.1297}
  \BibitemShut {NoStop}%
\bibitem [{\citenamefont {Zeng}\ and\ \citenamefont {Bian}(2020)}]{Zheng2020}%
  \BibitemOpen
  \bibfield  {author} {\bibinfo {author} {\bibfnamefont {A.-W.}\ \bibnamefont
  {Zeng}}\ and\ \bibinfo {author} {\bibfnamefont {X.-B.}\ \bibnamefont
  {Bian}},\ }\bibfield  {title} {\bibinfo {title} {Impact of statistical
  fluctuations on high harmonic generation in liquids},\ }\href
  {https://doi.org/10.1103/PhysRevLett.124.203901} {\bibfield  {journal}
  {\bibinfo  {journal} {Phys. Rev. Lett.}\ }\textbf {\bibinfo {volume} {124}},\
  \bibinfo {pages} {203901} (\bibinfo {year} {2020})}\BibitemShut {NoStop}%
\bibitem [{\citenamefont {Neufeld}\ \emph
  {et~al.}(2022{\natexlab{a}})\citenamefont {Neufeld}, \citenamefont
  {Nourbakhsh}, \citenamefont {Tancogne-Dejean},\ and\ \citenamefont
  {Rubio}}]{Ofer2022}%
  \BibitemOpen
  \bibfield  {author} {\bibinfo {author} {\bibfnamefont {O.}~\bibnamefont
  {Neufeld}}, \bibinfo {author} {\bibfnamefont {Z.}~\bibnamefont {Nourbakhsh}},
  \bibinfo {author} {\bibfnamefont {N.}~\bibnamefont {Tancogne-Dejean}},\ and\
  \bibinfo {author} {\bibfnamefont {A.}~\bibnamefont {Rubio}},\ }\bibfield
  {title} {\bibinfo {title} {Ab initio cluster approach for high harmonic
  generation in liquids},\ }\href {https://doi.org/10.1021/acs.jctc.2c00235}
  {\bibfield  {journal} {\bibinfo  {journal} {Journal of Chemical Theory and
  Computation}\ }\textbf {\bibinfo {volume} {18}},\ \bibinfo {pages} {4117}
  (\bibinfo {year} {2022}{\natexlab{a}})}\BibitemShut {NoStop}%
\bibitem [{\citenamefont {Li}\ \emph {et~al.}(2024)\citenamefont {Li},
  \citenamefont {Chen}, \citenamefont {Ding}, \citenamefont {Liu},
  \citenamefont {Wang},\ and\ \citenamefont {Bian}}]{Li2024}%
  \BibitemOpen
  \bibfield  {author} {\bibinfo {author} {\bibfnamefont {Z.-L.}\ \bibnamefont
  {Li}}, \bibinfo {author} {\bibfnamefont {J.-X.}\ \bibnamefont {Chen}},
  \bibinfo {author} {\bibfnamefont {Z.-W.}\ \bibnamefont {Ding}}, \bibinfo
  {author} {\bibfnamefont {J.-Q.}\ \bibnamefont {Liu}}, \bibinfo {author}
  {\bibfnamefont {Y.-B.}\ \bibnamefont {Wang}},\ and\ \bibinfo {author}
  {\bibfnamefont {X.-B.}\ \bibnamefont {Bian}},\ }\bibfield  {title} {\bibinfo
  {title} {Linking high-order harmonic generation and radial distribution
  function in liquids},\ }\href {https://doi.org/10.1103/PhysRevA.110.043507}
  {\bibfield  {journal} {\bibinfo  {journal} {Phys. Rev. A}\ }\textbf {\bibinfo
  {volume} {110}},\ \bibinfo {pages} {043507} (\bibinfo {year}
  {2024})}\BibitemShut {NoStop}%
\bibitem [{\citenamefont {Alexander}\ \emph {et~al.}(2023)\citenamefont
  {Alexander}, \citenamefont {Barnard}, \citenamefont {Larsen}, \citenamefont
  {Avni}, \citenamefont {Jarosch}, \citenamefont {Ferchaud}, \citenamefont
  {Gregory}, \citenamefont {Parker}, \citenamefont {Galinis}, \citenamefont
  {Tofful}, \citenamefont {Garratt}, \citenamefont {Matthews},\ and\
  \citenamefont {Marangos}}]{Alexander2023}%
  \BibitemOpen
  \bibfield  {author} {\bibinfo {author} {\bibfnamefont {O.}~\bibnamefont
  {Alexander}}, \bibinfo {author} {\bibfnamefont {J.~C.~T.}\ \bibnamefont
  {Barnard}}, \bibinfo {author} {\bibfnamefont {E.~W.}\ \bibnamefont {Larsen}},
  \bibinfo {author} {\bibfnamefont {T.}~\bibnamefont {Avni}}, \bibinfo {author}
  {\bibfnamefont {S.}~\bibnamefont {Jarosch}}, \bibinfo {author} {\bibfnamefont
  {C.}~\bibnamefont {Ferchaud}}, \bibinfo {author} {\bibfnamefont
  {A.}~\bibnamefont {Gregory}}, \bibinfo {author} {\bibfnamefont
  {S.}~\bibnamefont {Parker}}, \bibinfo {author} {\bibfnamefont
  {G.}~\bibnamefont {Galinis}}, \bibinfo {author} {\bibfnamefont
  {A.}~\bibnamefont {Tofful}}, \bibinfo {author} {\bibfnamefont
  {D.}~\bibnamefont {Garratt}}, \bibinfo {author} {\bibfnamefont {M.~R.}\
  \bibnamefont {Matthews}},\ and\ \bibinfo {author} {\bibfnamefont {J.~P.}\
  \bibnamefont {Marangos}},\ }\bibfield  {title} {\bibinfo {title} {Observation
  of recollision-based high-harmonic generation in liquid isopropanol and the
  role of electron scattering},\ }\href
  {https://doi.org/10.1103/PhysRevResearch.5.043030} {\bibfield  {journal}
  {\bibinfo  {journal} {Phys. Rev. Res.}\ }\textbf {\bibinfo {volume} {5}},\
  \bibinfo {pages} {043030} (\bibinfo {year} {2023})}\BibitemShut {NoStop}%
\bibitem [{\citenamefont {Gong}\ \emph {et~al.}(2022)\citenamefont {Gong},
  \citenamefont {Heck}, \citenamefont {Jelovina}, \citenamefont {Perry},
  \citenamefont {Zinchenko}, \citenamefont {Lucchese},\ and\ \citenamefont
  {W{\"o}rner}}]{gong2022attosecond}%
  \BibitemOpen
  \bibfield  {author} {\bibinfo {author} {\bibfnamefont {X.}~\bibnamefont
  {Gong}}, \bibinfo {author} {\bibfnamefont {S.}~\bibnamefont {Heck}}, \bibinfo
  {author} {\bibfnamefont {D.}~\bibnamefont {Jelovina}}, \bibinfo {author}
  {\bibfnamefont {C.}~\bibnamefont {Perry}}, \bibinfo {author} {\bibfnamefont
  {K.}~\bibnamefont {Zinchenko}}, \bibinfo {author} {\bibfnamefont
  {R.}~\bibnamefont {Lucchese}},\ and\ \bibinfo {author} {\bibfnamefont
  {H.~J.}\ \bibnamefont {W{\"o}rner}},\ }\bibfield  {title} {\bibinfo {title}
  {Attosecond spectroscopy of size-resolved water clusters},\ }\href@noop {}
  {\bibfield  {journal} {\bibinfo  {journal} {Nature}\ }\textbf {\bibinfo
  {volume} {609}},\ \bibinfo {pages} {507} (\bibinfo {year}
  {2022})}\BibitemShut {NoStop}%
\bibitem [{\citenamefont {Neufeld}\ \emph
  {et~al.}(2022{\natexlab{b}})\citenamefont {Neufeld}, \citenamefont
  {Nourbakhsh}, \citenamefont {Tancogne-Dejean},\ and\ \citenamefont
  {Rubio}}]{neufeld2022}%
  \BibitemOpen
  \bibfield  {author} {\bibinfo {author} {\bibfnamefont {O.}~\bibnamefont
  {Neufeld}}, \bibinfo {author} {\bibfnamefont {Z.}~\bibnamefont {Nourbakhsh}},
  \bibinfo {author} {\bibfnamefont {N.}~\bibnamefont {Tancogne-Dejean}},\ and\
  \bibinfo {author} {\bibfnamefont {A.}~\bibnamefont {Rubio}},\ }\bibfield
  {title} {\bibinfo {title} {Ab initio cluster approach for high harmonic
  generation in liquids},\ }\href@noop {} {\bibfield  {journal} {\bibinfo
  {journal} {Journal of Chemical Theory and Computation}\ }\textbf {\bibinfo
  {volume} {18}},\ \bibinfo {pages} {4117} (\bibinfo {year}
  {2022}{\natexlab{b}})}\BibitemShut {NoStop}%
\bibitem [{\citenamefont {Yin}\ \emph {et~al.}(2020)\citenamefont {Yin},
  \citenamefont {Luu},\ and\ \citenamefont {W{\"o}rner}}]{yin2020}%
  \BibitemOpen
  \bibfield  {author} {\bibinfo {author} {\bibfnamefont {Z.}~\bibnamefont
  {Yin}}, \bibinfo {author} {\bibfnamefont {T.~T.}\ \bibnamefont {Luu}},\ and\
  \bibinfo {author} {\bibfnamefont {H.~J.}\ \bibnamefont {W{\"o}rner}},\
  }\bibfield  {title} {\bibinfo {title} {Few-cycle high-harmonic generation in
  liquids: in-operando thickness measurement of flat microjets},\ }\href@noop
  {} {\bibfield  {journal} {\bibinfo  {journal} {Journal of Physics:
  Photonics}\ }\textbf {\bibinfo {volume} {2}},\ \bibinfo {pages} {044007}
  (\bibinfo {year} {2020})}\BibitemShut {NoStop}%
\bibitem [{\citenamefont {Chang}\ \emph {et~al.}(2022)\citenamefont {Chang},
  \citenamefont {Yin}, \citenamefont {Balciunas}, \citenamefont {W{\"o}rner},\
  and\ \citenamefont {Wolf}}]{chang2022}%
  \BibitemOpen
  \bibfield  {author} {\bibinfo {author} {\bibfnamefont {Y.-P.}\ \bibnamefont
  {Chang}}, \bibinfo {author} {\bibfnamefont {Z.}~\bibnamefont {Yin}}, \bibinfo
  {author} {\bibfnamefont {T.}~\bibnamefont {Balciunas}}, \bibinfo {author}
  {\bibfnamefont {H.~J.}\ \bibnamefont {W{\"o}rner}},\ and\ \bibinfo {author}
  {\bibfnamefont {J.-P.}\ \bibnamefont {Wolf}},\ }\bibfield  {title} {\bibinfo
  {title} {Temperature measurements of liquid flat jets in vacuum},\
  }\href@noop {} {\bibfield  {journal} {\bibinfo  {journal} {Structural
  Dynamics}\ }\textbf {\bibinfo {volume} {9}},\ \bibinfo {pages} {014901}
  (\bibinfo {year} {2022})}\BibitemShut {NoStop}%
\bibitem [{\citenamefont {Buttersack}\ \emph {et~al.}(2023)\citenamefont
  {Buttersack}, \citenamefont {Haak}, \citenamefont {Bluhm}, \citenamefont
  {Hergenhahn}, \citenamefont {Meijer},\ and\ \citenamefont
  {Winter}}]{buttersack2022}%
  \BibitemOpen
  \bibfield  {author} {\bibinfo {author} {\bibfnamefont {T.}~\bibnamefont
  {Buttersack}}, \bibinfo {author} {\bibfnamefont {H.}~\bibnamefont {Haak}},
  \bibinfo {author} {\bibfnamefont {H.}~\bibnamefont {Bluhm}}, \bibinfo
  {author} {\bibfnamefont {U.}~\bibnamefont {Hergenhahn}}, \bibinfo {author}
  {\bibfnamefont {G.}~\bibnamefont {Meijer}},\ and\ \bibinfo {author}
  {\bibfnamefont {B.}~\bibnamefont {Winter}},\ }\bibfield  {title} {\bibinfo
  {title} {{Imaging temperature and thickness of thin planar liquid water jets
  in vacuum}},\ }\href {https://doi.org/10.1063/4.0000188} {\bibfield
  {journal} {\bibinfo  {journal} {Structural Dynamics}\ }\textbf {\bibinfo
  {volume} {10}},\ \bibinfo {pages} {034901} (\bibinfo {year}
  {2023})}\BibitemShut {NoStop}%
\bibitem [{\citenamefont {Svoboda}\ \emph {et~al.}(2021)\citenamefont
  {Svoboda}, \citenamefont {Yin}, \citenamefont {Luu},\ and\ \citenamefont
  {W{\"o}rner}}]{svoboda2021polarization}%
  \BibitemOpen
  \bibfield  {author} {\bibinfo {author} {\bibfnamefont {V.}~\bibnamefont
  {Svoboda}}, \bibinfo {author} {\bibfnamefont {Z.}~\bibnamefont {Yin}},
  \bibinfo {author} {\bibfnamefont {T.~T.}\ \bibnamefont {Luu}},\ and\ \bibinfo
  {author} {\bibfnamefont {H.~J.}\ \bibnamefont {W{\"o}rner}},\ }\bibfield
  {title} {\bibinfo {title} {Polarization measurements of deep-to
  extreme-ultraviolet high harmonics generated in liquid flat sheets},\
  }\href@noop {} {\bibfield  {journal} {\bibinfo  {journal} {Optics Express}\
  }\textbf {\bibinfo {volume} {29}},\ \bibinfo {pages} {30799} (\bibinfo {year}
  {2021})}\BibitemShut {NoStop}%
\bibitem [{\citenamefont {Mondal}\ \emph
  {et~al.}(2023{\natexlab{b}})\citenamefont {Mondal}, \citenamefont {Waser},
  \citenamefont {Balciunas}, \citenamefont {Neufeld}, \citenamefont {Yin},
  \citenamefont {Tancogne-Dejean}, \citenamefont {Rubio},\ and\ \citenamefont
  {W{\"o}rner}}]{mondal2023few}%
  \BibitemOpen
  \bibfield  {author} {\bibinfo {author} {\bibfnamefont {A.}~\bibnamefont
  {Mondal}}, \bibinfo {author} {\bibfnamefont {B.}~\bibnamefont {Waser}},
  \bibinfo {author} {\bibfnamefont {T.}~\bibnamefont {Balciunas}}, \bibinfo
  {author} {\bibfnamefont {O.}~\bibnamefont {Neufeld}}, \bibinfo {author}
  {\bibfnamefont {Z.}~\bibnamefont {Yin}}, \bibinfo {author} {\bibfnamefont
  {N.}~\bibnamefont {Tancogne-Dejean}}, \bibinfo {author} {\bibfnamefont
  {A.}~\bibnamefont {Rubio}},\ and\ \bibinfo {author} {\bibfnamefont {H.~J.}\
  \bibnamefont {W{\"o}rner}},\ }\bibfield  {title} {\bibinfo {title}
  {High-harmonic generation in liquids with few-cycle pulses: effect of
  laser-pulse duration on the cut-off energy},\ }\href@noop {} {\bibfield
  {journal} {\bibinfo  {journal} {Optics Express}\ }\textbf {\bibinfo {volume}
  {31}},\ \bibinfo {pages} {34348} (\bibinfo {year}
  {2023}{\natexlab{b}})}\BibitemShut {NoStop}%
\bibitem [{\citenamefont {Hardy}\ and\ \citenamefont
  {Cottington}(1949)}]{d2oviscosity}%
  \BibitemOpen
  \bibfield  {author} {\bibinfo {author} {\bibfnamefont {R.~C.}\ \bibnamefont
  {Hardy}}\ and\ \bibinfo {author} {\bibfnamefont {R.~L.}\ \bibnamefont
  {Cottington}},\ }\bibfield  {title} {\bibinfo {title} {Viscosity of deuterium
  oxide and water in the range 5 to 125 c},\ }\href@noop {} {\bibfield
  {journal} {\bibinfo  {journal} {J. Res. Natl. Bur. Stand}\ }\textbf {\bibinfo
  {volume} {42}},\ \bibinfo {pages} {573} (\bibinfo {year} {1949})}\BibitemShut
  {NoStop}%
\bibitem [{\citenamefont {Baker}\ \emph
  {et~al.}(2006{\natexlab{a}})\citenamefont {Baker}, \citenamefont {Robinson},
  \citenamefont {Haworth}, \citenamefont {Teng}, \citenamefont {Smith},
  \citenamefont {Chirilă}, \citenamefont {Lein}, \citenamefont {Tisch},\ and\
  \citenamefont {Marangos}}]{Baker2006}%
  \BibitemOpen
  \bibfield  {author} {\bibinfo {author} {\bibfnamefont {S.}~\bibnamefont
  {Baker}}, \bibinfo {author} {\bibfnamefont {J.~S.}\ \bibnamefont {Robinson}},
  \bibinfo {author} {\bibfnamefont {C.~A.}\ \bibnamefont {Haworth}}, \bibinfo
  {author} {\bibfnamefont {H.}~\bibnamefont {Teng}}, \bibinfo {author}
  {\bibfnamefont {R.~A.}\ \bibnamefont {Smith}}, \bibinfo {author}
  {\bibfnamefont {C.~C.}\ \bibnamefont {Chirilă}}, \bibinfo {author}
  {\bibfnamefont {M.}~\bibnamefont {Lein}}, \bibinfo {author} {\bibfnamefont
  {J.~W.~G.}\ \bibnamefont {Tisch}},\ and\ \bibinfo {author} {\bibfnamefont
  {J.~P.}\ \bibnamefont {Marangos}},\ }\bibfield  {title} {\bibinfo {title}
  {Probing proton dynamics in molecules on an attosecond time scale},\ }\href
  {https://doi.org/10.1126/science.1123904} {\bibfield  {journal} {\bibinfo
  {journal} {Science}\ }\textbf {\bibinfo {volume} {312}},\ \bibinfo {pages}
  {424} (\bibinfo {year} {2006}{\natexlab{a}})}\BibitemShut {NoStop}%
\bibitem [{\citenamefont {Marques}\ \emph {et~al.}(2012)\citenamefont
  {Marques}, \citenamefont {Maitra}, \citenamefont {Nogueira}, \citenamefont
  {Gross},\ and\ \citenamefont {Rubio}}]{marques2012fundamentals}%
  \BibitemOpen
  \bibinfo {editor} {\bibfnamefont {M.~A.~L.}\ \bibnamefont {Marques}},
  \bibinfo {editor} {\bibfnamefont {N.~T.}\ \bibnamefont {Maitra}}, \bibinfo
  {editor} {\bibfnamefont {F.~M.~S.}\ \bibnamefont {Nogueira}}, \bibinfo
  {editor} {\bibfnamefont {E.~K.~U.}\ \bibnamefont {Gross}},\ and\ \bibinfo
  {editor} {\bibfnamefont {A.}~\bibnamefont {Rubio}},\ eds.,\ \href
  {https://doi.org/10.1007/978-3-642-23518-4} {\emph {\bibinfo {title}
  {Fundamentals of Time-Dependent Density Functional Theory}}},\ \bibinfo
  {edition} {1st}\ ed.,\ \bibinfo {series} {Lecture Notes in Physics}, Vol.\
  \bibinfo {volume} {837}\ (\bibinfo  {publisher} {Springer Berlin,
  Heidelberg},\ \bibinfo {year} {2012})\ pp.\ \bibinfo {pages} {XXXII, 559},\
  \bibinfo {note} {published: 20 January 2012 (Softcover), 21 January 2012
  (eBook)}\BibitemShut {NoStop}%
\bibitem [{\citenamefont {Tancogne-Dejean}\ \emph {et~al.}(2020)\citenamefont
  {Tancogne-Dejean}, \citenamefont {Oliveira}, \citenamefont {Andrade},
  \citenamefont {Appel}, \citenamefont {Borca}, \citenamefont {Le~Breton},
  \citenamefont {Buchholz}, \citenamefont {Castro}, \citenamefont {Corni},
  \citenamefont {Correa}, \citenamefont {De~Giovannini}, \citenamefont
  {Delgado}, \citenamefont {Eich}, \citenamefont {Flick}, \citenamefont {Gil},
  \citenamefont {Gomez}, \citenamefont {Helbig}, \citenamefont {H\"{u}bener},
  \citenamefont {Jest\"{a}dt}, \citenamefont {Jornet-Somoza}, \citenamefont
  {Larsen}, \citenamefont {Lebedeva}, \citenamefont {L\"{u}ders}, \citenamefont
  {Marques}, \citenamefont {Ohlmann}, \citenamefont {Pipolo}, \citenamefont
  {Rampp}, \citenamefont {Rozzi}, \citenamefont {Strubbe}, \citenamefont
  {Sato}, \citenamefont {Sch\"{a}fer}, \citenamefont {Theophilou},
  \citenamefont {Welden},\ and\ \citenamefont {Rubio}}]{Tancogne-Dejean2020}%
  \BibitemOpen
  \bibfield  {author} {\bibinfo {author} {\bibfnamefont {N.}~\bibnamefont
  {Tancogne-Dejean}}, \bibinfo {author} {\bibfnamefont {M.~J.~T.}\ \bibnamefont
  {Oliveira}}, \bibinfo {author} {\bibfnamefont {X.}~\bibnamefont {Andrade}},
  \bibinfo {author} {\bibfnamefont {H.}~\bibnamefont {Appel}}, \bibinfo
  {author} {\bibfnamefont {C.~H.}\ \bibnamefont {Borca}}, \bibinfo {author}
  {\bibfnamefont {G.}~\bibnamefont {Le~Breton}}, \bibinfo {author}
  {\bibfnamefont {F.}~\bibnamefont {Buchholz}}, \bibinfo {author}
  {\bibfnamefont {A.}~\bibnamefont {Castro}}, \bibinfo {author} {\bibfnamefont
  {S.}~\bibnamefont {Corni}}, \bibinfo {author} {\bibfnamefont {A.~A.}\
  \bibnamefont {Correa}}, \bibinfo {author} {\bibfnamefont {U.}~\bibnamefont
  {De~Giovannini}}, \bibinfo {author} {\bibfnamefont {A.}~\bibnamefont
  {Delgado}}, \bibinfo {author} {\bibfnamefont {F.~G.}\ \bibnamefont {Eich}},
  \bibinfo {author} {\bibfnamefont {J.}~\bibnamefont {Flick}}, \bibinfo
  {author} {\bibfnamefont {G.}~\bibnamefont {Gil}}, \bibinfo {author}
  {\bibfnamefont {A.}~\bibnamefont {Gomez}}, \bibinfo {author} {\bibfnamefont
  {N.}~\bibnamefont {Helbig}}, \bibinfo {author} {\bibfnamefont
  {H.}~\bibnamefont {H\"{u}bener}}, \bibinfo {author} {\bibfnamefont
  {R.}~\bibnamefont {Jest\"{a}dt}}, \bibinfo {author} {\bibfnamefont
  {J.}~\bibnamefont {Jornet-Somoza}}, \bibinfo {author} {\bibfnamefont {A.~H.}\
  \bibnamefont {Larsen}}, \bibinfo {author} {\bibfnamefont {I.~V.}\
  \bibnamefont {Lebedeva}}, \bibinfo {author} {\bibfnamefont {M.}~\bibnamefont
  {L\"{u}ders}}, \bibinfo {author} {\bibfnamefont {M.~A.~L.}\ \bibnamefont
  {Marques}}, \bibinfo {author} {\bibfnamefont {S.~T.}\ \bibnamefont
  {Ohlmann}}, \bibinfo {author} {\bibfnamefont {S.}~\bibnamefont {Pipolo}},
  \bibinfo {author} {\bibfnamefont {M.}~\bibnamefont {Rampp}}, \bibinfo
  {author} {\bibfnamefont {C.~A.}\ \bibnamefont {Rozzi}}, \bibinfo {author}
  {\bibfnamefont {D.~A.}\ \bibnamefont {Strubbe}}, \bibinfo {author}
  {\bibfnamefont {S.~A.}\ \bibnamefont {Sato}}, \bibinfo {author}
  {\bibfnamefont {C.}~\bibnamefont {Sch\"{a}fer}}, \bibinfo {author}
  {\bibfnamefont {I.}~\bibnamefont {Theophilou}}, \bibinfo {author}
  {\bibfnamefont {A.}~\bibnamefont {Welden}},\ and\ \bibinfo {author}
  {\bibfnamefont {A.}~\bibnamefont {Rubio}},\ }\bibfield  {title} {\bibinfo
  {title} {Octopus, a computational framework for exploring light-driven
  phenomena and quantum dynamics in extended and finite systems},\ }\href
  {https://doi.org/10.1063/1.5142502} {\bibfield  {journal} {\bibinfo
  {journal} {The Journal of Chemical Physics}\ }\textbf {\bibinfo {volume}
  {152}},\ \bibinfo {pages} {124119} (\bibinfo {year} {2020})}\BibitemShut
  {NoStop}%
\bibitem [{\citenamefont {M\"oller}\ \emph {et~al.}(2012)\citenamefont
  {M\"oller}, \citenamefont {Cheng}, \citenamefont {Khan}, \citenamefont
  {Zhao}, \citenamefont {Zhao}, \citenamefont {Chini}, \citenamefont {Paulus},\
  and\ \citenamefont {Chang}}]{Max2012}%
  \BibitemOpen
  \bibfield  {author} {\bibinfo {author} {\bibfnamefont {M.}~\bibnamefont
  {M\"oller}}, \bibinfo {author} {\bibfnamefont {Y.}~\bibnamefont {Cheng}},
  \bibinfo {author} {\bibfnamefont {S.~D.}\ \bibnamefont {Khan}}, \bibinfo
  {author} {\bibfnamefont {B.}~\bibnamefont {Zhao}}, \bibinfo {author}
  {\bibfnamefont {K.}~\bibnamefont {Zhao}}, \bibinfo {author} {\bibfnamefont
  {M.}~\bibnamefont {Chini}}, \bibinfo {author} {\bibfnamefont {G.~G.}\
  \bibnamefont {Paulus}},\ and\ \bibinfo {author} {\bibfnamefont
  {Z.}~\bibnamefont {Chang}},\ }\bibfield  {title} {\bibinfo {title}
  {Dependence of high-order-harmonic-generation yield on driving-laser
  ellipticity},\ }\href {https://doi.org/10.1103/PhysRevA.86.011401} {\bibfield
   {journal} {\bibinfo  {journal} {Phys. Rev. A}\ }\textbf {\bibinfo {volume}
  {86}},\ \bibinfo {pages} {011401} (\bibinfo {year} {2012})}\BibitemShut
  {NoStop}%
\bibitem [{\citenamefont {Ferr{\'{e}}}\ \emph {et~al.}(2015)\citenamefont
  {Ferr{\'{e}}}, \citenamefont {Boguslavskiy}, \citenamefont {Dagan},
  \citenamefont {Blanchet}, \citenamefont {Bruner}, \citenamefont {Burgy},
  \citenamefont {Camper}, \citenamefont {Descamps}, \citenamefont {Fabre},
  \citenamefont {Fedorov}, \citenamefont {Gaudin}, \citenamefont {Geoffroy},
  \citenamefont {Mikosch}, \citenamefont {Patchkovskii}, \citenamefont {Petit},
  \citenamefont {Ruchon}, \citenamefont {Soifer}, \citenamefont {Staedter},
  \citenamefont {Wilkinson}, \citenamefont {Stolow}, \citenamefont {Dudovich},\
  and\ \citenamefont {Mairesse}}]{Ferre2015}%
  \BibitemOpen
  \bibfield  {author} {\bibinfo {author} {\bibfnamefont {A.}~\bibnamefont
  {Ferr{\'{e}}}}, \bibinfo {author} {\bibfnamefont {A.~E.}\ \bibnamefont
  {Boguslavskiy}}, \bibinfo {author} {\bibfnamefont {M.}~\bibnamefont {Dagan}},
  \bibinfo {author} {\bibfnamefont {V.}~\bibnamefont {Blanchet}}, \bibinfo
  {author} {\bibfnamefont {B.~D.}\ \bibnamefont {Bruner}}, \bibinfo {author}
  {\bibfnamefont {F.}~\bibnamefont {Burgy}}, \bibinfo {author} {\bibfnamefont
  {A.}~\bibnamefont {Camper}}, \bibinfo {author} {\bibfnamefont
  {D.}~\bibnamefont {Descamps}}, \bibinfo {author} {\bibfnamefont
  {B.}~\bibnamefont {Fabre}}, \bibinfo {author} {\bibfnamefont
  {N.}~\bibnamefont {Fedorov}}, \bibinfo {author} {\bibfnamefont
  {J.}~\bibnamefont {Gaudin}}, \bibinfo {author} {\bibfnamefont
  {G.}~\bibnamefont {Geoffroy}}, \bibinfo {author} {\bibfnamefont
  {J.}~\bibnamefont {Mikosch}}, \bibinfo {author} {\bibfnamefont
  {S.}~\bibnamefont {Patchkovskii}}, \bibinfo {author} {\bibfnamefont
  {S.}~\bibnamefont {Petit}}, \bibinfo {author} {\bibfnamefont
  {T.}~\bibnamefont {Ruchon}}, \bibinfo {author} {\bibfnamefont
  {H.}~\bibnamefont {Soifer}}, \bibinfo {author} {\bibfnamefont
  {D.}~\bibnamefont {Staedter}}, \bibinfo {author} {\bibfnamefont
  {I.}~\bibnamefont {Wilkinson}}, \bibinfo {author} {\bibfnamefont
  {A.}~\bibnamefont {Stolow}}, \bibinfo {author} {\bibfnamefont
  {N.}~\bibnamefont {Dudovich}},\ and\ \bibinfo {author} {\bibfnamefont
  {Y.}~\bibnamefont {Mairesse}},\ }\bibfield  {title} {\bibinfo {title}
  {{Multi-channel electronic and vibrational dynamics in polyatomic resonant
  high-order harmonic generation}},\ }\href
  {https://doi.org/10.1038/ncomms6952} {\bibfield  {journal} {\bibinfo
  {journal} {Nature Communications}\ }\textbf {\bibinfo {volume} {6}},\
  \bibinfo {pages} {5952} (\bibinfo {year} {2015})}\BibitemShut {NoStop}%
\bibitem [{\citenamefont {Yu}\ \emph {et~al.}(2022)\citenamefont {Yu},
  \citenamefont {Saalmann},\ and\ \citenamefont {Rost}}]{PhysRevA.105.L041101}%
  \BibitemOpen
  \bibfield  {author} {\bibinfo {author} {\bibfnamefont {C.}~\bibnamefont
  {Yu}}, \bibinfo {author} {\bibfnamefont {U.}~\bibnamefont {Saalmann}},\ and\
  \bibinfo {author} {\bibfnamefont {J.~M.}\ \bibnamefont {Rost}},\ }\bibfield
  {title} {\bibinfo {title} {High-order harmonics from backscattering of
  delocalized electrons},\ }\href
  {https://doi.org/10.1103/PhysRevA.105.L041101} {\bibfield  {journal}
  {\bibinfo  {journal} {Phys. Rev. A}\ }\textbf {\bibinfo {volume} {105}},\
  \bibinfo {pages} {L041101} (\bibinfo {year} {2022})}\BibitemShut {NoStop}%
\bibitem [{\citenamefont {Prendergast}\ \emph {et~al.}(2005)\citenamefont
  {Prendergast}, \citenamefont {Grossman},\ and\ \citenamefont
  {Galli}}]{10.1063/1.1940612}%
  \BibitemOpen
  \bibfield  {author} {\bibinfo {author} {\bibfnamefont {D.}~\bibnamefont
  {Prendergast}}, \bibinfo {author} {\bibfnamefont {J.~C.}\ \bibnamefont
  {Grossman}},\ and\ \bibinfo {author} {\bibfnamefont {G.}~\bibnamefont
  {Galli}},\ }\bibfield  {title} {\bibinfo {title} {The electronic structure of
  liquid water within density-functional theory},\ }\href
  {https://doi.org/10.1063/1.1940612} {\bibfield  {journal} {\bibinfo
  {journal} {The Journal of Chemical Physics}\ }\textbf {\bibinfo {volume}
  {123}},\ \bibinfo {pages} {014501} (\bibinfo {year} {2005})},\ \Eprint
  {https://arxiv.org/abs/https://pubs.aip.org/aip/jcp/article-pdf/doi/10.1063/1.1940612/15369007/014501\_1\_online.pdf}
  {https://pubs.aip.org/aip/jcp/article-pdf/doi/10.1063/1.1940612/15369007/014501\_1\_online.pdf}
  \BibitemShut {NoStop}%
\bibitem [{\citenamefont {Soper}(2013)}]{soper2013radial}%
  \BibitemOpen
  \bibfield  {author} {\bibinfo {author} {\bibfnamefont {A.~K.}\ \bibnamefont
  {Soper}},\ }\bibfield  {title} {\bibinfo {title} {The radial distribution
  functions of water as derived from radiation total scattering experiments: is
  there anything we can say for sure?},\ }\href@noop {} {\bibfield  {journal}
  {\bibinfo  {journal} {International Scholarly Research Notices}\ }\textbf
  {\bibinfo {volume} {2013}},\ \bibinfo {pages} {279463} (\bibinfo {year}
  {2013})}\BibitemShut {NoStop}%
\bibitem [{\citenamefont {Baker}\ \emph
  {et~al.}(2006{\natexlab{b}})\citenamefont {Baker}, \citenamefont {Robinson},
  \citenamefont {Haworth}, \citenamefont {Teng}, \citenamefont {Smith},
  \citenamefont {Chirila}, \citenamefont {Lein}, \citenamefont {Tisch},\ and\
  \citenamefont {Marangos}}]{baker06a}%
  \BibitemOpen
  \bibfield  {author} {\bibinfo {author} {\bibfnamefont {S.}~\bibnamefont
  {Baker}}, \bibinfo {author} {\bibfnamefont {J.~S.}\ \bibnamefont {Robinson}},
  \bibinfo {author} {\bibfnamefont {C.~A.}\ \bibnamefont {Haworth}}, \bibinfo
  {author} {\bibfnamefont {H.}~\bibnamefont {Teng}}, \bibinfo {author}
  {\bibfnamefont {R.~A.}\ \bibnamefont {Smith}}, \bibinfo {author}
  {\bibfnamefont {C.~C.}\ \bibnamefont {Chirila}}, \bibinfo {author}
  {\bibfnamefont {M.}~\bibnamefont {Lein}}, \bibinfo {author} {\bibfnamefont
  {J.~W.~G.}\ \bibnamefont {Tisch}},\ and\ \bibinfo {author} {\bibfnamefont
  {J.~P.}\ \bibnamefont {Marangos}},\ }\bibfield  {title} {\bibinfo {title}
  {Probing proton dynamics in molecules on an attosecond time scale},\
  }\href@noop {} {\bibfield  {journal} {\bibinfo  {journal} {Science}\ }\textbf
  {\bibinfo {volume} {312}},\ \bibinfo {pages} {424} (\bibinfo {year}
  {2006}{\natexlab{b}})}\BibitemShut {NoStop}%
\bibitem [{\citenamefont {Patchkovskii}(2009)}]{Patchkovskii2009}%
  \BibitemOpen
  \bibfield  {author} {\bibinfo {author} {\bibfnamefont {S.}~\bibnamefont
  {Patchkovskii}},\ }\bibfield  {title} {\bibinfo {title} {Nuclear dynamics in
  polyatomic molecules and high-order harmonic generation},\ }\href
  {https://doi.org/10.1103/PhysRevLett.102.253602} {\bibfield  {journal}
  {\bibinfo  {journal} {Phys. Rev. Lett.}\ }\textbf {\bibinfo {volume} {102}},\
  \bibinfo {pages} {253602} (\bibinfo {year} {2009})}\BibitemShut {NoStop}%
\bibitem [{\citenamefont {He}\ \emph {et~al.}(2018)\citenamefont {He},
  \citenamefont {Zhang}, \citenamefont {Lan}, \citenamefont {Cao},
  \citenamefont {Zhu}, \citenamefont {Zhai}, \citenamefont {Wang},
  \citenamefont {Shi}, \citenamefont {Li}, \citenamefont {Bian}, \citenamefont
  {Lu},\ and\ \citenamefont {Bandrauk}}]{He2018}%
  \BibitemOpen
  \bibfield  {author} {\bibinfo {author} {\bibfnamefont {L.}~\bibnamefont
  {He}}, \bibinfo {author} {\bibfnamefont {Q.}~\bibnamefont {Zhang}}, \bibinfo
  {author} {\bibfnamefont {P.}~\bibnamefont {Lan}}, \bibinfo {author}
  {\bibfnamefont {W.}~\bibnamefont {Cao}}, \bibinfo {author} {\bibfnamefont
  {X.}~\bibnamefont {Zhu}}, \bibinfo {author} {\bibfnamefont {C.}~\bibnamefont
  {Zhai}}, \bibinfo {author} {\bibfnamefont {F.}~\bibnamefont {Wang}}, \bibinfo
  {author} {\bibfnamefont {W.}~\bibnamefont {Shi}}, \bibinfo {author}
  {\bibfnamefont {M.}~\bibnamefont {Li}}, \bibinfo {author} {\bibfnamefont
  {X.-B.}\ \bibnamefont {Bian}}, \bibinfo {author} {\bibfnamefont
  {P.}~\bibnamefont {Lu}},\ and\ \bibinfo {author} {\bibfnamefont {A.~D.}\
  \bibnamefont {Bandrauk}},\ }\bibfield  {title} {\bibinfo {title} {Monitoring
  ultrafast vibrational dynamics of isotopic molecules with frequency
  modulation of high-order harmonics},\ }\href
  {https://doi.org/10.1038/s41467-018-03568-3} {\bibfield  {journal} {\bibinfo
  {journal} {Nature Communications}\ }\textbf {\bibinfo {volume} {9}},\
  \bibinfo {pages} {1108} (\bibinfo {year} {2018})}\BibitemShut {NoStop}%
\bibitem [{\citenamefont {Legrand}\ \emph {et~al.}(2002)\citenamefont
  {Legrand}, \citenamefont {Suraud},\ and\ \citenamefont
  {Reinhard}}]{Legrand2002}%
  \BibitemOpen
  \bibfield  {author} {\bibinfo {author} {\bibfnamefont {C.}~\bibnamefont
  {Legrand}}, \bibinfo {author} {\bibfnamefont {E.}~\bibnamefont {Suraud}},\
  and\ \bibinfo {author} {\bibfnamefont {P.-G.}\ \bibnamefont {Reinhard}},\
  }\bibfield  {title} {\bibinfo {title} {Comparison of
  self-interaction-corrections for metal clusters},\ }\href
  {https://doi.org/10.1088/0953-4075/35/4/333} {\bibfield  {journal} {\bibinfo
  {journal} {Journal of Physics B: Atomic, Molecular and Optical Physics}\
  }\textbf {\bibinfo {volume} {35}},\ \bibinfo {pages} {1115} (\bibinfo {year}
  {2002})}\BibitemShut {NoStop}%
\bibitem [{\citenamefont {Behler}\ and\ \citenamefont
  {Parrinello}(2007)}]{PhysRevLett.98.146401}%
  \BibitemOpen
  \bibfield  {author} {\bibinfo {author} {\bibfnamefont {J.}~\bibnamefont
  {Behler}}\ and\ \bibinfo {author} {\bibfnamefont {M.}~\bibnamefont
  {Parrinello}},\ }\bibfield  {title} {\bibinfo {title} {Generalized
  neural-network representation of high-dimensional potential-energy
  surfaces},\ }\href {https://doi.org/10.1103/PhysRevLett.98.146401} {\bibfield
   {journal} {\bibinfo  {journal} {Phys. Rev. Lett.}\ }\textbf {\bibinfo
  {volume} {98}},\ \bibinfo {pages} {146401} (\bibinfo {year}
  {2007})}\BibitemShut {NoStop}%
\bibitem [{\citenamefont {O’Neill}\ \emph {et~al.}(2024)\citenamefont
  {O’Neill}, \citenamefont {Shi}, \citenamefont {Fong}, \citenamefont
  {Michaelides},\ and\ \citenamefont {Schran}}]{o2024pair}%
  \BibitemOpen
  \bibfield  {author} {\bibinfo {author} {\bibfnamefont {N.}~\bibnamefont
  {O’Neill}}, \bibinfo {author} {\bibfnamefont {B.~X.}\ \bibnamefont {Shi}},
  \bibinfo {author} {\bibfnamefont {K.}~\bibnamefont {Fong}}, \bibinfo {author}
  {\bibfnamefont {A.}~\bibnamefont {Michaelides}},\ and\ \bibinfo {author}
  {\bibfnamefont {C.}~\bibnamefont {Schran}},\ }\bibfield  {title} {\bibinfo
  {title} {To pair or not to pair? machine-learned explicitly-correlated
  electronic structure for nacl in water},\ }\href@noop {} {\bibfield
  {journal} {\bibinfo  {journal} {The Journal of Physical Chemistry Letters}\
  }\textbf {\bibinfo {volume} {15}},\ \bibinfo {pages} {6081} (\bibinfo {year}
  {2024})}\BibitemShut {NoStop}%
\bibitem [{\citenamefont {Ceriotti}\ \emph {et~al.}(2010)\citenamefont
  {Ceriotti}, \citenamefont {Parrinello}, \citenamefont {Markland},\ and\
  \citenamefont {Manolopoulos}}]{ceriotti2010efficient}%
  \BibitemOpen
  \bibfield  {author} {\bibinfo {author} {\bibfnamefont {M.}~\bibnamefont
  {Ceriotti}}, \bibinfo {author} {\bibfnamefont {M.}~\bibnamefont
  {Parrinello}}, \bibinfo {author} {\bibfnamefont {T.~E.}\ \bibnamefont
  {Markland}},\ and\ \bibinfo {author} {\bibfnamefont {D.~E.}\ \bibnamefont
  {Manolopoulos}},\ }\bibfield  {title} {\bibinfo {title} {Efficient stochastic
  thermostatting of path integral molecular dynamics},\ }\href@noop {}
  {\bibfield  {journal} {\bibinfo  {journal} {The Journal of chemical physics}\
  }\textbf {\bibinfo {volume} {133}} (\bibinfo {year} {2010})}\BibitemShut
  {NoStop}%
\bibitem [{\citenamefont {Litman}\ \emph {et~al.}(2024)\citenamefont {Litman},
  \citenamefont {Kapil}, \citenamefont {Feldman}, \citenamefont {Tisi},
  \citenamefont {Begu{\v{s}}i{\'c}}, \citenamefont {Fidanyan}, \citenamefont
  {Fraux}, \citenamefont {Higer}, \citenamefont {Kellner}, \citenamefont {Li}
  \emph {et~al.}}]{litman2024pi}%
  \BibitemOpen
  \bibfield  {author} {\bibinfo {author} {\bibfnamefont {Y.}~\bibnamefont
  {Litman}}, \bibinfo {author} {\bibfnamefont {V.}~\bibnamefont {Kapil}},
  \bibinfo {author} {\bibfnamefont {Y.~M.}\ \bibnamefont {Feldman}}, \bibinfo
  {author} {\bibfnamefont {D.}~\bibnamefont {Tisi}}, \bibinfo {author}
  {\bibfnamefont {T.}~\bibnamefont {Begu{\v{s}}i{\'c}}}, \bibinfo {author}
  {\bibfnamefont {K.}~\bibnamefont {Fidanyan}}, \bibinfo {author}
  {\bibfnamefont {G.}~\bibnamefont {Fraux}}, \bibinfo {author} {\bibfnamefont
  {J.}~\bibnamefont {Higer}}, \bibinfo {author} {\bibfnamefont
  {M.}~\bibnamefont {Kellner}}, \bibinfo {author} {\bibfnamefont {T.~E.}\
  \bibnamefont {Li}}, \emph {et~al.},\ }\bibfield  {title} {\bibinfo {title}
  {i-pi 3.0: a flexible, efficient framework for advanced atomistic
  simulations},\ }\href@noop {} {\bibfield  {journal} {\bibinfo  {journal}
  {arXiv preprint arXiv:2405.15224}\ } (\bibinfo {year} {2024})}\BibitemShut
  {NoStop}%
\bibitem [{\citenamefont {Schran}\ \emph {et~al.}(2021)\citenamefont {Schran},
  \citenamefont {Thiemann}, \citenamefont {Rowe}, \citenamefont {M{\"u}ller},
  \citenamefont {Marsalek},\ and\ \citenamefont
  {Michaelides}}]{schran2021machine}%
  \BibitemOpen
  \bibfield  {author} {\bibinfo {author} {\bibfnamefont {C.}~\bibnamefont
  {Schran}}, \bibinfo {author} {\bibfnamefont {F.~L.}\ \bibnamefont
  {Thiemann}}, \bibinfo {author} {\bibfnamefont {P.}~\bibnamefont {Rowe}},
  \bibinfo {author} {\bibfnamefont {E.~A.}\ \bibnamefont {M{\"u}ller}},
  \bibinfo {author} {\bibfnamefont {O.}~\bibnamefont {Marsalek}},\ and\
  \bibinfo {author} {\bibfnamefont {A.}~\bibnamefont {Michaelides}},\
  }\bibfield  {title} {\bibinfo {title} {Machine learning potentials for
  complex aqueous systems made simple},\ }\href@noop {} {\bibfield  {journal}
  {\bibinfo  {journal} {Proceedings of the National Academy of Sciences}\
  }\textbf {\bibinfo {volume} {118}},\ \bibinfo {pages} {e2110077118} (\bibinfo
  {year} {2021})}\BibitemShut {NoStop}%
\bibitem [{\citenamefont {Blum}\ \emph {et~al.}(2009)\citenamefont {Blum},
  \citenamefont {Gehrke}, \citenamefont {Hanke}, \citenamefont {Havu},
  \citenamefont {Havu}, \citenamefont {Ren}, \citenamefont {Reuter},\ and\
  \citenamefont {Scheffler}}]{blum2009ab}%
  \BibitemOpen
  \bibfield  {author} {\bibinfo {author} {\bibfnamefont {V.}~\bibnamefont
  {Blum}}, \bibinfo {author} {\bibfnamefont {R.}~\bibnamefont {Gehrke}},
  \bibinfo {author} {\bibfnamefont {F.}~\bibnamefont {Hanke}}, \bibinfo
  {author} {\bibfnamefont {P.}~\bibnamefont {Havu}}, \bibinfo {author}
  {\bibfnamefont {V.}~\bibnamefont {Havu}}, \bibinfo {author} {\bibfnamefont
  {X.}~\bibnamefont {Ren}}, \bibinfo {author} {\bibfnamefont {K.}~\bibnamefont
  {Reuter}},\ and\ \bibinfo {author} {\bibfnamefont {M.}~\bibnamefont
  {Scheffler}},\ }\bibfield  {title} {\bibinfo {title} {Ab initio molecular
  simulations with numeric atom-centered orbitals},\ }\href@noop {} {\bibfield
  {journal} {\bibinfo  {journal} {Computer Physics Communications}\ }\textbf
  {\bibinfo {volume} {180}},\ \bibinfo {pages} {2175} (\bibinfo {year}
  {2009})}\BibitemShut {NoStop}%
\bibitem [{\citenamefont {Nourbakhsh}\ \emph {et~al.}(2022)\citenamefont
  {Nourbakhsh}, \citenamefont {Neufeld}, \citenamefont {Tancogne-Dejean},\ and\
  \citenamefont {Rubio}}]{2212.04177}%
  \BibitemOpen
  \bibfield  {author} {\bibinfo {author} {\bibfnamefont {Z.}~\bibnamefont
  {Nourbakhsh}}, \bibinfo {author} {\bibfnamefont {O.}~\bibnamefont {Neufeld}},
  \bibinfo {author} {\bibfnamefont {N.}~\bibnamefont {Tancogne-Dejean}},\ and\
  \bibinfo {author} {\bibfnamefont {A.}~\bibnamefont {Rubio}},\ }\href@noop {}
  {\bibinfo {title} {An ab initio supercell approach for high-harmonic
  generation in liquids}} (\bibinfo {year} {2022}),\ \Eprint
  {https://arxiv.org/abs/arXiv:2212.04177} {arXiv:2212.04177} \BibitemShut
  {NoStop}%
\bibitem [{\citenamefont {Xu}\ and\ \citenamefont {Meng}(2025)}]{xu2025high}%
  \BibitemOpen
  \bibfield  {author} {\bibinfo {author} {\bibfnamefont {J.}~\bibnamefont
  {Xu}}\ and\ \bibinfo {author} {\bibfnamefont {S.}~\bibnamefont {Meng}},\
  }\bibfield  {title} {\bibinfo {title} {High-harmonic generation and
  femtosecond-resolved ultrafast dynamics in liquid water},\ }\href@noop {}
  {\bibfield  {journal} {\bibinfo  {journal} {The Journal of Physical Chemistry
  Letters}\ }\textbf {\bibinfo {volume} {16}},\ \bibinfo {pages} {5295}
  (\bibinfo {year} {2025})}\BibitemShut {NoStop}%
\bibitem [{\citenamefont {Lewenstein}\ \emph
  {et~al.}(1994{\natexlab{b}})\citenamefont {Lewenstein}, \citenamefont
  {Balcou}, \citenamefont {Ivanov}, \citenamefont {L'Huillier},\ and\
  \citenamefont {Corkum}}]{PhysRevA.49.2117}%
  \BibitemOpen
  \bibfield  {author} {\bibinfo {author} {\bibfnamefont {M.}~\bibnamefont
  {Lewenstein}}, \bibinfo {author} {\bibfnamefont {P.}~\bibnamefont {Balcou}},
  \bibinfo {author} {\bibfnamefont {M.~Y.}\ \bibnamefont {Ivanov}}, \bibinfo
  {author} {\bibfnamefont {A.}~\bibnamefont {L'Huillier}},\ and\ \bibinfo
  {author} {\bibfnamefont {P.~B.}\ \bibnamefont {Corkum}},\ }\bibfield  {title}
  {\bibinfo {title} {Theory of high-harmonic generation by low-frequency laser
  fields},\ }\href {https://doi.org/10.1103/PhysRevA.49.2117} {\bibfield
  {journal} {\bibinfo  {journal} {Phys. Rev. A}\ }\textbf {\bibinfo {volume}
  {49}},\ \bibinfo {pages} {2117} (\bibinfo {year}
  {1994}{\natexlab{b}})}\BibitemShut {NoStop}%
\end{thebibliography}%

\begin{acknowledgments}
The authors thank Andreas Schneider and Mario Seiler for their contributions to the construction and improvements of the experiment, Nirit Dudovich for helpful discussions and George Trenins for providing the molecular dynamics trajectories used to calculate HOMO and LUMO states of liquid water.  

\end{acknowledgments}

\section*{Author Contributions}

All the authors participated in the discussion of the results and contributed to the manuscript.

\section*{Competing interests}
The authors declare no competing interests.

\section*{Extended Data Figures}

\begin{figure*} [t!]
\includegraphics[width=\textwidth]{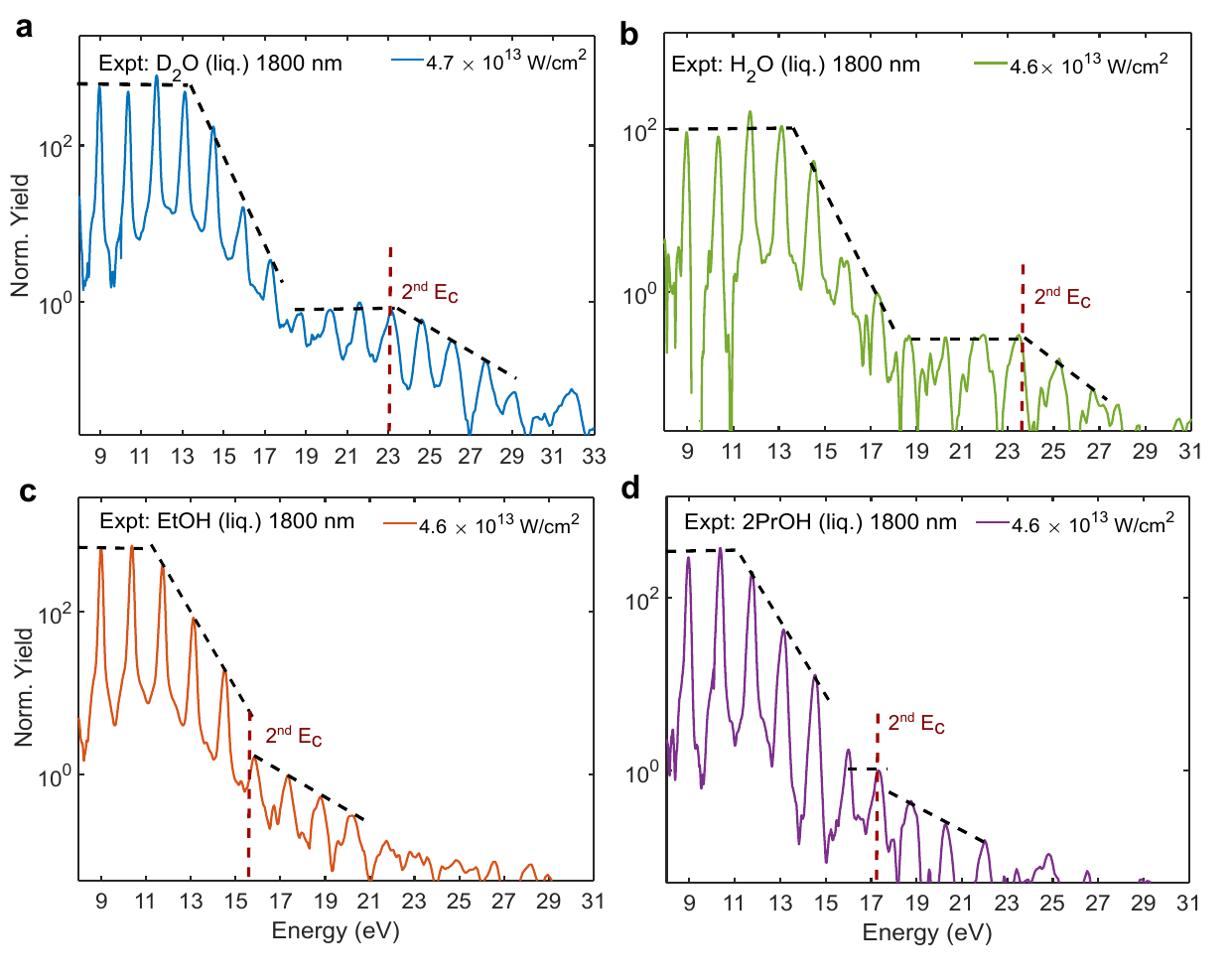}
{{\textbf{Extended Data Fig. 1}} Experimental HHG spectra from liquid H$_2$O, D$_2$O, ethanol and isopropanol at 1800 nm driving wavelength, all showing second plateaus, indicating the generality of the phenomenon.}
\end{figure*}
\begin{figure*} [t!]
\includegraphics[width=\textwidth]{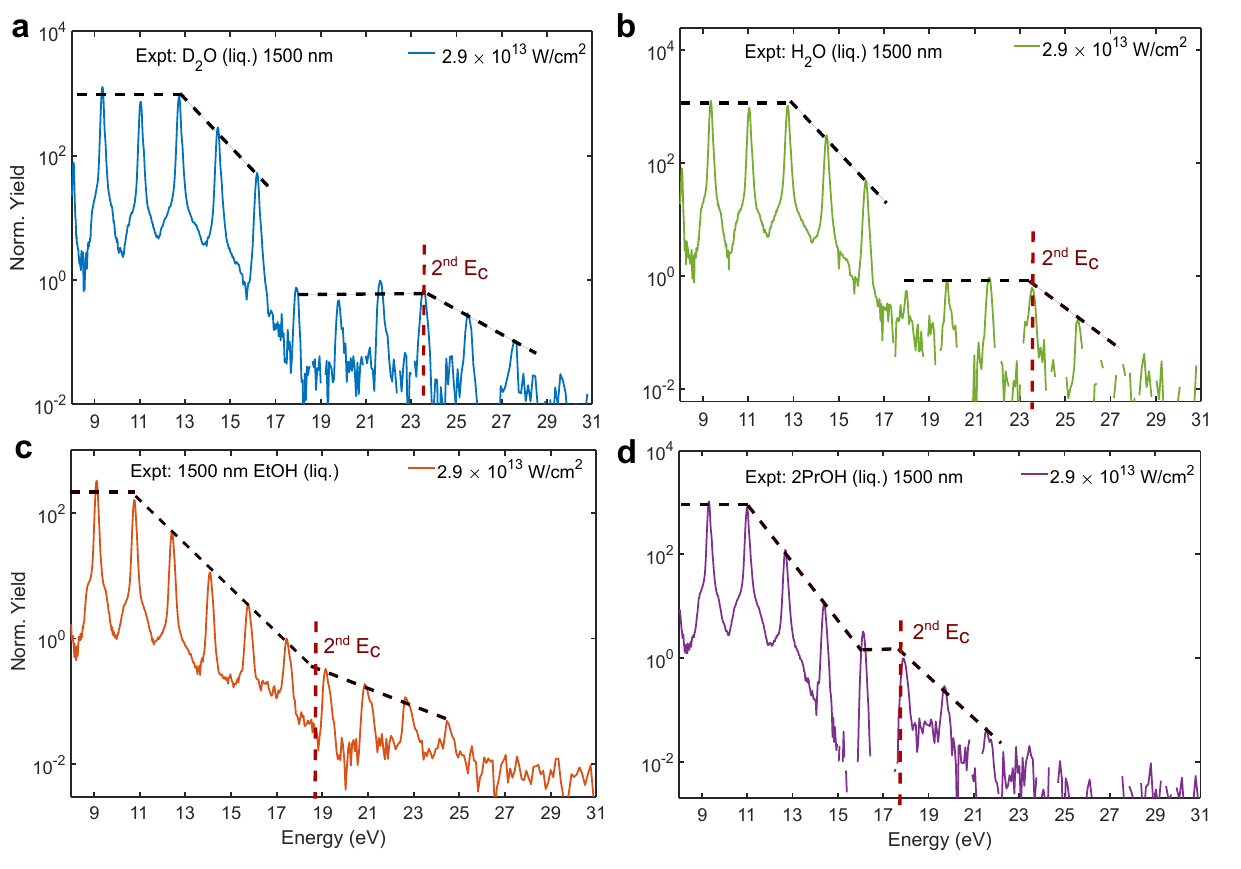}
{{\textbf{Extended Data Fig. 2}} Experimental HHG spectra from liquid H$_2$O, D$_2$O, ethanol and isopropanol at 1500 nm driving wavelength, all showing second plateaus, indicating the generality of the phenomenon.}
\end{figure*}
\begin{figure*} [h!]
\includegraphics[width=0.8\textwidth]{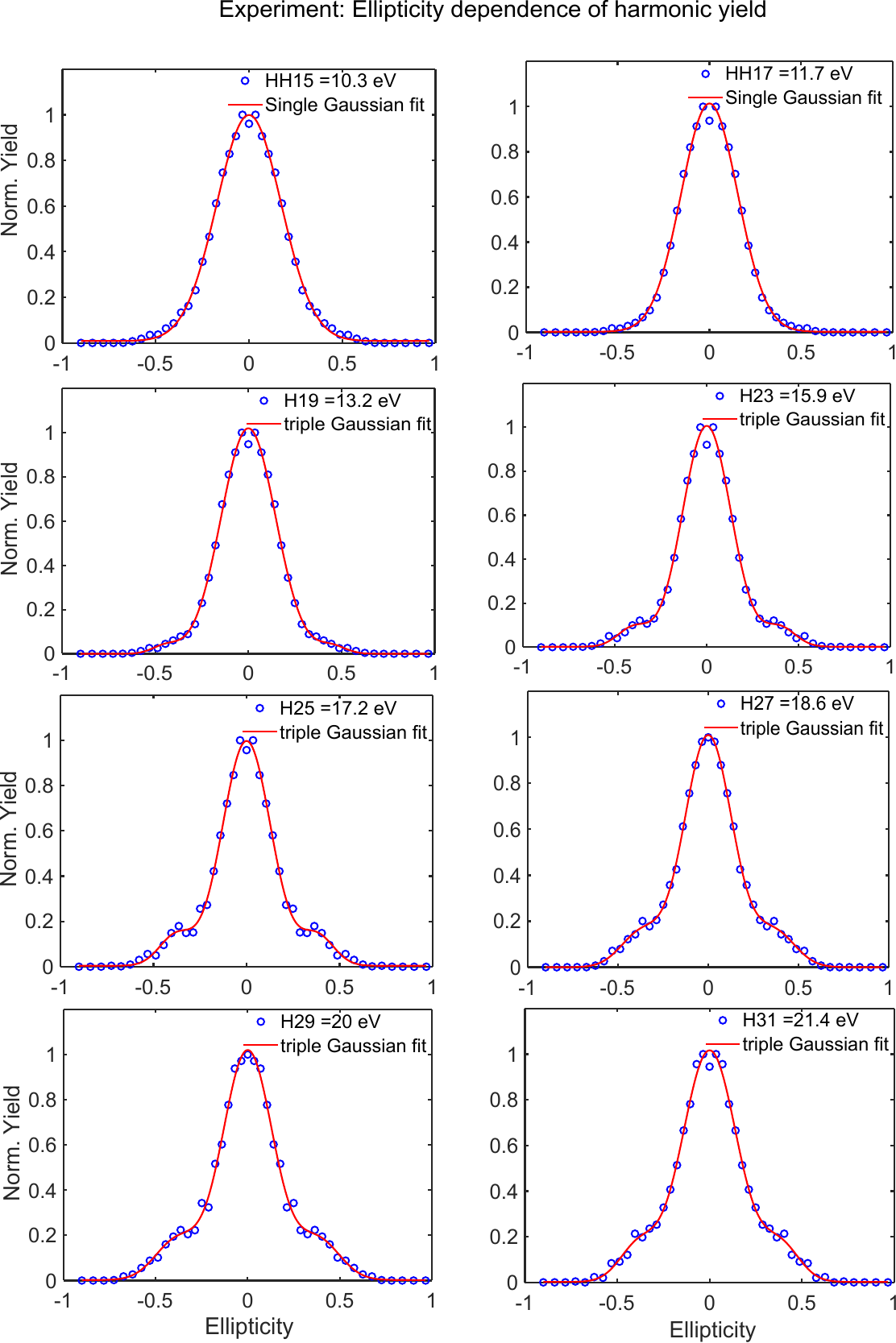}\\
{{\textbf{Extended Data Fig. 3}} Experimentally observed ellipticity dependence of harmonics for liquid ethanol using 1800 nm wavelength at 2.4$\times$10$^{13}$ W/cm$^2$.}
\end{figure*}
\begin{figure*} [t!]
\includegraphics[width=0.8\textwidth]{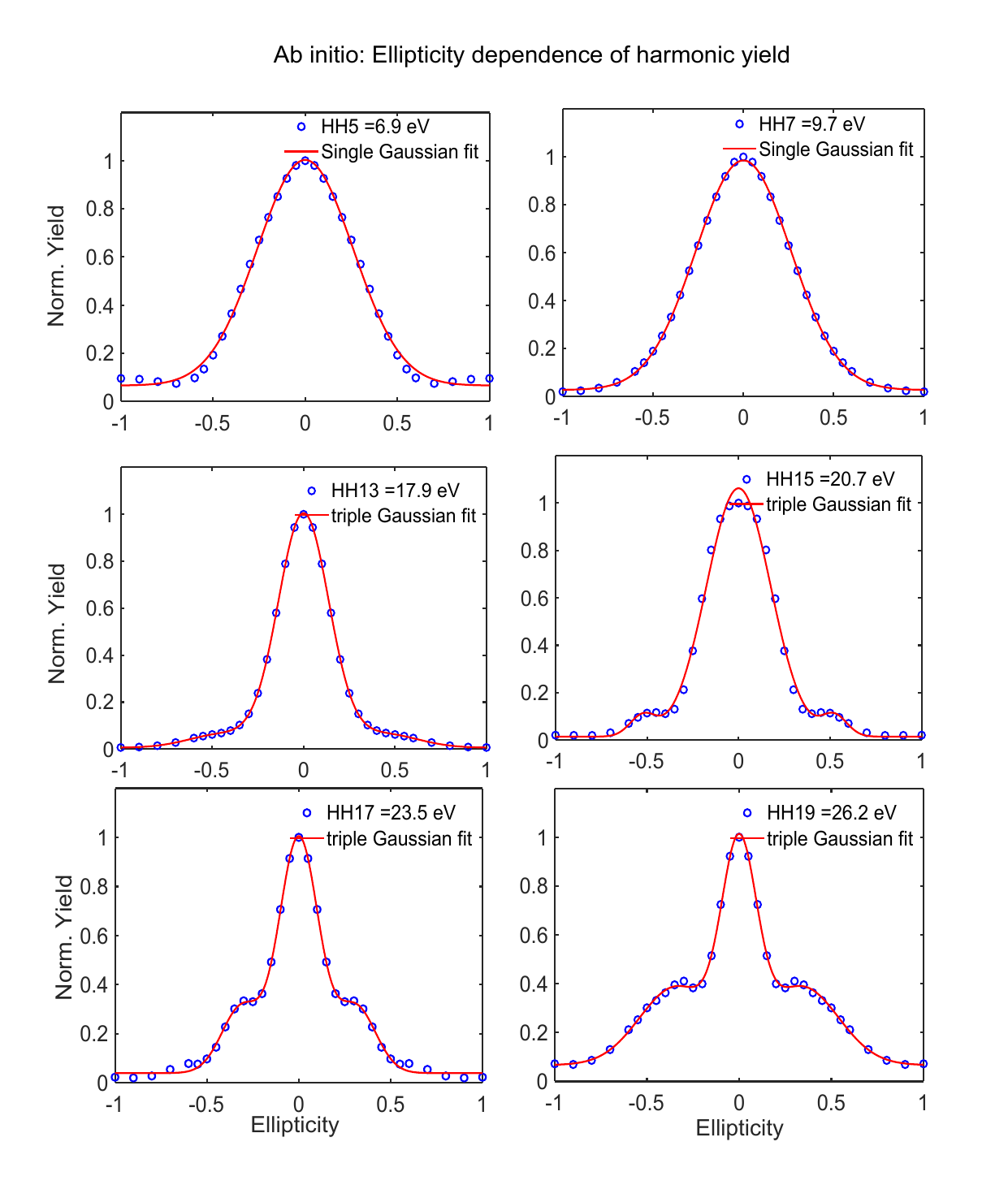}\\
{{\textbf{Extended Data Fig. 4}} Ellipticity dependence of harmonics obtained from ab initio simulations for liquid water at 900 nm at 4$\times$10$^{13}$ W/cm$^2$.}
\end{figure*}
\begin{figure*} [t!]
\includegraphics[width=\textwidth]{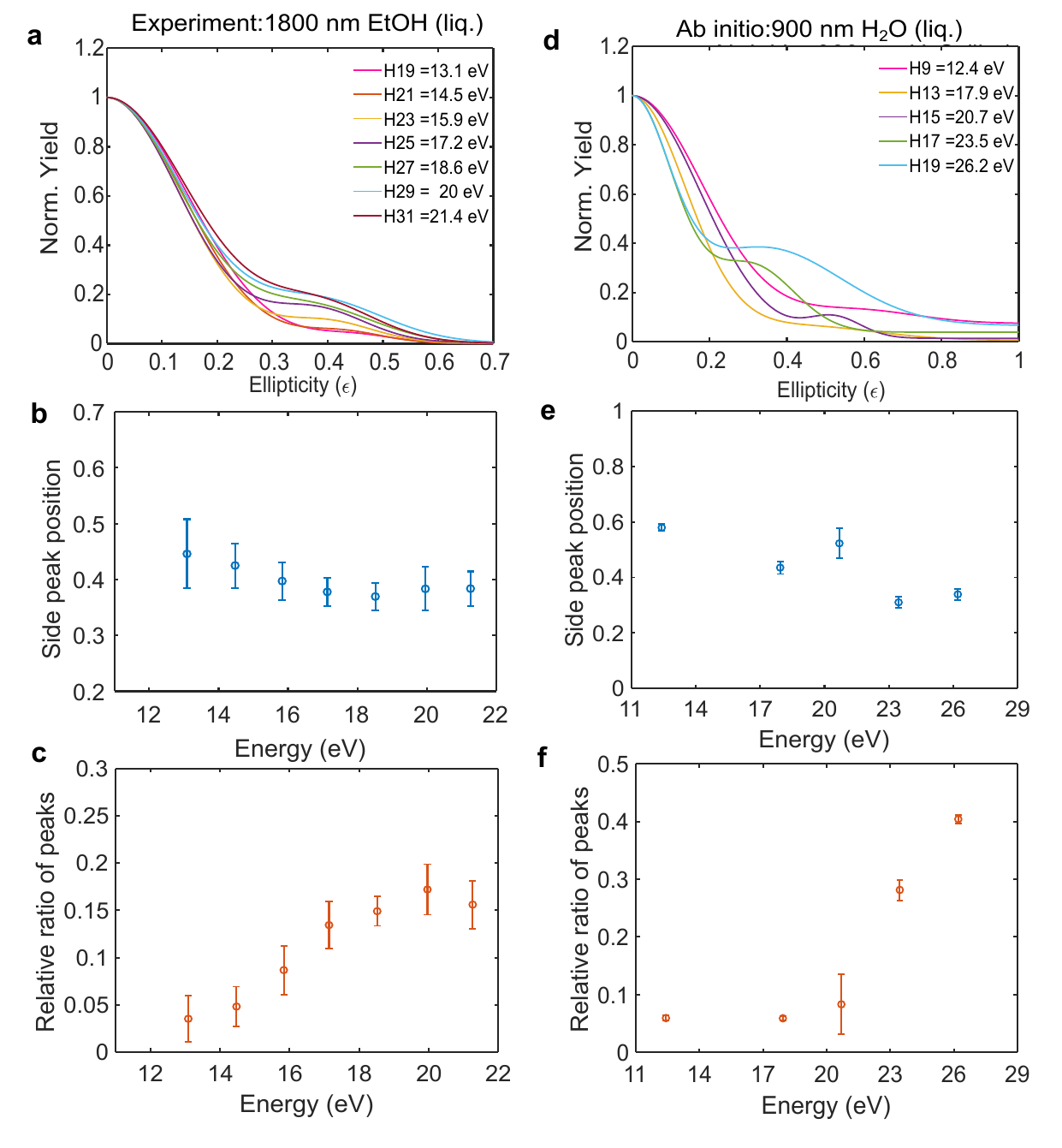}
{{\textbf{Extended Data Fig. 5}} 
Detailed analysis of the ellipticity dependence of measured (left) and calculated (right) second-plateau harmonics.
(a) Curves fitted to the experimental ellipticity dependence of harmonics beyond 1$^{st}$ cut-off energy ($\sim$11.5 eV) for liquid EtOH driven at 1800 nm at 2.4$\times$10$^{13}$ W/cm$^2$. (b) Extracted side-peak position and (c) ratio of the side-peak amplitude with respect to the central peak at ellipticity = 0 (linear polarization). (d) Curves fitted to the calculated ellipticity dependence of harmonics emitted from H$_2$O clusters driven at 900 nm at 4$\times$10$^{13}$ W/cm$^2$. (d) Extracted side-peak position and (e) ratio of the side-peak amplitude with respect to the central peak at ellipticity = 0 (linear polarization). The triple Gaussian fit did not converge for the calculated harmonic of energy 15.2 eV, therefore the data point at 15.2 eV have been excluded from Ext. Data Fig 4.}
\end{figure*}

\end{document}


\preprint{APS/123-QED}

\title{Supporting Information for \\
Multi-plateau high-harmonic generation in liquids driven by off-site recombination
}

\author{Angana Mondal}
\affiliation{Laboratorium für Physikalische Chemie, ETH Zürich, Zurich, Switzerland}
\affiliation{These authors contributed equally}
\author{Ofer Neufeld}
\email[]{ofern@technion.ac.il}
\affiliation{Technion Israel Institute of Technology, Faculty of Chemistry, Haifa 3200003, Israel}
\affiliation{Max Planck Institute for the Structure and Dynamics of Matter and Center for Free-Electron Laser Science, Luruper Chaussee 149, 22761 Hamburg, Germany}
\affiliation{These authors contributed equally}
\author{Tadas Balciunas}
\affiliation{Laboratorium für Physikalische Chemie, ETH Zürich, Zurich, Switzerland}
\author{Benedikt Waser}
\affiliation{Laboratorium für Physikalische Chemie, ETH Zürich, Zurich, Switzerland}
\author{Serge Müller}
\affiliation{Laboratorium für Physikalische Chemie, ETH Zürich, Zurich, Switzerland}
\author{Mariana Rossi}
\affiliation{Max Planck Institute for the Structure and Dynamics of Matter and Center for Free-Electron Laser Science, Luruper Chaussee 149, 22761 Hamburg, Germany}
\author{Zhong Yin}
\affiliation{Laboratorium für Physikalische Chemie, ETH Zürich, Zurich, Switzerland}
\affiliation{International Center for Synchrotron Radiation Innovation Smart, Tohoku University, Sendai, Japan}
\author{Angel Rubio}
\affiliation{Max Planck Institute for the Structure and Dynamics of Matter and Center for Free-Electron Laser Science, Luruper Chaussee 149, 22761 Hamburg, Germany}
\affiliation{Center for Computational Quantum Physics (CCQ), The Flatiron Institute, 162 Fifth Avenue, New York NY 10010, USA}
\author{Nicolas Tancogne-Dejean}
\email[]{nicolas.tancogne-dejean@mpsd.mpg.de}
\affiliation{Max Planck Institute for the Structure and Dynamics of Matter and Center for Free-Electron Laser Science, Luruper Chaussee 149, 22761 Hamburg, Germany}
\author{Hans Jakob Wörner}
\email[]{hwoerner@ethz.ch}
\affiliation{Laboratorium für Physikalische Chemie, ETH Zürich, Zurich, Switzerland}

\maketitle
\twocolumngrid

\section{Molecular dynamics simulations}
The dynamics of pure water at normal density (0.997 g/mL) and room temperature (300 K) were run with a committee model based on high-dimensional neural network potentials (HDNNP)~\cite{PhysRevLett.98.146401} published in Ref.~\cite{o2024pair}. We used the potentials trained on revPBE-D3 data. The water box contained 111 water molecules and was thermalized in the NVT ensemble with a stochastic velocity rescaling thermostat. The path-integral molecular dynamics were run with 32 beads and the PILE-L thermostat~\cite{ceriotti2010efficient}. All dynamics were run with the i-PI code~\cite{litman2024pi} interfaced with the CP2K module for HDNNP potentials~\cite{schran2021machine}. From these simulations, 50 uncorrelated configurational snapshots with classical nuclei and 50 uncorrelated snapshots with quantum nuclei were randomly selected and the HOMO and LUMO orbitals were then calculated from a single-point calculation with the PBE exchange-correlation functional and \textit{tight} species defaults in the FHI-aims code~\cite{blum2009ab}.

Following Ref.~\cite{gong2022attosecond}, we define the spread of the HOMO and LUMO states in terms of the first moment of their density,
\begin{equation}\label{eq:spread}
    M = \frac{\int d\mathbf{r} \rho_i(\mathbf{r}-\mathbf{r_i})|\mathbf{r}-\mathbf{r_i}|}{\int d\mathbf{r} \rho_i(\mathbf{r})}\,,
\end{equation}
where $\mathbf{r_i}$ is the center of mass of the corresponding state.

We obtain from our different simulations:
\begin{table}[ht]
\begin{tabular}{ |c|c|c|c|c|  }
 \hline
 & \multicolumn{2}{|c|}{Quantum nuclei} &  \multicolumn{2}{|c|}{Classical nuclei}\\
 \hline
 & Mean & Std. Error & Mean & Std. Error \\
 \hline
 HOMO  & 1.974  & 0.135 & 2.212 & 0.083 \\
 LUMO  &   9.494 &  0.126 & 9.682 & 0.109\\
 \hline
\end{tabular}
\caption{Values for the first moments (defined in Eq.\ref{eq:spread}) of the HOMO and LUMO orbitals averaged over 50 snapshots of room-temperature simulations with classical and quantum nuclei. Values are in \AA ngstroms.}
\end{table}

These results clearly show that the HOMO states of the various snapshots are well-localized in liquid water, while the LUMO states are fully delocalized, as clearly seen from the snapshots shown in Fig. 4 of the main text.

\section{Analytical cutoff law for nearest-neighbor recombination}

In this section, we derive a cutoff law for the nearest-neighbor recombination channel discussed in the main text (which also holds for the next-nearest-neighbor case). For this purpose, we start from the semiclassical equation of motion in one dimension, assuming a linearly polarized laser
\begin{eqnarray}
 x(t) &=& \frac{qE}{m\omega^2} [\cos(\omega t_i)-\cos(\omega t) - \sin(\omega t_i) \omega(t-t_i)]\,,\\
 v(t) &=& \frac{qE}{m\omega} [\sin(\omega t)-\sin(\omega t_i)] \,,
\end{eqnarray}
where we assumed that the trajectories start at $x(t_i)=0$ with a vanishing velocity ($v(t_i)=0$). Here $E$ is the electric field strength, $\omega$ its frequency, and $q$ and $m$ are, respectively, the charge and mass of the electron.
In the following, as well as in the next section, we assume that the parent ions are not moving on the timescale of one optical cycle. This is justified by the simulation results obtained from real-time TDDFT coupled with Ehrenfest dynamics~\cite{2212.04177} or coupled to non-adiabatic dynamics~\cite{xu2025high}.

In order to evaluate the link between the energy cutoff and the distance of the nearest neighbor, we need to solve the recombination condition $x(t_r) = \pm d_{\rm NN}$, where $d_{\rm NN}$ is the distance between the parent ion and the recombination center.
From the equation of motion, we have the condition linking the ionization time $t_i$ to the recombination time $t_r$
\begin{equation}
  \pm d_{\rm NN} =  \frac{qE}{m\omega^2} [\cos(\omega t_i)-\cos(\omega t_r) - \sin(\omega t_i) \omega(t_r-t_i)]\,.
  \label{eq:tr}
\end{equation}
From the recombination time, we obtain the gain in kinetic energy for a trajectory initiated at $t_i$ and returning at $t_r$
\begin{equation}
 \Delta E_K(t_i) = 2 U_p \Big[ \sin(\omega t_r) - \sin(\omega t_i) \Big]^2\,.
 \label{eq:kin}
\end{equation}
However, compared to the usual case, there are some subtleties. It is clear that here not all trajectories are allowed in the sense that not all initiated trajectories will reach a NN site. In order to determine which trajectories contribute, we first need to determine for a given trajectory what is its maximum excursion distance. \\
The maximum excursion for the first half cycle is reached with the condition $\partial_t x(t) = v(t) = 0$. This gives us a general condition defining the time of maximal excursion $t^*$:
\begin{equation}
 \frac{qE}{m\omega} [\sin(\omega t^*)-\sin(\omega t_i)] = 0,
\end{equation}
which has an analytic solution $\omega t^*=\pi-\omega t_i + 2n\pi$. We therefore have an exact expression for the excursion distance  $l_{1 exc}$ for the first half cycle of a trajectory initiated at a given ionization time:
\begin{equation}
l_{1 exc} = |x(\pi-\omega t_i)| = \frac{qE}{m\omega^2} [2\cos(\omega t_i)-\sin(\omega t_i) (\pi-2\omega t_i)]\,.
\end{equation}

\subsection{Direct NN recombination}
For every trajectory with $l_{1 exc} \ge d_{\rm NN}$, we can have direct recombination to the NN. These trajectories are denoted as direct NN trajectories, and are obtained by solving Eq. (\ref{eq:tr}) and evaluating the corresponding kinetic energy upon recombination.\\
As shown in Fig. S1, only short duration trajectories contribute to the direct NN pathway, and their energy is lower than the gas-phase on-site recombination cutoff of $3.17 U_p$, suggesting that these trajectories are not contributing to the second plateau.

\begin{figure}[h!]
  \begin{center}
    \includegraphics[width=0.9\columnwidth]{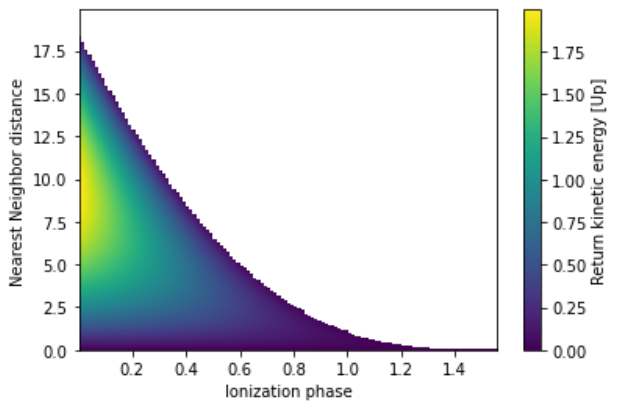}
 \caption{Plot obtained using $\lambda=900$\,nm and $I_0=7\times 10^{13}$W.cm$^{-2}$  for the numerical solution of all possible direct NN trajectories.}
 \end{center}
\end{figure}

\subsection{Indirect NN recombination}
However, it is possible to have another type of NN recombination. The next possibility (in terms of number of missed scattering events) is a recombination to a NN initiated at the second half cycle. For this, we now evaluate the second excursion distance, which is for the next half-cycle and given by $\omega t^{**}=3\pi-\omega t_i$.
We find that:
\begin{eqnarray}
l_{2 exc} = |x(3\pi-\omega t_i)| = l_{exc} + \frac{qE}{m\omega^2} \sin(\omega t_i) 2\pi  \,.
\end{eqnarray}
With this in mind, we consider the next set of trajectories, the indirect NN pathway, in which we have $l_{1 exc} < d_{\rm NN}$ and  $l_{2 exc} \ge d_{\rm NN}$.
This gives the relation shown in Fig. S2. 
\begin{figure}[h!]
  \begin{center}
    \includegraphics[width=0.9\columnwidth]{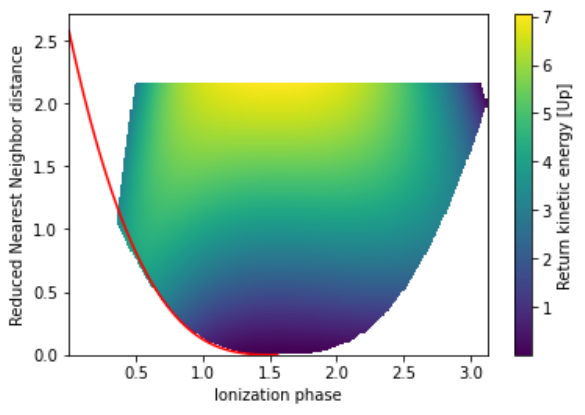}
 \caption{Plot obtained using $\lambda=900$\,nm and $I_0=7\times 10^{13}$W.cm$^{-2}$ and approximation for $f(\omega t_i)$ in red.}
 \end{center}
\end{figure}

From Fig. S2 we see that the cutoff energy is potentially larger than 3.17 $U_p$ (in agreement with the numerical simulations presented in the main text). Especially, the cutoff energy increases with the neighbor distance, as that provides the electron a longer time to classically accelerate under the effect of the laser field. 
Below a reduced distance $d_{NN}^{\rm red} = d_{NN}\frac{m\omega^2}{qE}$ of $\approx 0.8$, the maximum kinetic energy on the edge is given by the condition
\begin{equation}
 f(\omega t_i) = [2\cos(\omega t_i)-\sin(\omega t_i) (\pi-2\omega t_i)]= d_{NN}^{\rm red}.
\end{equation}
Solving this equation gives the ionization time of the most energetic trajectory $t_i^*(d_{NN}^{\rm red})$, from which we obtain the kinetic energy using Eq.~\ref{eq:kin}, based on the return time found by Eq.~\ref{eq:tr}.

In order to obtain an analytical approximate solution to this transcendental equation, we employ a Taylor expansion around $\pi/2$,
\begin{equation}
 \lim_{x\to\frac{\pi}{2}}f(x) = -\frac{2}{3} (x-\frac{\pi}{2})^3 + \mathcal{O}(x^3)\,, 
\end{equation}
yielding an approximate expression for $f$ for values $x<\pi/2$:
\begin{equation}
 f(x) \approx \frac{2}{3} (\frac{\pi}{2}-x)^3\,.
\end{equation}
While this overshoots the expression in the limit of vanishing ionization time (where the limit is $2-\pi \omega t_i + \mathcal(O)((\omega t_i)^2)$), for the region of interest (i.e. $d_{NN}^{\rm red}<0.8$), the Taylor expansion provides a very good approximation for the recombining trajectories and kinetic energies.
This result is shown in Fig. S2 for the comparison to the fully numerical solution discussed in the main text.

Lastly, we can obtain an approximate link between the ionization time for the most energetic indirect NN recombination and the reduced distance:
\begin{equation}
 \omega t_i^* \approx \frac{\pi}{2} - \Big(\frac{3}{2}d_{NN}^{\rm red}\Big)^{1/3}.
\end{equation}

We can now numerically compute the returning cutoff energy along the red line and plot it as a function of the reduced NN distance (see Fig. S3).

\begin{figure}[h!]
  \begin{center}
    \includegraphics[width=0.9\columnwidth]{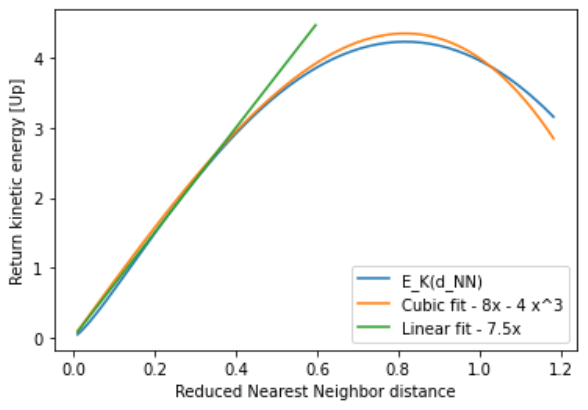}
 \caption{Returning kinetic energy as a function of $d_{NN}^{\rm red}$ along the critical path, together with a linear and cubic fit}
 \end{center}
\end{figure}

With the fit obtained, we write the approximate cutoff law
\begin{equation}
 \Delta E_K(d_{NN}) \approx U_p (8  d_{NN}^{\rm red} -4 (d_{NN}^{\rm red})^3) =  2 q E d_{NN} -  \frac{m^2\omega^4d_{NN}^3}{qE}\,.
\end{equation}
We found that the cutoff does not depend on the wavelength to first order, and that is linearly depends on the electric-field strength, in agreement with the measurements and the \textit{ab-initio} simulations discussed in the main text.

Employing similar conditions to Fig. 3 in the main text ($\lambda=900$\,nm, $I_0=7\times 10^{13}$W.cm$^{-2}$, $d_{NN}=5.5\AA$), with the analytical approximation we find $d_{NN}^{\rm red}\approx0.6$. In this case, we find from the linear fit that the cutoff is $\sim 4.5 U_p$ and from the cubic fit that it is $\sim 3.94 U_p$.

\section{Extended Lewenstein-like model}

We now investigate the impact of hole delocalization on liquid HHG in the context of elliptical driving fields used in the main text. For this, we need to compute the harmonic yield vs. ellipticity. We follow the work of Lewenstein et al.~\cite{PhysRevA.49.2117} and extend it to more complex ground-state wavefunctions that incorporate multiple possible ionization and recombination sites.
Based on the molecular-dynamics results, we can conclude that (i) the HOMO state is delocalized over a few molecules, (ii) the LUMO state is fully delocalized over the entire liquid sample. This motivates employing the same approximation as typically used in the gas phase, assuming in particular that the electron evolves as a Volkov state.
Subsequently, we can employ the stationary phase method to compute the integration over the 
initial momentum $\mathbf{p}$ of the ionized electron, and obtain the usual expression
\begin{widetext}
\begin{equation}
 \mathbf{d}(t) = \frac{1}{\sqrt{i}} \int_{-\infty}^t dt' \mathbf{E}(t') \mathbf{d}(\mathbf{p}(t,t')-q\mathbf{A}(t')) \frac{(2\pi)^{3/2}}{(t-t'-i\epsilon)^{3/2}}
 e^{-iS(\mathbf{p},t,t')} \mathbf{d}^*(\mathbf{p}(t,t')-q\mathbf{A}(t))+ c.c.\,,
 \label{eq:dipole_lewenstein}
\end{equation}
with the stationary quasi-classical action
\begin{equation}
 S(\mathbf{p},t,t') = \int_{t_i}^t dt' \Big(\frac{1}{2}[\mathbf{p}(t,t_i)-\mathbf{A}(t')]^2 + I_p \Big)\,. 
\end{equation}
\end{widetext}
This model can be solved numerically, provided that we readily have proper dipole matrix elements $\mathbf{d}$. 
These are usually taken from analytic expressions for a Gaussian or hydrogenic model of atomic-like orbitals. Here, we use a different model that allows on-site recombination, as well as off-site.

We consider a Gaussian $s$-orbital-like wavefunction of the form
\begin{equation}
 \Psi(\mathbf{r}) = \left(\frac{\alpha}{\pi}\right)^{3/4} e^{-\alpha (\mathbf{r}-\mathbf{r_0})^2/2}\,,
\end{equation}
with a parameter $\alpha$ that determines the width of the Gaussian centered around $\mathbf{r_0}$. The dipole matrix element is obtained by assuming transitions to and from plane waves, which is a typical approximation and is valid at higher energies. After some algebra, we obtain:
\begin{equation}
 \mathbf{d}(\mathbf{p}) 
 = i\left(\frac{1}{\pi\alpha}\right)^{3/4} e^{-i\mathbf{p}\cdot\mathbf{r_0}} e^{-\frac{\mathbf{p}^2}{2\alpha}} \Big[  \mathbf{r_0} + \frac{\mathbf{p}}{\alpha}  \Big]\,, 
\end{equation}
which reduces to the usual gas-phase SFA result if we take the center of the wavefunctions at the origin.

We now assume that the bound state, here the HOMO of the liquid, is composed of multiple Gaussians, with different weights and spreads, still using normalized wavefunctions. More precisely, we assume one main central Gaussian at the origin, and $N$ surrounding identical centers with a lesser weight and a smaller spread, all located at the same distance, mimicking a liquid solvation shell.
Denoting the relative weight $\beta$ and the relative spread $\sigma$, this corresponds to a wavefunction of the form
\begin{equation}
  \Psi(\mathbf{r}) = w_0\Big[ e^{-\alpha \mathbf{r}^2/2} + \beta \sum_{i=1}^N e^{-\alpha\sigma (\mathbf{r}-\mathbf{r_i})^2/2}\Big]\,,
  \label{eq:multigaussian_wavefunction}
\end{equation}
where $w_0$ is an overall normalization factor.

From the previous result of the shifted Gaussian model, and using the linearity of the integral, we can express the dipole elements for this model wavefunction as
\begin{eqnarray}
\mathbf{d}(\mathbf{p}) &=& iw_0\left(\frac{1}{\pi\alpha}\right)^{3/4} \Bigg( e^{-\frac{\mathbf{p}^2}{2\alpha}}\frac{\mathbf{p}}{\alpha} \nonumber\\
&&+  \frac{\beta}{\sigma^{3/4}} e^{-\frac{\mathbf{p}^2}{2\alpha\sigma}}\sum_{i=1}^Ne^{-i\mathbf{p}\cdot\mathbf{r_i}}\Big[  \mathbf{r_i} + \frac{\mathbf{p}}{\alpha\sigma}  \Big]\Bigg)\,.
\end{eqnarray}

This expression can now be used to perform SFA simulations with the delocalized HOMO state.
In order to determine the value of $\alpha$, we impose that the first moment of the central Gaussian is the same as that of a hydrogenic model for which the width is given by $I_p$, which we take as 10eV to mimic liquid water.
The values of $N$, $\beta$ and $\sigma$ are given below.
\begin{figure}[h!]
  \begin{center}
    \includegraphics[width=0.9\columnwidth]{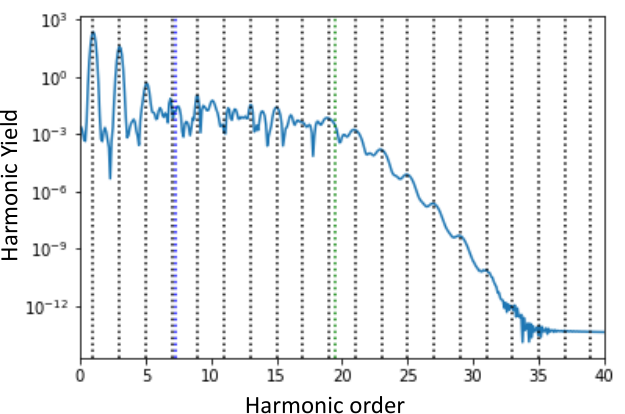}
 \caption{HHG spectrum for the single Gaussian model.}
 \end{center}
\end{figure}
Let us first compute the ellipticity dependence of a single Gaussian center. The width of the Gaussian is taken such that it gives the same first moment as a hydrogenic model, with an ionization potential of 10 eV.
We choose here $\lambda=900$\,nm, an intensity of $I_0 = 7\times10^{13}$ W.cm$^{-2}$, and a 6 cycle laser pulse.
The HHG is computed here numerically, from the Fourier transform of the time-dependent dipole obtained from Eq.~(\ref{eq:dipole_lewenstein}).\\
The HHG spectrum for the linearly polarized light is shown in Fig. S4.

Varying the ellipticity, we obtain the result shown in Fig. S5.
\begin{figure}[h!]
  \begin{center}
    \includegraphics[width=0.9\columnwidth]{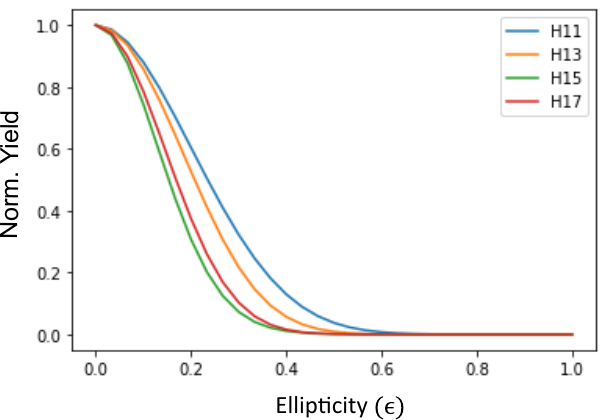}
 \caption{Ellipticity dependence of the harmonic yield for a single-Gaussian initial state}
 \end{center}
\end{figure}

We now perform calculations for the multi-Gaussian model.
We place here 3 surrounding Gaussian at a distance of 5.5 \AA~around the central Gaussian, corresponding to the second solvation shell of liquid water, with $\beta = 0.01$, and $\sigma = 0.025$, which means a width of the surrounding Gaussian of 1\% compared to the central one, and a FWHM divided by approximately 6.3.
The positions of the Gaussian are randomly selected positions on a circle, and the time-dependent dipole is obtained from averaging over 100 random configurations of the neighbors, in order to recover the isotropic limit that is otherwise lost for an individual multi-Gaussian wavefunction of Eq.~\ref{eq:multigaussian_wavefunction}.
Figure S6 shows that the delocalization of the hole leads to side peaks, as well as a broadening of the Gaussian profile of the ellipticity dependence. This result thus mimicks the experimentally observed data in Fig. 4 in the main text, as well as \textit{ab-initio} simulations, and is in-line with our intuition for NN and NNN trajectories. We note that this model does not include electron scattering and mean-free path effects, as we assumed here a free propagation in vacuum, and hence cannot lead to a multiple plateau structure as observed in the experiment and extensive simulations.

\begin{figure}[h!]
  \begin{center}
    \includegraphics[width=0.9\columnwidth]{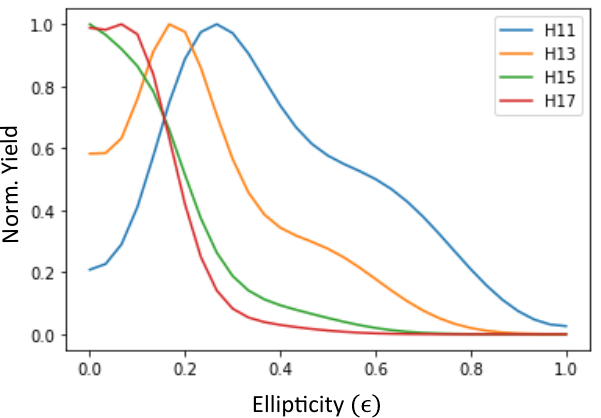}
 \caption{Ellipticity dependence of the harmonic yield for a multi-Gaussian initial state}
 \end{center}
\end{figure}

\section{Time-frequency analysis for liquid ammonia}

We further performed simulations of HHG from liquid ammonia, and performed a time-frequency analysis shown in Fig. S7, similar to what we present in the main text (Fig. 3) for liquid water and liquid methane. This result complements the data given in the main text and validates the generality of the physical mechanisms involved. 

\begin{figure} [h!]
\begin{center}
    \includegraphics[width=\columnwidth]{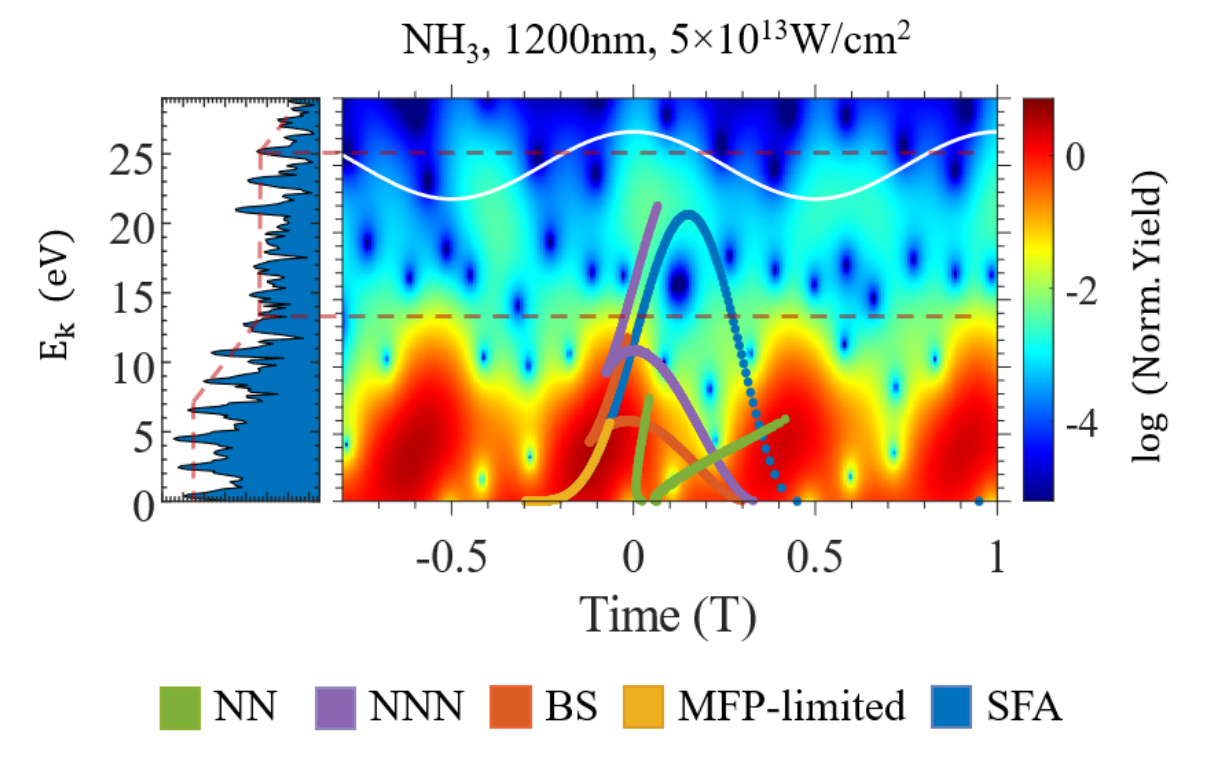}
    \caption{Time-frequency analysis showing the showing the temporal dependence of harmonic energies for emission from liquid ammonia. Predictions from various semi-classical trajectory simulations described in the main text are superimposed, as in Fig. 3 of the main text.}
\end{center}
\end{figure}

\section{Wavelength and Intensity dependence of second cut-off}
We demonstrate the experimental intensity independence of the second cut-off at multiple wavelengths (1200 nm, 1500 nm, and 1800 nm) for liquid D$_2$O in Fig. S8 (a-c). In all cases, the maximum intensity is capped below the plasma generation threshold. Figure S8 (d), shows the intensity independence for 900 nm H$_2$O from ab initio simulations. The results further confirm that the second plateau feature persists across the entire parameter range, and that the position of the second cut-off remains independent of the laser wavelength and only weakly dependent on the driving intensity, reinforcing the robustness and universality of this feature across a broad range of excitation conditions in both experiment and theory.

\begin{figure*}
\begin{center}
    \includegraphics[width=\textwidth]{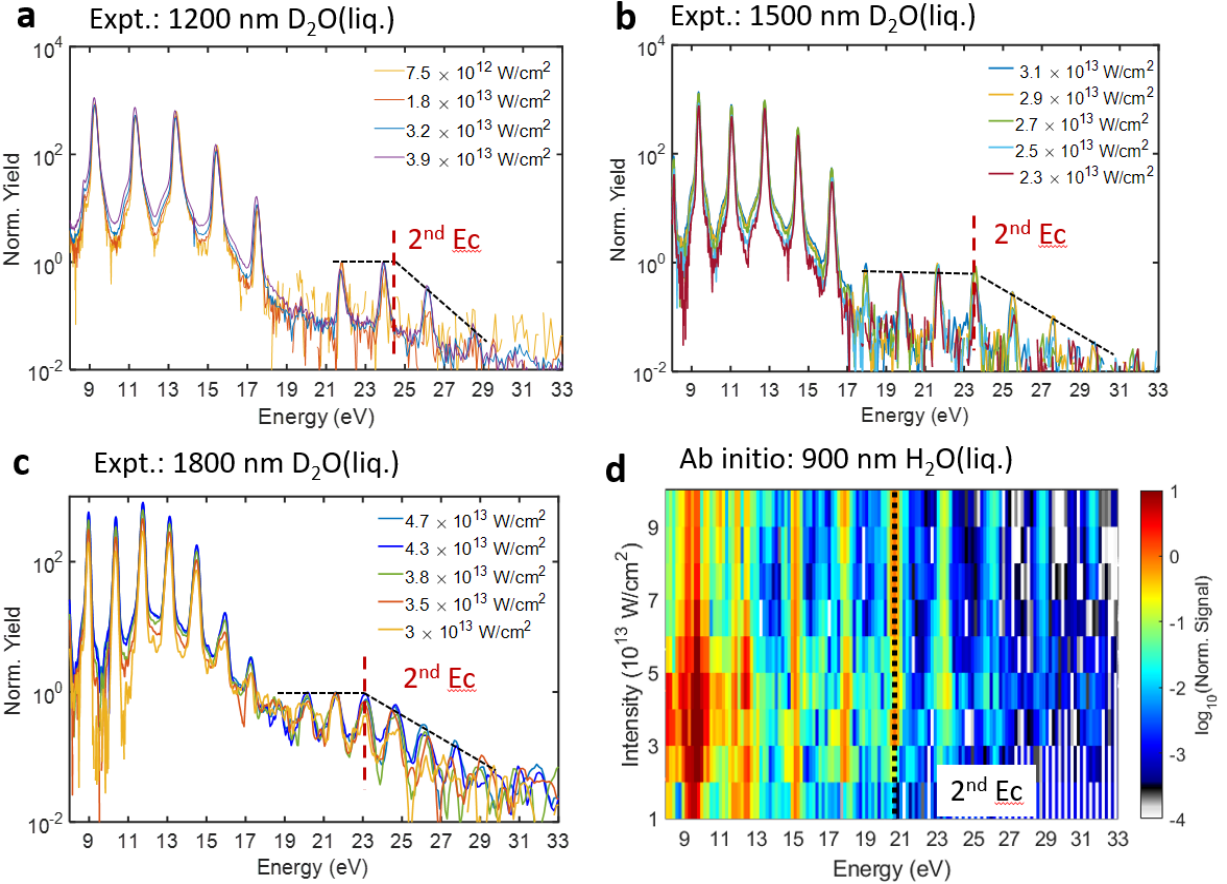}
    \caption{ Normalized HHG spectra measured in liquid D$_2$O at varying intensities for three different driving wavelengths: \textbf{a} 1200 nm, \textbf{b} 1500 nm, and \textbf{c} 1800 nm. Despite changes in intensity, the position of the second plateau cut-off (2nd Ec), marked by the red dashed line, remains essentially unchanged across all cases. This confirms the weak intensity dependence of the second cut-off energy. The black dashed lines are included to guide the eye through the second plateau and post-cutoff regions. \textbf{d} Ab initio simulation of HHG in liquid H$_2$O at 900 nm, showing the log-scaled normalized harmonic yield as a function laser intensity. The 2nd Ec is indicated by the black dashed line, beyond which the signal drops sharply across all intensities.} \end{center}
\end{figure*}

\section{Experimental and ab initio ellipticity dependence of harmonics in liquid water}
The appearance of multiple Gaussian in the ellipticity dependence of the harmonic yield is observed both experimentally and theoretically in ab initio cluster simulations. For both cases (Figure S9-10), the typical gas phase single gaussian behavior of the ellipticity dependence of the harmonic yield describes the first plateau harmonics appropriately, However, we see clear appearance of multiple gaussians (beyond the single guassian feature) for second plateau harmonics.
A comparison of the FWHM dependence on the harmonic energy of the liquid, derived from the single gaussian fits (Figs. S9-10), demonstrate clear deviation in behavior for the second plateau harmonics in liquids from those of their first plateau and their respective gas phasees.
\begin{figure*} [t!]
\begin{center}
    \includegraphics[width =\textwidth]{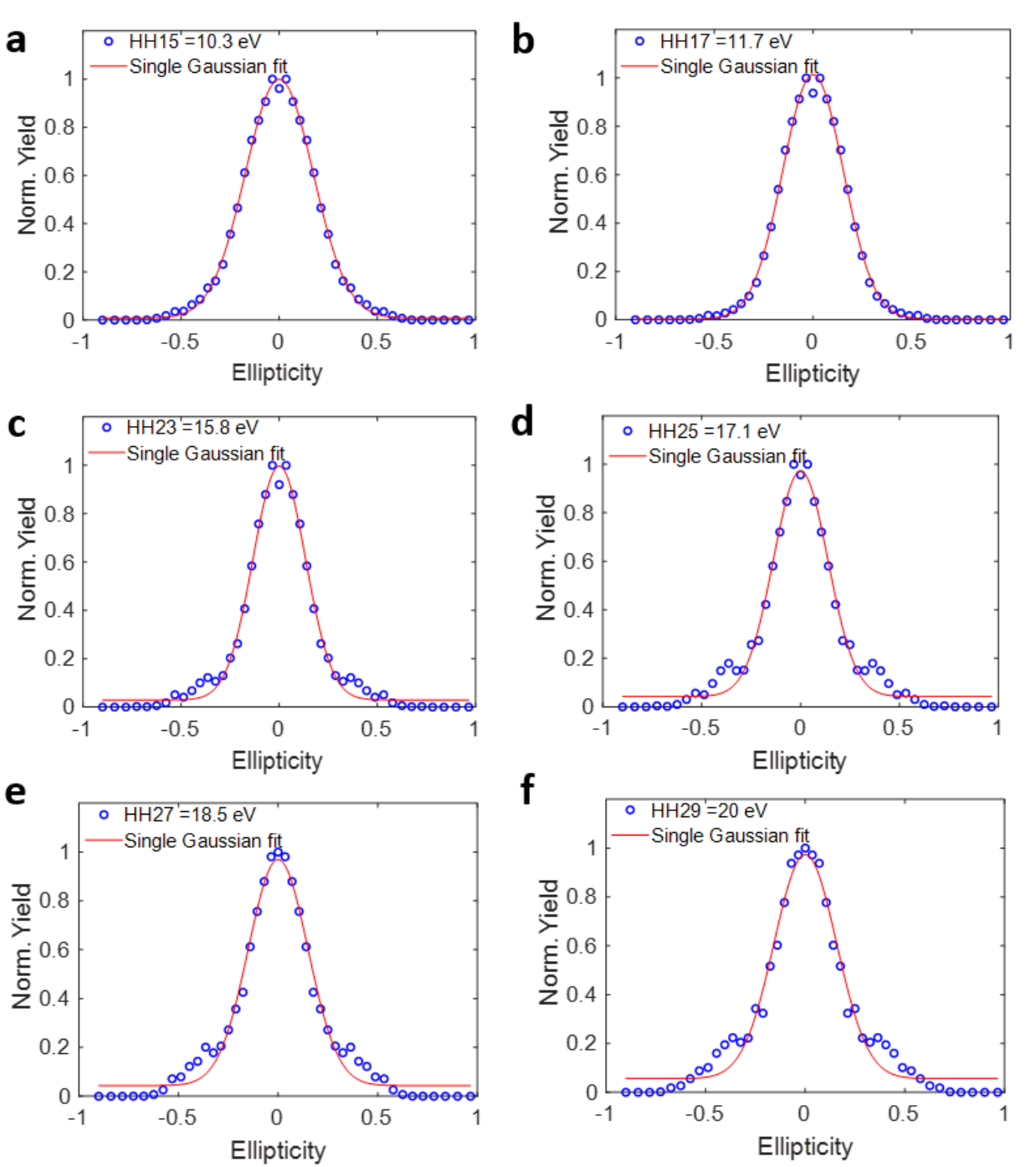}
    \caption{Experimentally observed ellipticity dependence of harmonics for liquid ethanol using 1800 nm wavelength and their single gaussian fits.}
\end{center}
\end{figure*}

\begin{figure*} [t!]
\begin{center}
\includegraphics[width =\textwidth]{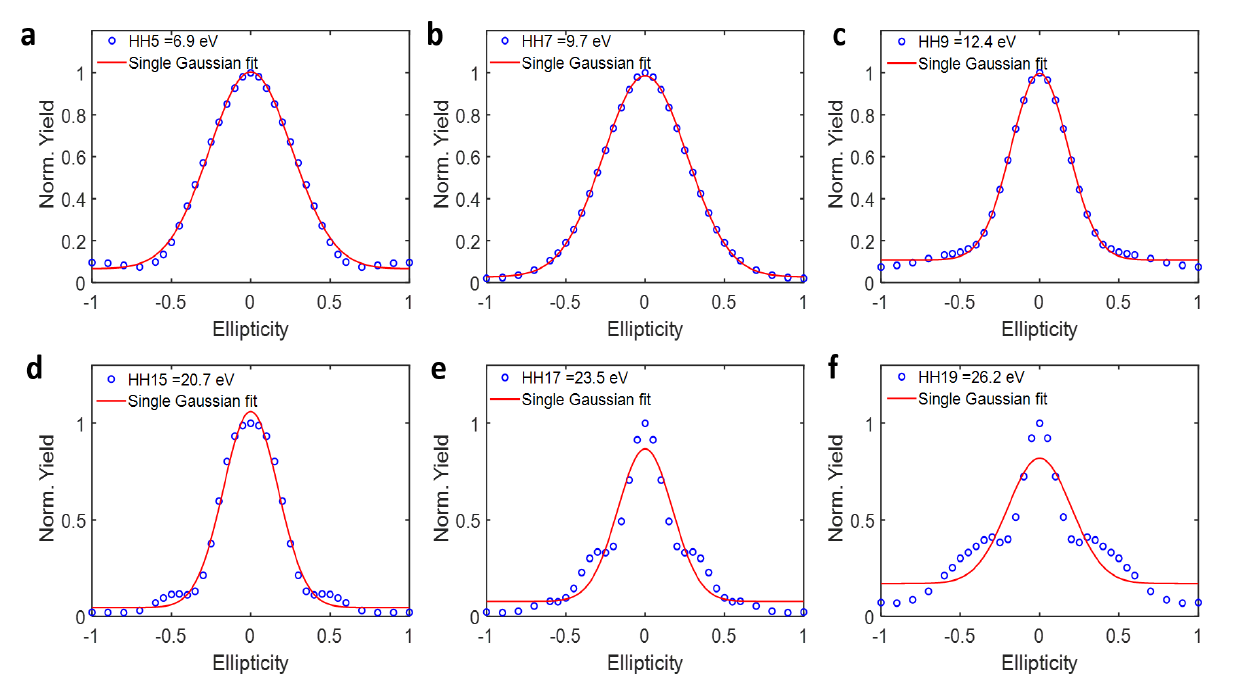}
    \caption{Ellipticity dependence of harmonics obtained from ab initio simulations for liquid water at 900 nm and their single gaussian fits.}
\end{center}
\end{figure*}

\begin{figure*} [t!]
\begin{center}
\includegraphics[width =\textwidth]{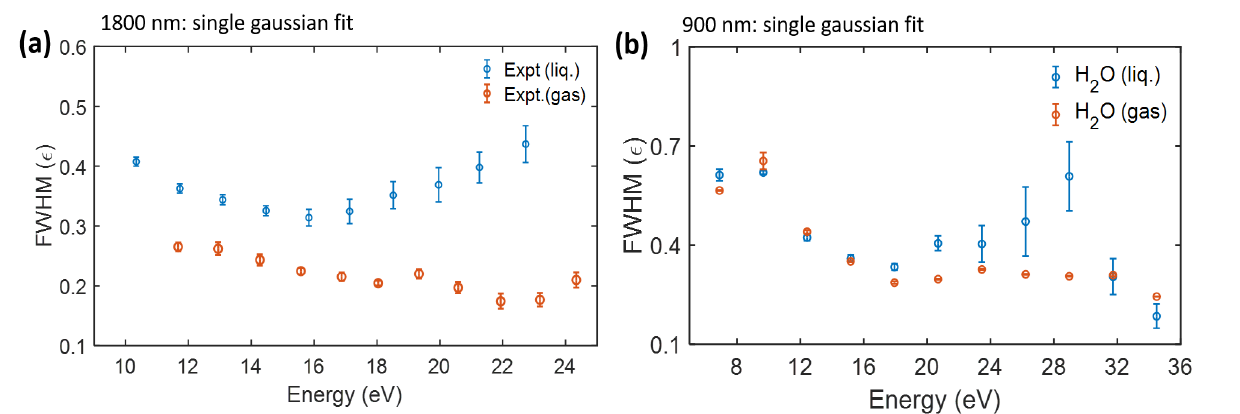}
    \caption{Ellipticity-dependent of FWHM of harmonic yields from single-Gaussian fits. (a) Experimental results at 1800 nm comparing gas-phase (blue) and liquid-phase (orange) ethanol. (b) Ab initio simulation results at 900 nm for H2O in gas (orange) and liquid (blue) phases.}
\end{center}
\end{figure*}

\section{Methodology for determination of second cut-off energy}
To accurately determine the second plateau cut-off energy (2$^{nd}$ E$_c$) in liquids, we apply two complementary methods:
\begin{enumerate}
    \item A derivative-based approach, and
    \item A piecewise linear fit intersection approach.
\end{enumerate}
This dual strategy accounts for variations in plateau shapes across different liquids and ensures consistency in identifying the cut-off even when the plateau is diffuse or compressed. In gas-phase HHG, a cut-off is conventionally defined as the transition between a flat plateau and an exponential decay. However, in condensed-phase systems, especially solids and liquids, the concept of a “plateau” becomes more nuanced. Strong scattering, local disorder, and sample-dependent properties often result in plateau regions that exhibit shallow decay rather than true flatness.

\noindent For example, in solid noble gases like argon and krypton, the extent and shape of the second plateau vary with driving intensity and sample morphology \cite{Ndabashimiye2016}. In $\alpha$-quartz, a narrow and flat second plateau exists (21–25 eV), whereas crystalline quartz displays a more complex “double hump” structure where a precise second cut-off is ambiguous \cite{luu18b}. In polycrystalline and amorphous solids, this second plateau is often absent altogether.

\noindent Liquids, being structurally disordered like amorphous solids, face similar challenges. Variations in density, molecular orientation, and scattering make the plateau features highly sample-dependent. Accordingly, our methodology focuses not only on absolute yield levels but on changes in slope behavior to identify the second plateau and its termination. To illustrate the determination of the second plateau cut off energy two limiting cases of second plateau characteristics, we refer to two representative examples.
\subsection{Liquid D$_2$O (Also representative of Water)}
\noindent For D$_2$O, a clear second plateau is observed between 18–23 eV, followed by an exponential decay. In the derivative-based approach, the logarithm of the harmonic yield is interpolated with a 0.25 eV step size. The first derivative (Fig. S11 b and e) reveals a steep slope in the first plateau ($<$18 eV), a flattened region in the second plateau (18–22 eV), and two local minimum at 22.5 eV and $\sim$ 24.4 eV, respectively. To consistently identify the second cut-off energy, we define it as the deeper minimum in the derivative curve that occurs after the maximum slope value (i.e., after the shallow region). This is indicated by the red dashed lines in Figs. S11 a, b, d and e. \\
\noindent In the piecewise linear fit approach, the intersection of the two regions: (1) the average of the second plateau yield (purple dashed lines in Fig. S12 c and f), and (2) the linear fit of post-plateau decay (green dashed lines in Fig. S12 c and f),  defines the second cut-off (black dashed lines in Fig. S12 a, c, d, and f). 

\noindent Across 1800 nm and 1500 nm driving wavelengths, both methods yield consistent cut-off energies (22.5–24.4 eV). The final value is reported, as the average of the four cut-off energy values, at 23.5 $\pm$ one harmonic order, accounting for resolution and spectral discreteness.

\begin{figure*} [t!]
\begin{center}
\includegraphics[width =\textwidth]{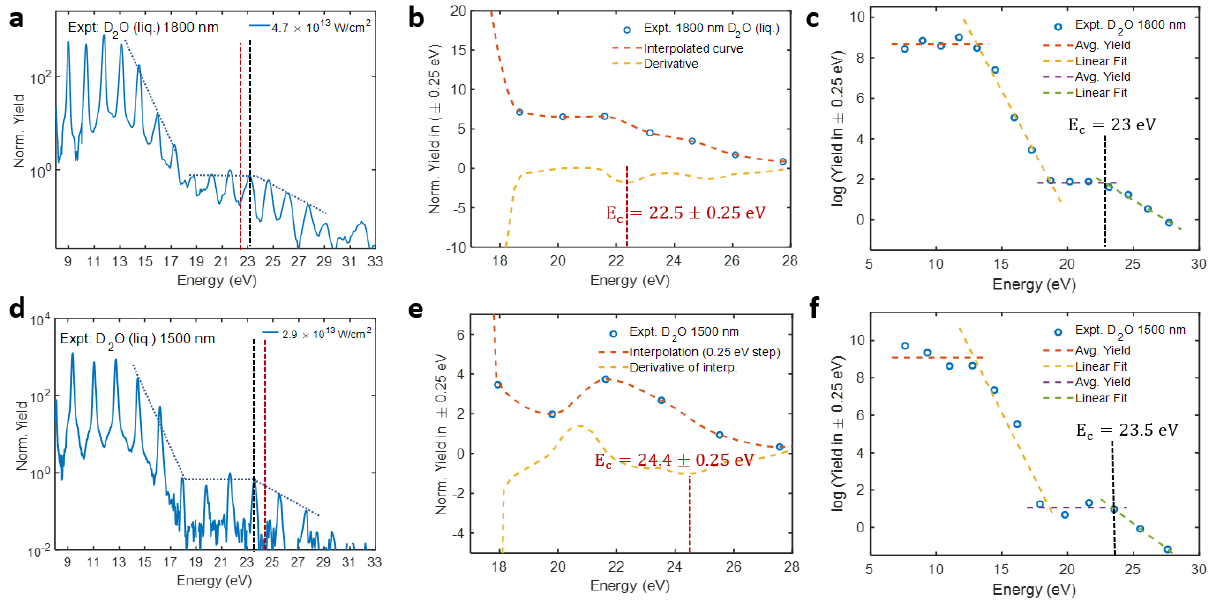}
    \caption{Determination of the second plateau cut-off energy for liquid D2O at 1800 nm and 1500 nm. \textbf{a} Experimental HHG spectrum of liquid D2O at 1800 nm, as also shown in Extended Data Fig. 1 a. The red dashed line marks the second cut-off energy determined using the derivative-based method, while the black dashed line marks the cut-off determined via the piecewise linear intersection method. \textbf{b} Derivative-based analysis: blue circles indicate the interpolated log(harmonic yield) at 0.25 eV intervals, the red dashed line shows the interpolated curve, and the yellow dashed line represents the first derivative. The deeper minimum following the initial maximum in the derivative curve is identified as the second cut-off energy (E$_c$ = 22.5 $\pm$ 0.25 eV). \textbf{c} Piecewise linear fit analysis: blue circles denote the same interpolated log(yield) points as in (b); the purple and green dashed lines represent linear fits to the first and second decay regions, respectively. Their intersection defines the second cut-off energy (E$_c$ = 23 eV). (\textbf{d–f)} Same as panels \textbf{a-c} , but for the 1500 nm driving wavelength (Extended Data Fig. 2 a). The cut-off energies obtained are E$_c$ = 24.4 $\pm$ 0.25 eV from the derivative method \textbf{e}, and E$_c$ = 23.5 eV from the linear intersection method \textbf{f}.}
\end{center}
\end{figure*} 
\subsection{Liquid Ethanol (Representative of Alcohols)}
\noindent In ethanol, a second plateau with constant yield is not clearly observable. Instead, the harmonic spectrum shows a transition in slope—from a steep decay after the first cut-off ($<$15 eV) to a more gradual decay beyond 16 eV (as shown in Fig. S13 a and d). In the derivative method, the global minimum following the first maximum in the slope is taken as the second cut-off, yielding 17.6–17.9 eV across 1800 nm (red dashed line in Fig. S12 b) and 1500 nm (red dashed in Fig. S13 b). In the linear fit method, the second cut-off is defined by the intersection of two exponential decay slopes—one before and one after $\sim$ 16 eV. The resulting second cut-off energies are 16.7 eV for 1800 nm and 18.8 eV for 1500 nm, as shown by that black dashed lines in Fig. S13(c) and Fig. S13(f), respectively.\\
\noindent The second cut-off energy for ethanol is therefore identified, corresponding to the average of the four values,  as 17.9 eV $\pm$ one harmonic order. This methodology was applied consistently across additional liquids such as H$_2$O and 2-propanol, and the cut-off energies were determined to be 23.4 eV $\pm$ one visible harmonic order and 17.0 eV $\pm$ one visible harmonic order, respectively. In all cases, we emphasize that assigning a cut-off energy with sub-harmonic resolution would be scientifically unjustified, as it exceeds both the experimental resolution and the discrete nature of the HHG spectrum. Thus, all reported cut-off values carry an uncertainty of $\pm$ one visible harmonic order, ensuring both rigor and physical relevance.\\
\begin{figure*} [t!]
\begin{center}
\includegraphics[width =\textwidth]{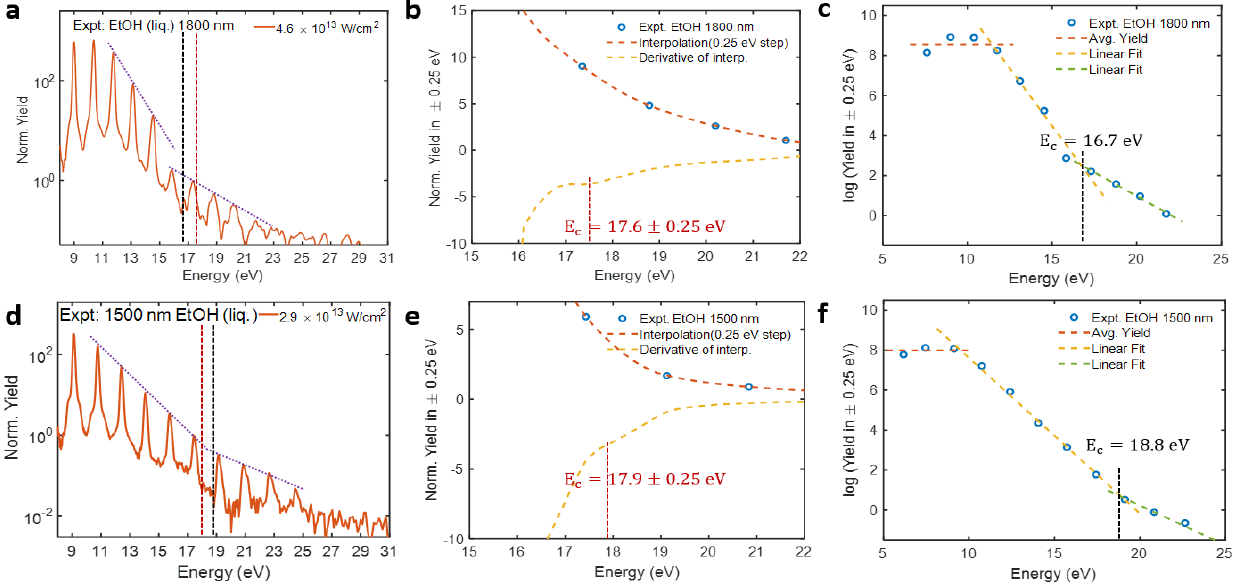}
    \caption{ Determination of the second plateau cut-off energy for liquid ethanol at 1800 nm and 1500 nm. \textbf{a} Experimental HHG spectrum of liquid ethanol at 1800 nm, as also shown in Extended Data Fig. 1 c. The red dashed line marks the second cut-off energy determined using the derivative-based method, while the black dashed line marks the cut-off determined via the piecewise linear intersection method. \textbf{b} Derivative-based analysis: blue circles indicate the interpolated log(harmonic yield) at 0.25 eV intervals, the red dashed line shows the interpolated curve, and the yellow dashed line represents the first derivative. The global minimum following the initial maximum in the derivative curve is identified as the second cut-off energy (E$_c$ = 17.6 $\pm$ 0.25 eV). (c) Piecewise linear fit analysis: blue circles denote the same interpolated log (yield) points as in \textbf{b}; the yellow and green dashed lines represent linear fits to the first and second decay regions, respectively. Their intersection defines the second cut-off energy (E$_c$ = 16.7 eV). \textbf{(d–f)} Same as panels \textbf{(a–c)} , but for the 1500 nm driving wavelength (Extended Data Fig. 2c). The cut-off energies obtained are E$_c$ = 17.9 $\pm$ 0.25 eV from the derivative method \textbf{e}, and E$_c$ = 18.8 eV from the linear intersection method (f).}
\end{center}
\end{figure*}

\bibliography{ref}